\documentclass[fleqn,twoside]{report}

\usepackage{longtable}
\usepackage{amsmath}
\usepackage{fancyhdr}
\usepackage[T1]{fontenc}
\usepackage[latin1]{inputenc}
\usepackage{array}
\usepackage[format=hang, indention=-2cm, font=small,labelfont=bf]{caption}
\usepackage{wasysym}
\usepackage{amsfonts}
\usepackage{amssymb}
\usepackage{hyperref}
\usepackage{xspace}
\usepackage[mathscr]{eucal}
\usepackage{graphics}
\usepackage{graphicx}
\usepackage{listings}

\fancypagestyle{plain} 
{
    \fancyhf{}

    \fancyfoot[RO,LE]{\thepage} 
}

        \renewcommand{\footnoterule}{\rule{0pt}{0pt}\vspace{0pt}} %Fu?zeilenlinie wegmachen

\newcommand\La{\mathscr{L}}
\newcommand\bra{\langle}
\newcommand\ket{\rangle}
\newcommand\Y{\mathscr{Y}}
\newcommand{\DRbar}{{\ensuremath{\overline{\mathrm{DR}}}}}
\newcommand{\MSbar}{{\ensuremath{\overline{\mathrm{MS}}}}}
\newcommand{\ttt}[1]{\texttt{#1}}

\newcommand{\mrm}[1]{\ensuremath{\mathrm{#1}}}

\newcommand{\e}{\mathrm{e}}

\newcommand{\p}{\mathrm{p}}

\newcommand{\Z}{{Z}}

\newcommand{\GeV}{\mathrm{GeV}}

\newcommand{\numentry}[2]{
\begin{minipage}[t]{1.3cm}\flushright\ttt{#1}\end{minipage}
\hspace{5mm}: 
\begin{minipage}[t]{13cm}\noindent
#2\end{minipage}\\[2mm]
}
\newcommand{\snumentry}[2]{
\begin{minipage}[t]{1.2cm}\flushright\ttt{#1}\end{minipage}
\hspace{2mm}:
\begin{minipage}[t]{11cm}\noindent
#2\end{minipage}\\[1mm]
}

\newcommand\SARAH{{\tt SARAH}\xspace}
\newcommand\HB{{\tt HiggsBounds}\xspace}
\newcommand\FeynArts{{\tt FeynArts}\xspace}
\newcommand\FormCalc{{\tt FormCalc}\xspace}
\newcommand\CalcHep{{\tt CalcHep}\xspace}
\newcommand\CompHep{{\tt CompHep}\xspace}
\newcommand\SPheno{{\tt SPheno}\xspace}
\newcommand\WHIZARD{{\tt WHIZARD}\xspace}
\newcommand\Mathematica{{\tt Mathematica}\xspace}
\newcommand\OMEGA{{\tt Omega}\xspace}
\newcommand\micrOmegas{{\tt micrOMEGAs}\xspace}
\newcommand\UFO{{\tt UFO}\xspace}
\newcommand\Madgraph{{\tt Madgraph}\xspace}
\newcommand\DR{{\overline{\text{DR}}}}

\makeatletter
\def\thickhrulefill{\leavevmode \leaders \hrule height 1pt\hfill \kern \z@}
\renewcommand{\maketitle}{\begin{titlepage}%
    \let\footnotesize\small
    \let\footnoterule\relax
    \parindent \z@
    \reset@font
    \null\vfil
    \vspace{-3cm}
    \begin{flushleft}       \vspace{2cm}       \huge \@title      \end{flushleft}
    \par
    \hrule height 1pt
    \par
    \begin{flushright}       \large \@author \par    \end{flushright}
    \begin{center}
    \vskip 1.cm
    \Large Version 3.2 \\
    \vskip 14.0cm
    {\Large \@date\par}%
    \end{center}
    \vfil\null
  \end{titlepage}%
   \setcounter{footnote}{0}%
}

\title{SARAH}
\author{A Model Builder's Tool}
\date{
Florian Staub,\\
fnstaub@th.physik.uni-bonn.de
\\
\vskip 0.5cm
\today
 }

\begin{document}

%\vspace{-9cm}
%\hspace{-0.5cm} 
%\parbox{17.5cm}{ \maketitle}
%\newpage
\maketitle

{\bf Disclaimer} \\
\verb"SARAH" is published under the GNU library public license \footnote{http://www.fsf.org/copyleft/lgpl.html}, this means that you can use it for free. We have tested this software and its result, but we can't guarantee that this software works correctly or that the physical results derived using this software are correct. \\ 
 \\
\vspace{0.5cm}
We have the following requests:
\begin{itemize}
\item If you find any bug, please inform us by eMail: fnstaub@physik.uni-bonn.de
\item If you have any suggestions, what is missing or can be improved, please let us know.
\item If you use \SARAH, please cite the appropriate references: % \verb"arXiv:0806.0538", \verb"arXiv:0909.2863", \verb"arXiv:1002.0840"
\begin{itemize}
%\cite{Staub:2012pb}
\item%{Staub:2012pb}
{\bf ``Linking SARAH and MadGraph using the UFO format''}
  \\{}F.~Staub.
  \\{}arXiv:1207.0906 [hep-ph]
 %(Jul 2012)
%\href{http://inspirehep.net/record/1121136}{HEP entry}
%\cite{Staub:2011dp}
\item%{Staub:2011dp}
{\bf ``A Tool Box for Implementing Supersymmetric Models''}
  \\{}F.~Staub, T.~Ohl, W.~Porod and C.~Speckner.
  \\{}arXiv:1109.5147 [hep-ph]
\\{}Comput.\ Phys.\ Commun.\  {\bf 183}, 2165 (2012) %(Sep 2011)
%\href{http://inspirehep.net/record/928307}{HEP entry}
%\cite{Staub:2010ty}
\item%{Staub:2010ty}
{\bf ``The Electroweak sector of the NMSSM at the one-loop level''}
  \\{}F.~Staub, W.~Porod and B.~Herrmann.
  \\{}arXiv:1007.4049 [hep-ph]
\\{}JHEP {\bf 1010}, 040 (2010) %(Jul 2010)
%\href{http://inspirehep.net/record/862413}{HEP entry}
%\cite{Staub:2010jh}
\item%{Staub:2010jh}
{\bf ``Automatic Calculation of supersymmetric Renormalization Group Equations and Self Energies''}
  \\{}F.~Staub.
  \\{}arXiv:1002.0840 [hep-ph]
\\{}Comput.\ Phys.\ Commun.\  {\bf 182}, 808 (2011) %(Feb 2010)
%\href{http://inspirehep.net/record/845241}{HEP entry}
%\cite{Staub:2008uz}
\item%{Staub:2008uz}
{\bf ``Sarah''}
  \\{}F.~Staub.
  \\{}arXiv:0806.0538 [hep-ph]
 %(Jun 2008)
%\href{http://inspirehep.net/record/787204}{HEP entry}
\end{itemize}
\end{itemize}

\newpage
\newpage

\tableofcontents

\chapter{Introduction}
Supersymmetry (SUSY) is one of the most popular extensions of the standard model
(SM) of particle physics \cite{susy1,susy2,susy3,susy4,susy5,susy6}: it solves
the hierarchy problem \cite{hierarchie1,hierarchie2}, leads to unification  of
the three gauge couplings \cite{uni1,uni2,uni3,uni4} and offers often a
candidate for dark matter \cite{dm}. \\
The minimal supersymmetric standard model (MSSM) is nowadays well studied. Every
event generator or diagram calculator can handle the MSSM out of the box.
Unfortunately, there remains a lot of work if somebody wants to change the
supersymmetric model, e.g. extend the particle content, add new gauge groups or
add new interactions to the superpotential. First, it must be checked that the
new model is free from gauge anomalies. As second step, the full Lagrangian must
be derived and all interactions have to be extracted. This is complicated by the
fact that the fields in gauge eigenstates have to be rotated to new mass
eigenstates: the rotations must be incorporated, mass matrices have to be
calculated and diagonalized. Finally, the tadpole equations are needed to find
the minimum of the potential. All these steps are needed just to get a rough
impression of the new model. If also phenomenological studies should be made by
using one of the existing programs, model files must be created. Moreover, for
the embedding of the model in a GUT theory, the Renormalization Group Equations
(RGEs) are needed. Furthermore, often loop corrections to the masses are
demanded.  \\
This is exactly that kind of work \SARAH was written for. \SARAH just needs the
gauge structure, particle content and superpotential to produce all information
about the gauge eigenstates. As gauge groups, all \(SU(N)\) groups can be
handled and the superfields can transform as any arbitrary, irreducible
representation of these groups. 
Breaking of gauge symmetries and mixings of particles can easily be added. Also
the gauge fixing are automatically derived, and the corresponding ghost
interactions are calculated. The two-Loop RGEs for the
superpotential parameters, the gauge couplings and the soft-breaking parameters
are derived. In addition, the self energies are calculated at one-loop level.
\SARAH can  write all information about the model to \LaTeX files, or create
model files for \FeynArts \cite{feynarts}, {\tt WHIZARD} \cite{Kilian:2007gr} and
\CalcHep/\CompHep \cite{calchep,comphep}, which can also be used for dark matter
studies using {\tt MicrOmegas} \cite{micromegas}. In addition, also the UFO format 
is supported which can be used for instance with {\tt MadGraph 5} \cite{Degrande:2011ua}.\\
Starting with the third version, \SARAH is also supposed to be the first
'spectrum-generator-generator': it uses all analytical expressions to generate
source code for \SPheno \cite{Porod:2003um}. The source code can be used to
calculate the mass spectrum with \SPheno for a new model using 2-loop RGEs and
1-loop corrections to the masses. In addition, the necessary routines for two-
and three-body decays are written. \\
The intention by the development of \SARAH was to make it very flexible: there
is a big freedom for the matter and gauge sector which can be handled. The work
with \SARAH should be easy: every information \SARAH needs are specified in an
easy to modify model file. Nevertheless, \SARAH is also fast: an existing model
can be changed within minutes, and the needed time for doing all necessary
calculations and writing a model file is normally less than 10 minutes. \\
% This manual is structured as follows: in the next chapter we explain the
% installation and the general setup of \SARAH. In the third chapter, the commands
% of \SARAH are explained before in two following chapters all possibilities of
% defining models and parameters are shown in detail. The last two chapters
% discuss the possible output and the usage of numerical values. 

\chapter{Quick start}
\section{Download and installation}
\SARAH is a package for Mathematica \footnote{Mathematica is a protected product
by Wolfram Research} and was tested with versions 5.2, 7 and 8. \\  
\SARAH can be downloaded from
\begin{verbatim}
http://projects.hepforge.org/sarah/
\end{verbatim}
The package archive contains the following directories:
\begin{enumerate}
\item \verb"Models": Definition of the different models
\item \verb"Package": All package files.
\item \verb"LaTeX-Packages": \LaTeX packages, which might be needed for the
output of \SARAH
\end{enumerate}
During the work, also the directory
\begin{verbatim}
Output 
\end{verbatim}
is created. It will contain all files written by \SARAH. \\
In addition, the root directory of \SARAH contains this manual ({\tt
sarah.pdf}), an overview of all models included in the package ({\tt
models.pdf}), a short introduction ({\tt Readme.txt}) as well as an example for
the evaluation of the MSSM ({\tt Example.nb}), an example for creating model files for 
\WHIZARD {\tt Example\_WHIZARD.nb} or in the UFO format {\tt Example\_UFO.nb} as
well as an example how to obtain the \SPheno source code  {\tt Example\_SPheno.nb}.\\
The package should be extracted to the application directory of Mathematica.
This directory is Linux
\begin{verbatim}
home/user/.Mathematica/Applications/
\end{verbatim}
and
\begin{verbatim}
Mathematica-Directory\AddOns\Applications\
\end{verbatim}
in Windows. \\
\section{Run \SARAH}
After the installation, the package is loaded in Mathematica via
\begin{verbatim}
<<"[SARAH Directory]/SARAH.m"
\end{verbatim} 
and a supersymmetric model is initialized by 
\begin{verbatim}
Start["Modelname"];
\end{verbatim} 
Here, \verb"Modelname" is the name of the corresponding model file, e.g. for the
minimal supersymmetric standard model the command would read
\begin{verbatim}
Start["MSSM"];
\end{verbatim} 
or for the next-to-minimal supersymmetric standard model in CKM basis
\begin{verbatim}
Start["NMSSM","CKM"];
\end{verbatim}
is used. In the following, we refer for all given examples the model file of the
MSSM. Our conventions concerning the fields definitions and rotations in the
MSSM are given in app.~\ref{sec:MSSM}.
\section{What happens automatically}
When a model is initialized using the {\tt Start} command, this model is first
checked for gauge anomalies and charge conservation. If not all checks are
fulfilled, a warning is printed. More information about the different checks is
given in app.~\ref{sec:GaugeAnomaly}. Afterwards, the calculation of the
complete Lagrangian at tree level starts. The performed steps are presented in
app.~\ref{sec:Lagrangian}. \\
The next steps are to accomplish all necessary rotations and redefinition of
fields: if a gauge symmetry is broken, the fields responsible for the symmetry
breaking are getting a vacuum expectation value (VEV) and the gauge fields are
rotated. Afterwards, the matter particles are rotated to the new mass eigenbasis
and the tadpole equations are derived. These steps can be repeated if more
rotations or symmetry breakings are necessary.  \\
During this evaluation some more things might done automatically: particles are
integrated out to get an effective theory, the ghost interactions are derived,
the mass matrices and tadpole equations are calculated at tree level. 
At the end, \SARAH splits the Lagrangian in
different pieces to increase the speed of following calculations. Additionally,
if numerical values for all parameters are provided, e.g. in the input files or
in a LesHouches spectrum file \cite{LesHouches}, \SARAH calculates the
eigenvalues of the mass matrices and the rotation matrices. \\
\begin{figure}[ht]
\centering
\includegraphics[scale=0.65]{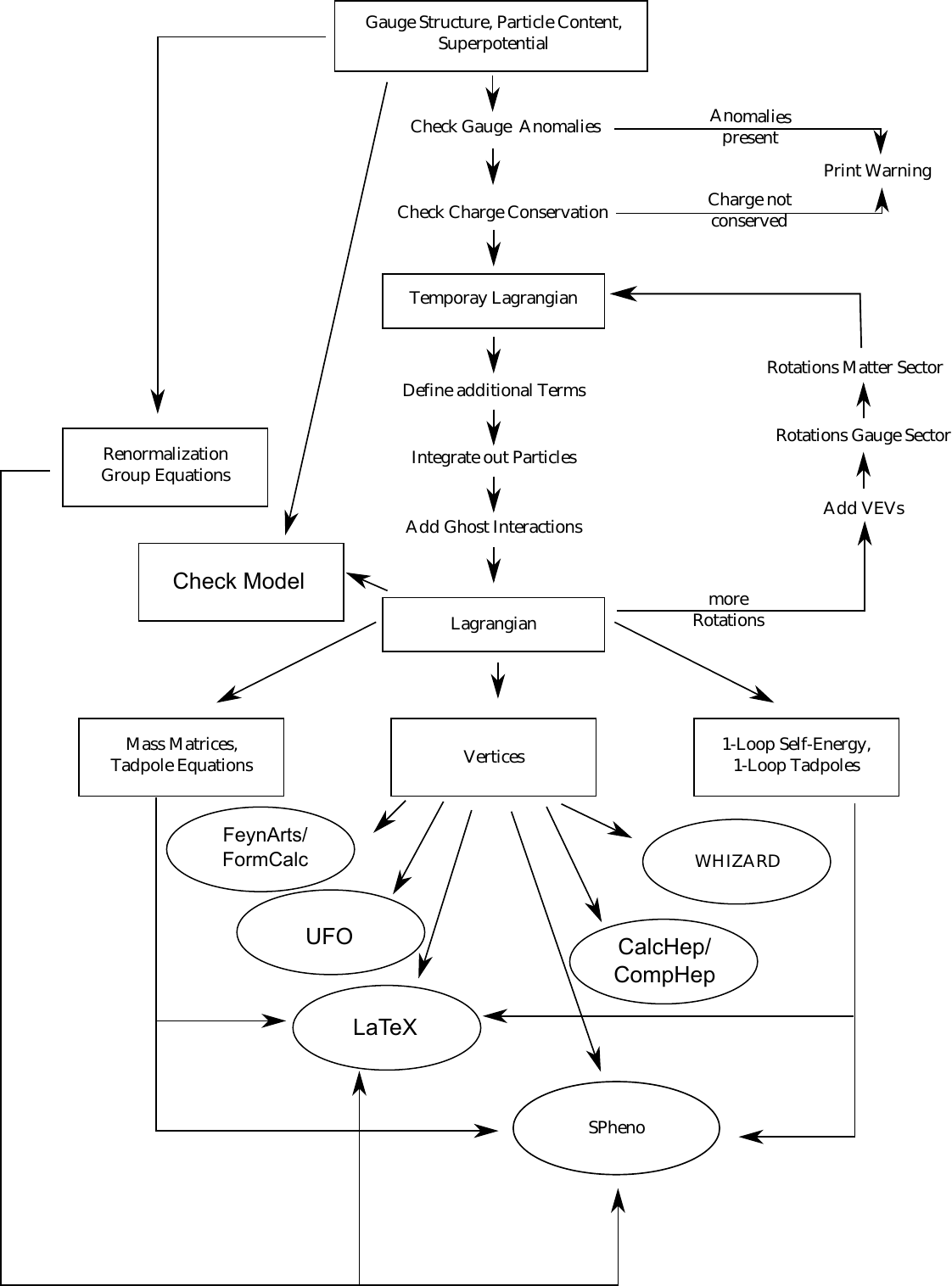}
\caption{Setup of \SARAH}
\end{figure}
\section{Commands}
The most important commands to work with \SARAH are:
\begin{enumerate}
 \item \verb"ShowModels": Shows a list with all installed models
 \item \verb"SARAH`FirsSteps": Shows a short introduction to \SARAH
 \item \verb"CheckModel": Performs several checks of the implementation of the current model
 \item \verb"Vertex[Fields, Options]": Calculates a vertex for given fields
 \item \verb"MassMatrices[$EIGENSTATES]": Shows all mass matrices for given
eigenstates \verb"$EIGENSTATES"
 \item \verb"TadpoleEquations[$EIGENSTATES]": Shows all tadpole equations for given
eigenstates \verb"$EIGENSTATES"
 \item \verb"MassMatrix[Field]": Shows the mass matrix of the field \verb"Field"
 \item \verb"TadpoleEquation[X]": Shows the tadpole equation corresponding to a vev or a scalar particle \verb"X"
to the VEV \verb"VEV"
 \item \verb"CalcRGEs[Options]": Calculates the RGEs
 \item \verb"CalcLoopCorrections[Options]": Calculates one-loop one and two-points functions for
given eigenstates \verb"$EIGENSTATES"
 \item \verb"ModelOutput[$EIGENSTATES,Options]": Create output defined by options  for
given eigenstates \verb"$EIGENSTATES".
 \item \verb"MakeVertexList[$EIGENSTATES,Options]": Calculates all vertices  for given
eigenstates \verb"$EIGENSTATES"
 \item \verb"MakeSPheno[Options]": Writes source code for \SPheno
 \item \verb"MakeTeX[Options]": Writes \LaTeX files 
 \item \verb"MakeCHep[Options]": Writes \CalcHep/\CompHep model files
 \item \verb"MakeFeynArts[Options]": Writes \FeynArts model file
 \item \verb"MakeWHIZARD[Options]": Writes model files for \WHIZARD and \OMEGA
 \item \verb"MakeUFO[Options]": Writes model files in the UFO format
 \item \verb"MakeLHPCstyle[$EIGENSTATES]": Writes steering files for the LHPC spectrum plotter
 \item \verb"MakeAll[Options]": Generates the output for \SPheno, \WHIZARD, \CalcHep, \FeynArts and in the UFO format as well as the \LaTeX\ files
\end{enumerate}

\chapter{Working with \SARAH}
\section{Definition of particles}
\label{nomenclature}
Before we explain the different functions of \SARAH, we have to clarify the nomenclature. All gauge eigenstates are named as follows:
\begin{center}
\begin{verbatim}
ParticleType <> Basis <> [Indices]
\end{verbatim}
\end{center}
\paragraph*{Type}
Here, \verb"ParticleType" is just one letter and indicates the type of a field. The convention is as follows:
\begin{enumerate}
\item \verb"F" for fermionic component of chiral superfield
\item \verb"S" for scalar component of chiral superfield
\item \verb"f" for fermionic component of vector superfield
\item \verb"V" for  bosonic component of vector superfield
\item \verb"g" for ghost field
\end{enumerate}
In addition, there are two types of auxiliary fields. This auxiliary field are not related in any way to the auxiliary components of the superfields in SUSY theories, but they are needed only for writing a \CalcHep/\CompHep model file (see sec.~\ref{auxCalc}). 
The type indicating letters are:
\begin{enumerate}
\item \verb"A" for an auxiliary scalar
\item \verb"a" for an auxiliary vector boson
\end{enumerate}
\begin{center}  
\fbox{
\parbox{12cm}{
After rotating gauge fields to new mass eigenstates, there are no longer constraints concerning the names of fermions and scalars. However, vector bosons must still begin with {\tt V} because the corresponding ghosts are automatically added!}    
}
\end{center} 
\paragraph*{Basis}
The \verb"Basis" of a particle is the name of the underlying superfield. In the MSSM for example, this can be \verb"dL" for the superfield of the left down quarks and squarks or \verb"G" for the vector superfield transforming under the strong interaction.
\paragraph*{Indices}
\verb"Indices" is the list of the indices which the particle carries. There are three different kinds of indices:
\begin{enumerate}
\item \verb"generation": For all particles which appear in more than one generation
\item \verb"lorentz:" For all particles carrying a Lorentz index
\item \verb"charge:" For all components of a chiral superfield charged under a non-Abelian gauge group if this indices are not implicit 
\item \verb"adjoint:" For all all components of a vector field for a non-Abelian gauge group if this indices are not implicit 
\end{enumerate} 
\paragraph*{Examples}
To clarify the above definitions, here some examples:
\begin{enumerate}
\item \verb"VB[{lorentz}]": B-Boson with one Lorentz index 
\item \verb"fB": Bino 
\item \verb"SHd0": Neutral down Higgs
\item \verb"FHd0": Neutral down Higgsino
\item \verb"VG[{adjcolor,Lorentz}]": Gluon with one index for the adjoint of the color group and one Lorentz index 
\item \verb"fG[{adjcolor}]": Gluino with one adjoint index 
\item \verb"gG[{adjcolor}]": Gluon ghost with one adjoint index 
\item \verb"SdL[{generation,color}]": Left handed d-squark with one generation and one color index
\item \verb"FdL[{generation,color}]": Left handed d-quark with one generation and one color index
\item \verb"hh[{generation}]": Neutral, CP-even Higgs (light and heavy Higgs) after EWSB
\end{enumerate}
A comprehensive overview about all models defined in \\SARAH is given in the file {\tt Models.pdf}. Furthermore, the MSSM is discussed in app.~\ref{sec:MSSM}.
\subsection{Antiparticles}
There are two functions to assign antiparticles: depending on the type of the particle \verb"conj" or \verb"bar" are used. 
\begin{enumerate}
\item Scalar, vector boson and Weyl spinor: \verb"conj", e.g. \verb"conj[SdR]" or \verb"conj[VWm]".
\item Dirac fermion and ghost: \verb"bar", e.g \verb"bar[Fd]" or \verb"bar[gG]".
\end{enumerate}
\SARAH checks if a particle is a real scalar or vector bosons respectively a Majorana fermion. In these cases it simplifies the expressions by using
\begin{verbatim}
conj[RP] := RP
bar[MF] := MF
\end{verbatim}
for a real particle \verb"RP"  or also real parameter, and for a Majorana fermion
\verb"MF". The names of all Majorana fermions of the current model are saved in
the list \verb"MajoranaPart", and all real parameters and particles are listed
in \verb"realVar". \\ 
Note, that the head \verb"bar" is overloaded: it is either interpreted as
hermitian or complex conjugated depending on the position of the fermions in a
Dirac chains in order to build up Lorentz scalars. \\
Some words about the necessity of \verb"conj": The function \verb"conj" is very similar to the existing function \verb"Conjugate" of Mathematica. But at least with Mathematica 5.2 there are some problems concerning this function: it is not possible to 
calculate the derivative with respect to a complex conjugate variable, and \verb"Conjugate" is a numerical function. Hence, it is in some cases too slow in handling big analytical expressions. This was improved in Mathematica 7.0 but we want to stay compatible also with version 5.2 and keep therefore \verb"conj".  
\section{Different eigenstates of one model}
\label{eigenstates} 
While calculating the Lagrangian of a model, \SARAH saves the information of the different eigenstates of the model. The name of the eigenstates can be defined by the user in the model file, see sec.~\ref{NameOfStates}. For the model file included in the official package, the eigenstates are
\begin{enumerate}
\item \verb"GaugeES": gauge eigenstates without any rotation
\item \verb"EWSB": eigenstates after electroweak symmetry breaking
\item \verb"SCKM": eigenstates in Super-CKM basis
\item \verb"TEMP": auxiliary eigenstates with no physical meaning
\end{enumerate}
\section{Model information}
There is a lot of information automatically calculated and saved by \SARAH when initializing a model. In this section is shown, how this information can be accessed and used. 

\subsection{Particle content}
To get an overview of all particles of the different eigenstates, use
\begin{verbatim}
Particles[Eigenstates]
\end{verbatim}
e.g. \verb"Particles[GaugeES]" or \verb"Particles[EWSB]" for the gauge eigenstates or the eigenstates after electroweak symmetry breaking (EWSB), respectively. The output is a list with the following information about each particle: 
\begin{enumerate}
\item Name of the particle
\item Type of the particle (\verb"F" for fermion, \verb"S" for scalar, \verb"G" for ghosts, \verb"A" for auxiliary field)
\item Number of first generation (can be different from 1 in effective theories)
\item Number of last generation
\item Indices of the particle
\end{enumerate}
\paragraph*{Example} For instance, the entry for the gauge eigenstates of the left-down quark reads
\begin{verbatim}
{FdL, 1, 3, F, {generation, color}}
\end{verbatim}
\subsection{Masses of particles}
\SARAH automatically calculates the tree level masses of all particles. This information is saved in
\begin{verbatim}
Masses[$EIGENSTATES]
\end{verbatim}
It returns are a list with replacements rules. First, a statement \verb"Mass[]" with the name of the particle is given followed by the value. There are three possibilities for the value
\begin{enumerate}
\item Expression: if an expression is used for a mass, \SARAH has calculated the tree level mass depending on other parameters of the model.
\item \verb"MassGiven": this means that a numerical value for the mass was given in the particle file (see \ref{particleFile}).
\item \verb"MassRead": this means that the value of the mass was read in from a LesHouches input file. 
\end{enumerate}
More information of defining masses is given in section \ref{particleFile}. 
\paragraph*{Example}
The mass of the Z-Boson (\verb"VZ") after EWSB is saved in \verb"Masses[EWSB]". The corresponding entry is
\begin{verbatim}
Mass[VZ] -> ((vd^2 + vu^2)*(g2*Cos[ThetaW] + g1*Sin[ThetaW])^2/4
\end{verbatim}
\subsection{Mass Matrices}
\SARAH calculates automatically the mass matrices before rotating the fields to the new eigenstates and saves the information in arrays. The basis of the rotations can be seen by using
\begin{verbatim}
MixBasis[$EIGENSTATES]
\end{verbatim}
The matrices itself are saved in two arrays:
\begin{verbatim}
MassMatrices[$EIGENSTATES]
\end{verbatim}
and
\begin{verbatim}
MassMatricesFull[$EIGENSTATES]
\end{verbatim}
The difference between this two arrays is that in the first one, the different generations are written as indices, while in the second on the generation indices are explicitly inserted. This means, in the first case the basis for the mass matrix in the down squark sector is just
\begin{verbatim}
(SdL[{gn,cn}],SdL[{gm,cm}])
\end{verbatim} 
while in the second case the basis vector is
\begin{verbatim}
(SdL[{1,cn1}],SdL[{2,cn2}],SdL[{3,cn3}],SdL[{1,cm1}],SdL[{2,cm2}],SdL[{3,cm2}])
\end{verbatim} 
It is also possible to use the command
\begin{verbatim}
MassMatrix[Field]
\end{verbatim}
to obtain the mass matrix for a specific particle. 
\paragraph*{Example}
Let's have a look at the down-squark sector after EWSB.
\begin{verbatim}
MixBasis[EWSB][[1]]
\end{verbatim} 
returns the basis of the mass matrix:
\begin{verbatim}
{{SdL, SdR}, {conj[SdL], conj[SdR]}}
\end{verbatim}
The (1,1) element of the mass matrix is saved in
\begin{verbatim}
MassMatricesFull[EWSB][[1,1,1]]
\end{verbatim}
and looks like
\begin{verbatim}
-(g1^2*vd^2)/24 - (g2^2*vd^2)/8 + (g1^2*vu^2)/24 + 
   (g2^2*vu^2)/8 + mq2[1, 1] + (vd^2*Yd[1, 1]^2)/2
\end{verbatim}
The same result is obtained by 
\begin{verbatim}
MassMatrix[Sd][[1,1]]
\end{verbatim}
\subsection{Tadpole Equations}
Gauge symmetries are broken, if particles receive a VEVs. These VEVs \(v_i\) (or equivalently the real components $\phi$ of the scalar field) should minimize the potential and therefore fulfill the equations
\begin{equation}
\label{eq:TadEqu}
\frac{\partial V}{\partial v} = \frac{\partial V}{\partial \phi} = 0 
\end{equation}
\(V\) is the scalar potential of the model. If CP is violated in the Higgs sector, similar conditions arise for the pseudo-scalar components of the fields
\begin{equation}
 \frac{\partial V}{\partial \sigma} = 0
\end{equation}
Eq.~(\ref{eq:TadEqu}) are the so called tadpole equations which are cubic equations in the VEVs. \SARAH saves the left hand side of (\ref{eq:TadEqu}) in arrays named
\begin{verbatim}
TadpoleEquations[$EIGENSTATES]
\end{verbatim} 
The order of the tadpole equations in this array corresponds to the order of the definition of VEVs in the model file, see (\ref{VEV}). \\
There is also the shorter command 
\begin{verbatim}
 TadpoleEquation[X]
\end{verbatim}
to obtain the tadpole equations corresponding to a specific VEV or state.
\paragraph*{Example}
The tadpole equation for \(v_d\) after EWSB is saved in
\begin{verbatim}
TadpoleEquations[EWSB][[1]]
\end{verbatim}
and reads
\begin{verbatim}
mHd2*vd + (g1^2*vd^3)/8 + (g2^2*vd^3)/8 - (g1^2*vd*vu^2)/8 -
  (g2^2*vd*vu^2)/8 + vd*\[Mu]^2 - vu*B[\[Mu]]
\end{verbatim}
The same result can be obtained by
\begin{verbatim}
 TadpoleEquation[vd]
\end{verbatim}
or 
\begin{verbatim}
 TadpoleEquation[phid]
\end{verbatim}

\subsection{Parts of the Lagrangian}
All information about a model and it's interactions are encoded in the full Lagrangian. \SARAH calculates the full Lagrangian from the superpotential and the gauge sector by using the method explained in appendix \ref{sec:Lagrangian}. The final results, the Lagrangians for the different eigenstates are saved as
\begin{verbatim}
Lagrangian[$EIGENSTATES]
\end{verbatim}
For a realistic SUSY model, the Lagrangian is generally very lengthy. Therefore, \SARAH splits it in different parts in order to speed up some calculations. This splitting might be also helpful for analyzing the structure of interactions involving different kinds of fields. An overview of all names for the different parts of the Lagrangian is given in app.~\ref{PartsLag}. \\
Also the results of the different steps during then calculation of the Lagrangian in gauge eigenstates (e.g. F-Terms, D-Terms, kinetic parts) are saved. Thus, it is possible to have a detailed look at specific parts of the Lagrangian. The names of the parts are also given in app.~\ref{sec:Lagrangian}. 
\subsection{Writing all information about particles and parameters in an external file}
All information about particles and parameters of the considered model can by written in two files
\verb"ParticleInfo.m" and \verb"ParameterInfo", which are saved in 
\begin{verbatim}
../\SARAH/Output/"Model Name"/
\end{verbatim}
by using
\begin{verbatim}
ExportModelInformation;
\end{verbatim}
The file \verb"ParticleInfo.m" contains the following information for all eigenstates of the model
\begin{enumerate}
\item R-Parity
\item PDG, PDG.IX
\item \LaTeX\ name
\item Output name
\item \FeynArts number
\item Type
\item Self-conjugated or not
\item Number of generations
\item Indices
\item Electric charge
\item Description
\item Mass, Width
\end{enumerate}
while \verb"ParameterInfo.m" contains the following information about all parameters 
\begin{enumerate}
\item Dependence on other parameters
\item Real or complex
\item Numerical value
\item Position in LesHouches input
\item \LaTeX{} name
\end{enumerate}
\section{Calculating Vertices}
One of the main functions of \SARAH is to calculate the vertices for a model. In contrast to the most other calculations, vertices are not calculated automatically when initializing a model: it can last several minutes to calculate all vertices of a model and sometimes these calculations are not necessary. Of course, it is also possible to tell \SARAH that all vertices should automatically calculated, see section \ref{sec:allvertices}. In this section, we want to focus  on calculating vertices 'by hand'.\\  
\subsection{Calculating specific vertices}
Vertices are calculated by
\begin{verbatim}
Vertex[ParticleList,Options]
\end{verbatim}
\verb"ParticleList" is a list containing the involved fields. This list can consist of up to 6 particles if an effective theory is analyzed (see sec.~\ref{IntOutorDel}). \\  
The following \verb"Options" are supported by the \verb"Vertex" command:
\begin{enumerate}
\item \verb"Eigenstates", Value: \verb"$EIGENSTATES", Default: Last entry in \verb"NameOfState" \\ 
Fixes the considered eigenstates 
\item \verb"UseDependences": Value \verb"True" or \verb"False", Default: \verb"False" \\
Optional relations between the parameters (see section \ref{parameterFile}) will be used, if \verb"UseDependences" is set to \verb"True".   
\end{enumerate} 
The output of \verb"Vertex" is an array: 
\begin{verbatim}
{{ParticleList},{{Coefficient 1, Lorentz 1},{Coefficient 2, Lorentz 2},...} 
\end{verbatim}
First, the list of the involved particles is given and the indices are inserted. The second part consists of the value of the vertex and can be also a list, if different Lorentz structures are possible. In the part independent of any Lorentz index the following symbols can appear
\begin{enumerate}
\item \verb"Delta[a,b]": Kronecker delta \(\delta_{\alpha \beta}\)
\item \verb"ThetaStep[i,j]": Step function \(\Theta_{i j}\)
\item \verb"Lam[t,a,b]": Gell-Mann matrix \(\lambda^t_{\alpha\beta}\)
\item \verb"LambdaProd[x,y][a,b]": Matrix product of two Gell-Mann matrices \(\left(\lambda^x \lambda^y\right)_{\alpha\beta}\)
\item \verb"Sig[t,a,b]": Pauli matrix \(\sigma^t_{\alpha\beta}\)
\item \verb"SigmaProd[x,y][a,b]": Matrix product of two Pauli matrices \(\left(\sigma^x \sigma^y\right)_{\alpha\beta}\)
\item \verb"fSU3[i,j,k]": Structure constants of \(SU(3)\): \(f^{ijk}\) 
\item \verb"fSU2[i,j,k]": Structure constants of \(SU(2)\): \(\epsilon^{ijk}\)
\item \verb"FST[SU[N]][i,j,k]": Structure constants of \(SU(N)\): \(f_N^{ijk}\)
\item \verb"TA[SU[N]][a,i,j]": Generator of  \(SU(N)\): \(T^a_{ij}\)
\item Couplings, e.g.  \verb"g1", \verb"g2", \verb"g3", \verb"Ye[a,b]", \verb"Yd[a,b]", \verb"Yu[a,b]", \dots 
\item Mixing matrices, e.g. \verb"ZD[a,b]", \dots
\end{enumerate}
The part transforming under the Lorentz group can consist of
\begin{enumerate}
\item \verb"gamma[lor]": Gamma matrix \(\gamma_\mu\)
\item \verb"g[lor1,lor2]": Metric tensor \(g_{\mu\nu}\)
\item \verb"Mom[particle,lor]": Momentum \(p^\mu_P\) of particle \(P\)
\item \verb"PL", \verb"PR": Polarization operators \(P_L = \frac{1 - \gamma_5}{2}\), \(P_R = \frac{1+\gamma_5}{2}\)
\item \verb"1": If the vertex is a Lorentz scalar.
\item \verb"LorentzProduct[_,_]:" A non commutative product of terms transforming under the Lorentz group
\end{enumerate}
\paragraph*{Examples}
Some examples to clarify the usage and output of \verb"Vertex":
\begin{enumerate}
\item {\bf One possible Lorentz structure:} {\tt Vertex[{hh,Ah,Z}]} leads to the vertex of scalar and a pseudo scalar Higgs with a $Z$-boson
\begin{verbatim}
{{hh[{gt1}], Ah[{gt2}], VZ[{lt3}]}, 
 {((MA[gt2,1]*MH[gt1,1] - MA[gt2,2]*MH[gt1,2])*(g2*Cos[ThetaW]+g1*Sin[ThetaW]))/2, 
                Mom[Ah[{gt2}], lt3] - Mom[hh[{gt1}],lt3]}}
\end{verbatim}
The output is divided in two parts. First, the involved particles are given, second, the value of the vertex is given. This second part is again split in two parts: one is the Lorentz independent part and the second part defines the transformation under the Lorentz group.  
\item {\bf Several possible Lorentz structures} {\tt Vertex[{bar[Fd],Fd,hh}]} is the interaction between d-quarks and a Higgs:
\begin{verbatim}
{{bar[Fd[{gt1, ct1}]], Fd[{gt2, ct2}], hh[{gt3}]}, 
 {((-I)*Delta[ct1,ct2]*Delta[gt1,gt2]*MH[gt3,2]*Yd[gt2,gt1])/Sqrt[2],PL}, 
 {((-I)*Delta[ct1,ct2]*Delta[gt1,gt2]*MH[gt3,2]*Yd[gt1,gt2])/Sqrt[2],PR}}
\end{verbatim}
Obviously, there are three parts: one for the involved particles and two for the different Lorentz structures. \verb"PL" and \verb"PR" are the polarization projectors \(P_L = \frac{1}{2} (1 - \gamma_5), P_R = \frac{1}{2} (1 + \gamma_5)\).
\item {\bf Changing the considered eigenstates and using Weyl fermions} It is also possible to calculate the vertices for Weyl fermions and/or to consider the gauge eigenstates. For instance,
\begin{verbatim}
Vertex[{fB, FdL, conj[SdL]}, Eigenstates -> GaugeES]
\end{verbatim}
returns
\begin{verbatim}
{{fB, FdL[{gt2, ct2}], conj[SdL[{gt3, ct3}]]}, 
 {((-I/3)*g1*Delta[ct2, ct3]*Delta[gt2, gt3])/Sqrt[2],1}}
\end{verbatim}
\item {\bf Using dependences} With {\tt Vertex[\{conj[Se], Se, VP\}, UseDependences -> True]} \(g_1\) and \(g_2\) are replaced by the electric charge \(e\). This and similar relations can be defined in the parameters file (see sec.~\ref{sec:ParameterFile}). 
\begin{verbatim}
{{conj[Se[{gt1}]], Se[{gt2}], VP[{lt3}]}, 
{(-I)*e*Delta[gt1,gt2],-Mom[conj[Se[{gt1}]],lt3]+Mom[Se[{gt2}],lt3]}}
\end{verbatim}
\item {\bf Fixing the generations} It is possible to give the indices of the involved particles already as input
\begin{verbatim}
Vertex[{hh[{1}], hh[{1}], Ah[{2}], Ah[{2}]}]
\end{verbatim} 
leads to
\begin{verbatim}
{{hh[{1}], hh[{1}], Ah[{2}], Ah[{2}]}, 
 {(-I/4)*(g1^2 + g2^2)*Cos[2*\[Alpha]]*Cos[2*\[Beta]], 1}}
\end{verbatim}
Obviously, the given definition of the mixing matrices for the Higgs fields were automatically inserted. If the indices  are fixed by a replacement, the definition of the mixing matrix wouldn't be used
\begin{verbatim}
Vertex[{hh, hh, Ah, Ah}] /. {gt1->1, gt2->1,gt3->2, gt3->2}
\end{verbatim} 
returns
\begin{verbatim}
{{hh[{1}], hh[{1}], Ah[{2}], Ah[{gt4}]}, 
 {(-I/4)*(g1^2 + g2^2)*(conj[ZA[2, 1]]*conj[ZA[gt4, 1]] - 
   conj[ZA[2, 2]]*conj[ZA[gt4, 2]])*(conj[ZH[1, 1]]^2 - conj[ZH[1, 2]]^2), 1}}
\end{verbatim}
However,
\begin{verbatim}
Vertex[{hh, hh, Ah, Ah}] /. {gt1->1, gt2->1,gt3->2, gt3->2} /.subAlways
\end{verbatim}
leads to the former expression using the mixing angle $\beta$.  
\item {\bf Effective operators} In effective theories also interactions between two fermions and two scalars are possible. As example an effective vertex for a model in which the gluino was integrated out:
\begin{verbatim}
Vertex[{Fd, Fd, conj[Sd], conj[Sd]}]
\end{verbatim}
Returns
\begin{verbatim}
{{Fd[{gt1, ct1}], Fd[{gt2, ct2}], conj[Sd[{gt3, ct3}]], conj[Sd[{gt4, ct4}]]}, 
{-(g3^2*(sum[j1, 1, 8, (Lam[j1, ct3, ct2]*Lam[j1, ct4, ct1])/Mass[fG][j1]]*
           ZD[gt3, gt2]*ZD[gt4, gt1] + 
         sum[j1, 1, 8, (Lam[j1, ct3, ct1]*Lam[j1, ct4, ct2])/Mass[fG][j1]]*
           ZD[gt3, gt1]*ZD[gt4, gt2])),
    LorentzProduct[PL, PL]}, {0, LorentzProduct[PR, PL]}, 
{g3^2*(sum[j1, 1, 8, (Lam[j1, ct2, ct3]*Lam[j1, ct4, ct1])/Mass[fG][j1]]*
       ZD[gt3, 3 + gt2]*ZD[gt4, gt1] + 
       sum[j1, 1, 8, (Lam[j1, ct2, ct4]*Lam[j1, ct3, ct1])/Mass[fG][j1]]*
       ZD[gt3, gt1]*ZD[gt4, 3 + gt2]),
    LorentzProduct[PL, PR]}, {0, LorentzProduct[PR, PR]}, 
{0, LorentzProduct[gamma,PL, PL]}, {0, LorentzProduct[gamma, PR, PL]}, 
{0, LorentzProduct[gamma, PL, PR]}, {0, LorentzProduct[gamma,PR, PR]}}
\end{verbatim}
Obviously, \SARAH checks the eight possible operators (4 different combination of polarization operators with and without a \(\gamma\) matrix) and returns the result for each operator. 
\end{enumerate}

\subsection{Calculating all vertices}
\label{sec:allvertices}
To calculate all vertices at once for a given model, use
\begin{verbatim}
MakeVertexList[Eigenstates,Options]
\end{verbatim}
First, the name of the eigenstates has to be given. The possible options are:
\begin{enumerate}
\item \verb"effectiveOperators", Values: \verb"True" or \verb"False", Default: \verb"False" \\
If also higher dimensional operators should be calculated. By default, this concerns only four point interactions. 
\item \verb"SixParticleInteractions", Values: \verb"True" or \verb"False", Default: \verb"False" \\
If also the six-point interactions should be calculated. 
\item \verb"GenericClasses", Values: \verb"All" or a list of generic types, Default: \verb"All"\\
Calculates the vertices only for the given types of interaction
\end{enumerate}
The results are saved in list named
\begin{verbatim}
 SA`VertexList[Type]
\end{verbatim}
with {\tt Type} = {\tt SSS,SSSS,SSVV,SSV,SVV,FFS,FFV,VVV,VVVV,GGS,GGV,ASS}.

\section{Renormalization Group Equations}
\label{RGEs}
\SARAH calculates the renormalization group equations (RGEs) for the parameters of the superpotential, the soft-breaking terms and the gauge couplings at one and two loop level. This is done by using the generic formulas of \cite{Martin:1993zk} extended by the results for several Abelian gauge groups \cite{Fonseca:2011vn} and Dirac mass terms for gauginos \cite{Goodsell:2012fm}. \\
 
The calculation is started via
\begin{verbatim}
CalcRGEs[Options]
\end{verbatim}
\paragraph*{Options} The different options are
\begin{enumerate}
\item \verb"TwoLoop", Value: \verb"True" or \verb"False", Default: \verb"True" \\
If also the two loop RGEs should be calculated.
\item \verb"ReadLists", Value: \verb"True" or \verb"False", Default: \verb"False" \\
If the RGEs have already be calculated, the results are saved in the output directory. The RGEs can be read from these files instead of doing the complete calculation again. 
\item \verb"VariableGenerations", Value: List of particles, Default: \verb"{}"\\
Some theories contain heavy superfields which should be integrated out above the SUSY scale. Therefore, it is possible to calculate the RGEs assuming the number of generations of specific superfields as free variable to make the dependence on these fields obvious. The new variable is named \verb"NumberGenertions[X]", where \verb"X" is the name of the superfield.
\item \verb"NoMatrixMultiplication", Values: \verb"True" or \verb"False", Default: \verb"False"\\
Normally, the \(\beta\)-functions are simplified by writing the sums over generation indices as matrix multiplication. This can be switched off using this option. 
\item \verb"IgnoreAt2Loop", Values: a list of parameters, Default: \verb"{}"\\
The calculation of 2-loop RGEs for models with many new interactions can be very
 time-consuming. However, often one is only interested in the dominant
 effects of the new contributions at the 1-loop level. Therefore, {\tt IgnoreAt2Loop -> \$LIST} can be used to neglect parameters at the two-loop level The entries of \$LIST can be superpotential or soft  SUSY-breaking parameters as well as gauge couplings. 
\end{enumerate}
The \(\beta\)-functions will be stored in the following arrays:
\begin{enumerate}
\item \verb"Gij": Anomalous dimensions of all chiral superfields
\item \verb"BetaWijkl": Quartic superpotential parameters
\item \verb"BetaYijk": Trilinear superpotential parameters
\item \verb"BetaMuij": Bilinear superpotential parameters
\item \verb"BetaLi": Linear superpotential parameters
\item \verb"BetaQijkl": Quartic soft-breaking parameters
\item \verb"BetaTijk": Trilinear soft-breaking parameters
\item \verb"BetaBij": Bilinear soft-breaking parameters
\item \verb"BetaSLi": Linear soft-breaking parameters
\item \verb"Betam2ij": Scalar squared masses
\item \verb"BetaMi": Majorana Gaugino masses
\item \verb"BetaGauge": Gauge couplings
\item \verb"BetaVEVs": VEVs
\item \verb"BetaDGi": Dirac gaugino mass terms
\end{enumerate}
These arrays are also saved in the directory
\begin{verbatim}
../\SARAH/Output/"ModelName"/RGE
\end{verbatim}
All entries of this arrays are three dimensional: The first entry is the name of the parameter, the second the one-loop \(\beta\)-function and the third one the two loop \(\beta\)-function. 
\paragraph*{GUT normalization} The gauge couplings of \(U(1)\) gauge groups are often normalized at the GUT scale with respect to a specific GUT group. Therefore, it is possible to define for each gauge coupling the GUT-normalization by the corresponding entry in the parameters file. See sec.~\ref{sec:ParameterFile} for more information. \\

Generally, the results contain sums over the generation indices of the particles in the loop. \SARAH always tries to write them as matrix multiplications, in order to shorten the expressions. Therefore, new symbols are introduced:
\begin{enumerate}
\item \verb"MatMul[A,B,C,...][i,j]": \((A B C \dots)_{i,j}\). Matrix multiplication, also used for vector-matrix and vector-vector multiplication.
\item \verb"trace[A,B,C,...]": \(\mbox{Tr}(A B C \dots)\). Trace of a matrix or of a product of matrices.
\item \verb"Adj[M]": \(M^\dagger\). Adjoint of a matrix
\item \verb"Tp[M]": \(M^T\). Transposed of a matrix  
\end{enumerate} 
\paragraph*{Remarks} Some remarks about the output:
\begin{enumerate}
\item To differ between generation and other indices during the calculation, \verb"Kronecker[i,j]" is used for generation indices instead of \verb"Delta[i,j]".  
\item The results for the scalar masses are simplified by using abbreviations for often appearing traces, see also Ref.~\cite{Martin:1993zk}. The definition of the traces are saved in the array \verb"TraceAbbr". 
\item If the model contains parameters with three indices, matrix multiplication is automatically switched off and the results are given as sum over the involved indices. In addition, these expressions are simplified by replacing a parameter with three indices by a sum of parameters with two indices. The $\beta$ function in this form a saved in {\tt NAME <> 3I} with {\tt NAME} stands for the standard array containing the RGEs. 
\end{enumerate}
\paragraph*{Examples}
\begin{enumerate}
\item {\bf \(\beta\)-function of Yukawa coupling} The Yukawa couplings of the MSSM are saved in \verb"BetaYijk". The first entry consists of
\begin{verbatim}
BetaYijk[[1,1]]:  Ye[i1,i2] ,
\end{verbatim}
i.e. this entry contains the \(\beta\)-functions for the electron Yukawa coupling. The corresponding one-loop \(\beta\)-function is
\begin{verbatim}
BetaYijk[[1,2]]:
(-9*g1^2*Ye[i1,i2])/5-3*g2^2*Ye[i1,i2]+3*trace[Yd,Adj[Yd]]*Ye[i1,i2]+ 
  trace[Ye,Adj[Ye]]*Ye[i1, i2]+3*MatMul[Ye,Adj[Ye],Ye][i1, i2]
\end{verbatim}
The two-loop \(\beta\)-function is saved in \verb"BetaYijk[[1,3]]" but we skip it here because of its length. 
\item {\bf \(\beta\)-function of soft-breaking masses and abbreviations for traces} The soft-breaking mass of the selectron is the first entry of \verb"Betam2ij"
\begin{verbatim}
 Betam2ij[[1,1]]:           me2[i1,i2]
\end{verbatim}
 and the one-loop \(\beta\)-function is saved in  \verb"Betam2ij[[1,2]]":
\begin{verbatim}
(-24*g1^2*MassB*conj[MassB]+10*g1^2*Tr1[1])*Kronecker[i1,i2]/5 + 
 4*mHd2*MatMul[Ye,Adj[Ye]][i1,i2]+4*MatMul[T[Ye],Adj[T[Ye]]][i1,i2] + 
  2*MatMul[me2,Ye,Adj[Ye]][i1,i2]+4*MatMul[Ye, ml2, Adj[Ye]][i1,i2] + 
  2*MatMul[Ye,Adj[Ye],me2][i1,i2]
\end{verbatim}
The definition of the element \verb"Tr1[1]" is saved in \verb"TraceAbbr[[1,1]]":
\begin{verbatim}
{Tr1[1], -mHd2 + mHu2 + trace[md2] + trace[me2] - trace[ml2] +
          trace[mq2] - 2*trace[mu2]}
\end{verbatim}
\item {\bf Number of generations as variable}: With
\begin{verbatim}
CalcRGEs[VariableGenerations -> {q}]
\end{verbatim}
the number of generations of the left-quark superfield is handled as variable. Therefore, the one-loop \(\beta\)-function of the hypercharge couplings reads 
\begin{verbatim}
 (63*g1^3)/10 + (g1^3*NumberGenerations[q])/10
\end{verbatim}
\item{\bf No matrix multiplication} Using matrix multiplication can be switched off by
\begin{verbatim}
CalcRGEs[NoMatrixMultiplication -> True]
\end{verbatim}
The one-loop \(\beta\)-function for the electron Yukawa coupling is now written as
\begin{verbatim}
  sum[j2,1,3,sum[j1,1,3,conj[Yd[j2,j1]]*Yu[i1,j1]]*Yd[j2,i2]] + 
2*sum[j2,1,3,sum[j1,1,3,conj[Yu[j1,j2]]*Yu[j1,i2]]*Yu[i1,j2]] + 
  sum[j2,1,3,sum[j1,1,3,conj[Yu[j2,j1]]*Yu[i1,j1]]*Yu[j2,i2]] + 
(3*sum[j2,1,3,sum[j1,1,3,conj[Yu[j1,j2]]*Yu[j1,j2]]]*Yu[i1,i2])/2 + 
(3*sum[j2,1,3,sum[j1,1,3,conj[Yu[j2,j1]]*Yu[j2,j1]]]*Yu[i1,i2])/2 - 
 (13*g1^2*Yu[i1,i2])/15-3*g2^2*Yu[i1,i2]-(16*g3^2*Yu[i1,i2])/3
\end{verbatim}
\item {\bf Ignoring parameters at two-loop} Using
\begin{verbatim}
CalcRGEs[IgnoreAt2Loop -> {T[L1],T[L2],L1,L2}]
\end{verbatim}
in the MSSM with trilinear $R$pV would ignore the $\lambda$ and $\lambda'$ coupling as well as their soft-breaking equivalents in the calculation of the 2-loop RGEs.
\end{enumerate}
\section{Loop Corrections}
\SARAH calculates the analytical expressions for the one-loop corrections to the tadpoles and the self energy of all particles. These calculations are performed in \(\overline{\text{DR}}\)-scheme and in the 't Hooft gauge. To command to start the calculation is
\begin{verbatim}
CalcLoopCorrections[Eigenstates,Options];
\end{verbatim}
As usual, \verb"Eigenstates" can be for instance in the case of the MSSM either \verb"GaugeES" for the gauge eigenstates or \verb"EWSB" for the eigenstates after EWSB. If the vertices for the given set of eigenstates were not calculated before, this is done before the calculation of the loop contributions begins.
As option a list with fields can be given ({\tt OnlyWith -> {Particle1,Particle2,...}}). Only corrections involving these fields as internal particles are included.
 \\
\paragraph*{Conventions} Using the conventions of Ref.~\cite{Pierce:1996zz}, the results will contain the  Passarino Veltman integrals listed in app.~\ref{sec:Integrals}. The involved couplings are abbreviated by
\begin{enumerate}
\item \verb"Cp[p1,p2,p3]" and \verb"Cp[p1,p2,p3,p4]" for non-chiral, three and four point interactions involving the particles \verb"p1" - \verb"p4".
\item \verb"Cp[p1,p2,p3][PL]" and \verb"Cp[p1,p2,p3][PR]" for chiral, three-point interactions involving the fields \verb"p1" - \verb"p3". 
\end{enumerate}
The self energies can be used for calculating the radiative corrections to masses and mass matrices, respectively. We have summarized the needed formulas for this purpose in app.~\ref{sec:OneLoopMass}. For calculating the loop corrections to a mass matrix, it is convenient to use unrotated, external fields, while the fields in the loop are rotated. Therefore, \SARAH adds to the symbols of the external particle in the interaction an \verb"U" for 'unrotated', e.g. \verb"Sd" \(\rightarrow\) \verb"USd". The mixing matrix associated to this field in the vertex has to be replaced by the identity matrix when calculating the correction to the mass matrix. 
\paragraph*{Results} The results for the loop corrections are saved in two different ways. First as list containing the different loop contribution for each particle. Every entry reads
\begin{verbatim}
 {Particles, Vertices, Type, Charge Factor, Symmetry Factor}
\end{verbatim}
and includes the following information
 \begin{enumerate}
 \item \verb"Particles": The particles in the loop. 
 \item \verb"Vertices": The needed Vertex for the correction is given.  
 \item \verb"Charge Factor": If several gauge charges of one particle are allowed in the loop, this factor will be unequal to one. In the case of the MSSM, only the a factor of 3 can appear because of the different colors. 
 \item \verb"Symmetry Factor": If the particles in the loop indistinguishable, the weight of the contribution is only half of the case of distinguishable particles. If two different charge flows are possible in the loop, the weight of the diagram is doubled, e.g. loop with charged Higgs and $W$-boson. The absolute value of the factor depends on the type of the diagram. 
\end{enumerate}

The information about the loop correction are also saved in the directory
\begin{verbatim}
../\SARAH/Output/"ModelName"/$EIGENSTATES/Loop
\end{verbatim}

\paragraph*{One Loop Tadpoles}
The complete results as sums of the different contributions are saved in the two dimensional array 
\begin{verbatim}
Tadpoles1LoopSums[$EIGENSTATES]
\end{verbatim}
The first column gives the name of the corresponding VEV, the second entry the one-loop correction. \\
A list of the different contributions, including symmetry and charge factors, is
\begin{verbatim}
Tadpoles1LoopList[$EIGENSTATES]; 
\end{verbatim}

\paragraph*{One Loop Self Energies}
The results are saved in the following two dimensional array 
\begin{verbatim}
SelfEnergy1LoopSum[$EIGENSTATES]
\end{verbatim}
The first column gives the name of the particle, the entry in the second column depends on the type of the field
\begin{enumerate}
\item Scalars: one-loop self energy \(\Pi(p^2)\)
\item Fermions: one-loop self energies for the different polarizations (\(\Sigma^L(p^2)\),\(\Sigma^R(p^2)\), \(\Sigma^S(p^2)\))	
\item Vector bosons: one-loop, transversal self energy \(\Pi^T(p^2)\)
\end{enumerate}
Also a list with the different contributions does exist:
\begin{verbatim}
SelfEnergy1LoopList[$EIGENSTATES]
\end{verbatim}

\paragraph*{Examples}
\begin{enumerate}
\item {\bf One-loop tadpoles} The correction of the tadpoles due to a chargino loop is saved in 
\begin{verbatim}
Tadpoles1LoopList[EWSB][[1]]; 
\end{verbatim}
and reads
\begin{verbatim}
 {bar[Cha],Cp[Uhh[{gO1}],bar[Cha[{gI1}]],Cha[{gI1}]],FFS,1,1/2}
\end{verbatim}
The meaning of the different entries is: (i) a chargino (\verb"Cha") is in the loop, (ii) the vertex with an external, unrotated Higgs (\verb"Uhh") with generation index \verb"gO1" and two charginos with index \verb"gI1" is needed, (iii) the generic type of the diagram is \verb"FFS", (iv) the charge factor is 1, (v) the diagram is weighted by a factor \(\frac{1}{2}\) with respect to the generic expression (see app.~\ref{sec:Integrals}).\\
The corresponding term in \verb"Tadpoles1LoopSum[EWSB]" is
\begin{verbatim}
4*sum[gI1,1,2, A0[Mass[bar[Cha[{gI1}]]]^2]*
   Cp[phid,bar[Cha[{gI1}]],Cha[{gI1}]]*Mass[Cha[{gI1}]]] 
\end{verbatim}
\item {\bf One-loop self-energies}
\begin{enumerate}
\item The correction to the down squark matrix due to  a four point interaction with a pseudo scalar Higgs is saved in
{\tt  SelfEnergy1LoopList[EWSB][[1, 12]]} and reads
\begin{verbatim}
{Ah,Cp[conj[USd[{gO1}]],USd[{gO2}],Ah[{gI1}],Ah[{gI1}]],SSSS,1,1/2}
\end{verbatim}
This has the same meaning as the term
\begin{verbatim}
-sum[gI1,1,2,A0[Mass[Ah[{gI1}]]^2]*
   Cp[conj[USd[{gO1}]],USd[{gO2}],Ah[{gI1}],Ah[{gI1}]]]/2
\end{verbatim}
in \verb"SelfEnergy1LoopSum[EWSB]". 
\item Corrections to the Z boson are saved in {\tt SelfEnergy1LoopList[EWSB][[15]]}. An arbitrary entry looks like
\begin{verbatim}
{bar[Fd], Fd, Cp[VZ, bar[Fd[{gI1}]], Fd[{gI2}]], FFV, 3, 1/2}
\end{verbatim}
and corresponds to
\begin{verbatim}
(3*sum[gI1, 1, 3, sum[gI2, 1, 3, 
     H0[p^2, Mass[bar[Fd[{gI1}]]]^2, Mass[Fd[{gI2}]]^2]*
 (conj[Cp[VZ,bar[Fd[{gI1}]],Fd[{gI2}]][PL]]*
     Cp[VZ,bar[Fd[{gI1}]],Fd[{gI2}]][PL] + 
  conj[Cp[VZ,bar[Fd[{gI1}]],Fd[{gI2}]][PR]]*
     Cp[VZ,bar[Fd[{gI1}]],Fd[{gI2}]][PR]) + 
 2*B0[p^2,Mass[bar[Fd[{gI1}]]]^2,Mass[Fd[{gI2}]]^2]*
     Mass[bar[Fd[{gI1}]]]*Mass[Fd[{gI2}]]*
  Re[Cp[VZ,bar[Fd[{gI1}]],Fd[{gI2}]][PL]*
     Cp[VZ,bar[Fd[{gI1}]],Fd[{gI2}]][PR])]])/2 
\end{verbatim}
in \verb"SelfEnergy1LoopListSum[EWSB]". 
\end{enumerate}
\end{enumerate}

\section{Output for \LaTeX and diagram calculators}
\label{modeloutput}
With
\begin{verbatim}
ModelOutput[Eigenstates,Options]
\end{verbatim}
\LaTeX-files and model files for \FeynArts and \CompHep/\CalcHep can be generated. 
Here, \verb"Eigenstates" specifies  the eigenstates which should be used, e.g. \verb"GaugeES" or \verb"EWSB". The options are the following
\begin{enumerate}
\item \verb"WriteTeX", values: \verb"True" or \verb"False", Default: \verb"False" \\
If a \LaTeX file containing all information about the model should be written. 
\item \verb"WriteFeynArts", values: \verb"True" or \verb"False", Default: \verb"False" \\
If a model file for \FeynArts should be written.
\item \verb"WriteCHep", values: \verb"True" or \verb"False", Default: \verb"False" \\
If a model file for \CompHep/\CalcHep should be written. 
\item \verb"WriteWHIZARD", values: \verb"True" or \verb"False", Default: \verb"False" \\
If a model file for WHIZARD should be written. 
\item \verb"WritUFO", values: \verb"True" or \verb"False", Default: \verb"False" \\
If model files in the UFO format should be written. 
\item \verb"effectiveOperators", Values: \verb"True" or \verb"False", Default: \verb"False" \\
If also higher dimensional operators should be calculated. By default, this concerns only four point interactions. 
\item \verb"SixParticleInteractions", Values: \verb"True" or \verb"False", Default: \verb"False" \\
If also the six-point interactions should be calculated. 
\item \verb"FeynmanDiagrams", Values: \verb"True" or \verb"False", Default: \verb"True"\\ 
If  Feynman diagrams for each vertex should be drawn in the \LaTeX  file. 
\item \verb"ReadLists", Values: \verb"True" or \verb"False", Default: \verb"False"\\
If the results of former calculations should be used to save time. 
\item \verb"IncludeRGEs", Values: \verb"True" or \verb"False", Default: \verb"True"\\ 
If the RGEs should be calculated.
\item \verb"TwoLoopRGEs", Values: \verb"True" or \verb"False", Default: \verb"True"\\ 
If the two loop RGEs should be calculated. (\verb"IncludeRGEs" must be set to \verb"True")
\item \verb"IncludeLoopCorrections", Values: \verb"True" or \verb"False", Default: \verb"True"\\
If the one-loop corrections to the self-energy and the tadpoles should be calculated. 
\end{enumerate}
The generated output will be saved in 
\begin{verbatim}
../\SARAH/Output/$MODEL/$EIGENSTATES/TeX/"
../\SARAH/Output/$MODEL/$EIGENSTATES/\FeynArts/"
../\SARAH/Output/$MODEL/$EIGENSTATES/CHep/"
../\SARAH/Output/$MODEL/$EIGENSTATES/UFO/"
../\SARAH/Output/$MODEL/$EIGENSTATES/WHIZARD/"
\end{verbatim}
%
% More details about the output is given in chapter sec.~\ref{output}. 
%

\chapter{The model files}
\label{modelfiles}
All information of the different models are saved in three different files which
have to be in one directory
\begin{verbatim}
.../\SARAH/Models/"ModelName"/
\end{verbatim} 
\begin{center}
\fbox{
\parbox{15cm}{
The directory name is equal to the name of the model and must contain a model
file with the same name!
}}
\end{center}
The three files are: one model file with the same name as the model directory
(\verb"ModelName.m"), a file containing additional information about the
particles of the model (\verb"particles.m") and a file containing additional
information about the parameters of the model (\verb"parameters.m"). Only the
first file is really necessary for calculating the Lagrangian and to get a first
impression of a model. However, for defining properties of parameters and
particles and for producing an appropriate output the other two files are
needed.  

\section{The model file}
The model file is the heart of \SARAH: the complete model is specified by the
model file. Thus, we will explain the general structure of the model file here,
and have a look at the different functions and its physical meaning in detail in
the next chapter. 
\subsection{Description}
The model file contains the following parts: First the gauge structure and the
particle content are given, and the matter interactions are defined by the
superpotential. These are general information needed for all eigenstates of the
model and must always be present. \\ 
This part is followed by the definition of the names for all eigenstates ({\tt
NameOfStates}). For these eigenstates can afterwards several properties be
defined using the corresponding {\tt DEFINITION} statements: decomposition of
scalars in scalar, pseudo scalar and VEV ({\tt DEFINITION[\$EIGENSTATES][VEVs]}),
rotations in the matter ({\tt DEFINITION[\$EIGENSTATES][MatterSector]}) and gauge
sector ({\tt DEFINITION[\$EIGENSTATES][GaugeSector]}). New couplings can be added
and existing couplings can be changed by hand ({\tt
DEFINITION[\$EIGENSTATES][Additional]}) and particles can be decomposed in the different flavors 
({\tt DEFINITION[\$EIGENSTATES][Flavors]})\\
Afterwards, the particles are defined, which should be integrated out or deleted.
At the end, the Dirac spinors have to be built out of Weyl spinors, a spectrum
file can be defined and a choice for automatically output can be made.  

\subsection{Schematic Structure}
The model file is structured as follows

\begin{center}
\underline{\bf{General information}}
\end{center}

\begin{enumerate}
\item Gauge structure of the model given by the vector superfields, see
sec.~\ref{vectorsuperfields}
\begin{verbatim}
Gauge[[1]] = {...
\end{verbatim}
\item Matter content given by the chiral superfields see
sec.~\ref{chiralsuperfields}
\begin{verbatim}
Fields[[1]] = { ...
\end{verbatim}

\item Superpotential, see sec.~\ref{superpotenialModel}
\begin{verbatim}
SuperPotential { ...
\end{verbatim}
\end{enumerate}
\begin{center}
\underline{\bf{Eigenstates}}
\end{center}

Names for the different eigenstates, see sec.~\ref{NameOfStates}
\begin{verbatim}
NameOfStates = { ...
\end{verbatim}

\begin{center}
\underline{\bf{Definition of properties for the different eigenstates}}
\end{center}

\begin{enumerate}
\item Vacuum expectation values, see sec.~\ref{VEV}
\begin{verbatim}
DEFINITION[$EIGENSTATES][VEVs]= { ...
\end{verbatim}
\item Rotations in the gauge sector, see sec.~\ref{mixGauge}
\begin{verbatim}
DEFINITION[$EIGENSTATES][GaugeSector]= { ...
\end{verbatim}
\item Expansion of flavors, see sec.~\ref{flavorExp}
\begin{verbatim}
DEFINITION[$EIGENSTATES][Flavors]= { ...
\end{verbatim}
\item Rotations in the matter sector, see sec.~\ref{mixPart}
\begin{verbatim}
DEFINITION[$EIGENSTATES][MatterSector]= { ...
\end{verbatim}
\item Gauge fixing terms: since version 3.1  no longer needed as input, see sec.~\ref{GaugeFixing}
% \begin{verbatim}
% DEFINITION[$EIGENSTATES][GaugeFixing]= { ...
% \end{verbatim}
\item Additional couplings or redefinition of existing couplings, see
sec.~\ref{ReDef}
\begin{verbatim}
DEFINITION[$EIGENSTATES][Additional]= { ...
\end{verbatim}
\item Definition of Dirac spinors, see sec.~\ref{diracSpinors}
\begin{verbatim}
DEFINITION[$EIGENSTATES][DiracSpinors]= { ...
\end{verbatim}
\end{enumerate}

\begin{center}
\underline{\bf{Additional, general Information}}
\end{center}

\begin{enumerate}
\item Integrating out or deleting  particles, see sec.~\ref{IntOutorDel}
\begin{verbatim}
IntegrateOut[[1]] = { ...
DeleteParticles[[1]] = { ...
\end{verbatim}
% \item Automatized output, see sec.~\ref{BatchMode}
% \begin{verbatim}
% makeOutput = { ...
% \end{verbatim}
\item Assigning a  spectrum file, see sec.~\ref{addNumValues}
\begin{verbatim}
SpectrumFile = ...
\end{verbatim}
\end{enumerate}
\section{Parameter file}
\label{parameterFile}
\subsection{General}
The information of the parameter file are needed for some in- and output
routines. Also, it is possible to define simplifying  assumptions for the
parameters. The parameter file consists of a list called
\verb"ParameterDefinition". This is an array with two columns: the first column
gives the name of the parameter, the second column defines the properties of the
parameter. These properties can be:
\begin{enumerate}
\item \verb"Description": Value: a string \\
A string for defining the parameter, see sec.~\ref{sec:GlobalDefinitions}
\item \verb"Real", Value: \verb"True" or \verb"False", Default: \verb"False" \\
  Defines, if a parameter is always taken to be real.
\item \verb"Form", Value: \verb"None", \verb"Diagonal" or \verb"Scalar",
Default: \verb"None" \\
\item \verb"Symmetry", Value: \verb"None", \verb"Symmetric", \verb"AntiSymmetric" or \verb"Hermitian",
Default: \verb"None" \\
For an explanation of the different options see sec.~\ref{assumptions}.
\item \verb"LaTeX", Value: \verb"None" or \LaTeX\ name \\ 
The name of the parameter used in the \LaTeX\ output. Standard \LaTeX\ language
should be used (\verb"\" has to be replaced by \verb"\\"). 
\item \verb"Dependence", Value: \verb"None" or a function \\
The parameter is always replaced by this definition, see
sec.~\ref{sec:ParameterFile}
\item \verb"DependenceOptional", Value: \verb"None" or a  function \\
It can be chosen during the work, if the parameter is replaced by the made
definition, see sec.~\ref{sec:ParameterFile}
\item \verb"DependenceNum", Value: \verb"None" or a  function \\
This definition is used in numerical calculations, see
sec.~\ref{sec:ParameterFile}
\item \verb"MatrixProduct", Value: \verb"None" or a list of two matrices\\
The parameter is defined as a product of two matrices, see
sec.~\ref{sec:ParameterFile}
\item \verb"LesHouches", Value: \verb"None" or the position in LesHouches input file
 \\
The numerical value of the parameter can be given by a LesHouches file, see
sec.~\ref{sec:ParameterFile}
\item \verb"Value", Value: \verb"None" or a number \\
A numerical value for the parameter can be chosen, see
sec.~\ref{sec:ParameterFile} 
\end{enumerate}
\subsection{Simplifying assumptions}
\label{assumptions}
\SARAH normally handles parameters in the most general way: most parameters
assumed to be complex and all tensors can have off diagonal values. This can be
changed by certain statements in the parameter file. First, it is possible to
define a parameter as real by setting
\begin{verbatim}
Real -> True
\end{verbatim}
The gauge couplings are by default assumed to be real. \\ 
The degrees of freedom for a tensor valued parameter \(T\) can be reduced by
using the \verb"Form" statement with the following options
\begin{enumerate}
\item \verb"Diagonal": only diagonal entries are assumed to be unequal from
zero: \(T_{ij} \rightarrow \delta_{ij} T_{ij}\)
\item \verb"Scalar" the tensor is replaced by a scalar: \(T_{ij} \rightarrow T\)
\end{enumerate}
Furthermore, some symmetries can be defined for a parameter by using the option
\verb"Symmetry"
\begin{enumerate}
\item \verb"Symmetric":  the tensor is assumed to be symmetric: \(T_{ij} =
T_{ji}\) for \(i>j\)
\item \verb"AntiSymmetric":  the tensor is assumed to be anti-symmetric: 
\(T_{ij} = - T_{ji}\) for \(i>j\)
\item \verb"Hermitian": the tensor is assumed to be hermitian:  \(T_{ij} =
T^*_{ji}\) for \(i>j\)
\end{enumerate}

\paragraph*{Examples}
\begin{enumerate} 
 \item CP and Flavor conserving Yukawa matrices are defined by
\begin{verbatim}
{Yu,  { Real -> True,
        Form -> Diagonal}}
\end{verbatim} 
\end{enumerate}

\subsection{Defining values and dependences for parameters}
\label{sec:ParameterFile}
\subsubsection{Dependences}
There are different possibilities to define dependences between parameters by
using the \verb"Dependence" statements. The difference between the three dependence
statements is the time at which the relations are used 
\begin{enumerate}
\item \verb"Dependence": The relations are always used. The corresponding
substitutions are saved in \verb"subAlways".
\item \verb"DependenceOptional": The relations are only used if the option
\verb"UseDependence" is set to \verb"True" as option for specific commands, e.g. when calculating vertices. The
substitutions are saved in \verb"subDependences".
\item \verb"DependenceNum": The dependences are only used when a numerical value
for the parameter is calculated. The substitutions are saved in \verb"subNum". 
\end{enumerate} 
The indices of vectors or tensors are implicitly assumed to be \verb"index1",
\verb"index2", \dots.  This in combination with  \verb"sum[index,start,final]"
can be used in the following way
\begin{verbatim}
{X, {Dependences -> sum[n1,1,3] sum[n2,1,3] A[index1,n1] Y[n1,n2] B[n2,index2]}}
\end{verbatim} 
and is interpreted as
\begin{equation}
X_{i_1 i_2} \rightarrow \sum_{n_1 = 1}^3 \sum_{n_2 = 1}^3 A_{i_1 n_1} Y_{n_1
n_2} B_{n_2 i_2}
\end{equation}

Matrix valued parameters can also defined as matrix product of other matrices by using
\verb"MatrixProduct". The argument must be a list consisting of two matrices of
same dimension:
\begin{verbatim}
{X, {MatrixProduct-> {A,B} }} 
\end{verbatim}
Using this definition, every appearance of a matrix product of \(A^\dagger B\)
is replaced by \(X\) and \(B^\dagger A\) by \(X^\dagger\) and can be used, for instance,
to define the CKM matrix. 

\paragraph*{Examples}
\begin{enumerate}
\item One dependence, which might be used always, is the parametrization of a
mixing matrix matrix by a mixing angle: the mixing of the charged Higgs (\verb"ZP") in
the MSSM can be parametrized by the mixing angle \(\beta\). This is defined in
\SARAH via
\begin{verbatim}
{ZP,  { Dependence -> {{-Cos[\[Beta]],Sin[\[Beta]]},
                       {Sin[\[Beta]],Cos[\[Beta]]}}
\end{verbatim}
\item The relation between the gauge couplings \(g_1\) and \(g_2\) and the
electric charge \(e\) is an example for an optional dependence. This relation is
defined by
\begin{verbatim}
{g1,  {DependenceOptional -> e/Cos[ThetaW]}}
\end{verbatim}
Now, the result for vertices can be expressed in terms of the electric charge by using
\begin{verbatim}
Vertex[List of Particles, UseDependences -> True]
\end{verbatim}
\item A relation, which might only be used for numerical calculations, is the
relation between the Weinberg angle and the gauge couplings:
\begin{verbatim}
{ThetaW, {DependenceNum ->  ArcCos[g2/Sqrt[g1^2+g2^2]]}
\end{verbatim}
\item The CKM matrix is defined as the product of two rotation matrices:
\begin{verbatim}
{CKM, {MatrixProduct -> {Vd,Vu} }}
\end{verbatim}
\end{enumerate}

\subsubsection{Numerical values}
If the considered parameter does not depend on other parameters, there are two
ways to assign a numerical value to this parameter:
\begin{enumerate}
\item \verb"Value": Adds directly a numerical value to the parameter definition
\item \verb"LesHouches": Defines the position of the numerical value for the
parameter in a LesHouches spectrum file (see also \cite{LesHouches}). The
statement has to has the following form:
\begin{enumerate}
\item If the dimension of the parameters is known by \SARAH, it is sufficient to
give just the name of the block, e.g. 
\begin{verbatim}
{Yu, { LesHouches -> Yu } };
\end{verbatim}
\item If the block appears several times in the LesHouches file, it is possible
to give the number of appearance as optional argument:
\begin{verbatim}
{g3,  {LesHouches -> {gauge,3}}}; 
\end{verbatim}
This reads the first block which are normally the GUT-scale values. However,
\begin{verbatim}
{g3,  {LesHouches -> {{gauge,3},2}}}; 
\end{verbatim}
reads the second block which are the values at the SUSY-scale. 
\end{enumerate}
\end{enumerate}

\paragraph*{Example}
\begin{enumerate}
\item The CKM matrix can be defined in the Wolfenstein parametrization as
\begin{verbatim}
{{Description ->"CKM Matrix", 
			 LaTeX -> "V^{CKM}",
             MatrixProduct -> {Vd,Vu},
             Dependence ->  None, 
             DependenceNum -> {{1-1/2*lWolf^2,lWolf,aWolf*lWolf^3*Sqrt[rWolf^2+nWolf^2]},
                               {-lWolf,1-1/2*lWolf^2, aWolf*lWolf^2},
                               {aWolf*lWolf^3*Sqrt[(1-rWolf)^2 +nWolf^2],-aWolf*lWolf^2,1}},
             LesHouches -> VCKM,
             DependenceSPheno -> Matmul[Transpose[conj[Vd]],Vu],
             OutputName-> VCKM      }}, 
\end{verbatim}
\item The Wolfenstein parameters are real and the experimental values are known
\begin{verbatim}
{{Description->"Wolfenstein Parameter eta", 
             Value -> 0.341,
             Real -> True,
             OutputName-> nWolf,
             LesHouches -> {WOLFENSTEIN,4} 	            }},
\end{verbatim} 
\end{enumerate}

\section{Particles File}
\label{particleFile}
The particle file  contains information about all fields of the model. This information is usually needed needed for
an appropriate output. 
\begin{enumerate}
\item \verb"Description": A string for defining the particle
\item \verb"RParity": The R-Parity of a particle: +1 or -1. If not defined, +1
is used.
\item \verb"PDG": The PDG number. Needed, if the written model files should be
readable by event generators or if masses are given by a LesHouches file. 
\item \verb"PDG.IX": Defines a nine-digit number of a particle supporting the proposal
Ref.~\cite{Brooijmans:2012yi,Basso:2012ew}. By default, the entries of {\tt PDG} are used. 
To switch to the new scheme,
 either at the beginning of a \SARAH session or in the model files, the
 following statement has to be added:
\begin{verbatim}
UsePDGIX = True;
\end{verbatim}
\item \verb"ElectricCharge": Defines in the electric charge of a particle in units of $e$. This
information is exported to the {\tt CalcHep} and UFO model files. 
\item \verb"Width": The width of the particle. If not defined, 0 is used. 
\item \verb"Mass": The options of defining the mass of a particle are:
\begin{enumerate}
\item Numerical Value: A numerical value for the mass of the particle is given.
\item \verb"Automatic": \SARAH derives the tree level expression for the mass
from the Lagrangian and calculates the value by using the values of the other
parameters.  
\item \verb"LesHouches:" \SARAH reads the mass from a LesHouches file. 
\end{enumerate}  
\item \verb"LaTeX": The name of the particle in \LaTeX\ files in standard \LaTeX\
language. If not defined, the Mathematica \verb"InputForm" of the particle name
is used. It is also possible to give a list with two entries corresponding to the name of
the particle and anti-particle.
\item \verb"FeynArtsNr": The number of the particle in a \FeynArts model file.
If not defined, the number will be generated automatically: the smallest free number is taken.
\item \verb"Output": A short form the particle name consisting of two letter
with no no-standard signs. Needed, to make sure that all programs outside \SARAH
can read the name correctly. If not defined, the Mathematica \verb"InputForm" is
used and potentially truncated.   
\item \verb"LHPC": Defines the column and color used for the particle in the steering file
for the  {\tt LHPC Spectrum Plotter}. All colors defined in {\tt gnuplot} can be used.
\end{enumerate}

This information must be given for all eigenstates in arrays named
\begin{verbatim}
ParticleDefinition[$EIGENSTATES]
\end{verbatim}
Only for the \LaTeX output also the names of the Weyl spinors and intermediate
states (like scalar and pseudo scalar components of Higgs) should be given to
improve the look and readability of the produced pdf file. 
\paragraph*{Example}
For the eigenstates after EWSB, an entry might look like
\begin{verbatim}
ParticleDefinitions[EWSB} = {
   ...
   {Sd ,  {  Description -> "Down Squark",
             RParity -> -1,
             PDG ->  {1000001,2000001,1000003,2000003,1000005,2000005},
             PDG.IX -> {-200890201,-200890202,-200890203,-200890204,-200890205,-200890206}
             Mass -> Automatic,
             FeynArtsNr -> 14,
             LaTeX -> "\\tilde{d}",,
             ElectricCharge -> -1/3,
             LHPC -> {7, "cyan"},
             OutputName -> "sd" }},
   ... }
\end{verbatim}
\section{Global definitions}
\label{sec:GlobalDefinitions}
It is also possible to define global properties for parameters or particles
which are present in more than one model file. These properties are afterwards
used for all models. The global definitions are saved in the files
\verb"particles.m" and \verb"parametes.m" directly in the main model directory.
For each parameter or particle, an entry like 
\begin{verbatim}
{{        Descriptions -> "Down Squark", 
          RParity -> -1,
          PDG ->  {1000002,2000002,1000004,2000004,1000006,2000006},
          Width -> Automatic,
          Mass -> Automatic,
          FeynArtsNr -> 13,   
          LaTeX -> "\\tilde{u}",
          OutputName -> "um" }},   
\end{verbatim}
can be added. In particular, the entry \verb"Description" is important. This
should be an unique identifier for each particle or parameter. This identifier
can later on be used in the different files of the different models, e.g.     
\begin{verbatim}
{Su ,  {  Descriptions -> "Down Squark"}},   
\end{verbatim}
Of course, it is also possible to overwrite some of the global definitions by
defining them locally, too. For instance, to use \verb"u" instead of \verb"um"
as output name in a specific model, the entry should be changed to  
\begin{verbatim}
{Su ,  {  Descriptions -> "Down Squark",
          OutputName -> "u" }},   
\end{verbatim}
in the corresponding particle file of the model.
\chapter{Definition of models}
\section{Particle content}
\subsection{Vector superfields}
\label{vectorsuperfields}
The vector superfields are defined by the array \verb"Gauge". An entry reads
\begin{verbatim}
Gauge[[i]]={Superfield Name, Dimension, Name of Gauge Group, Coupling, Expand};
\end{verbatim}
The different parts have the following meaning:
\begin{enumerate}
\item \verb"Superfield name":\\
This is the name for the vector superfield and also the basis of the names for
vector bosons and gauginos as explained in sec.~\ref{nomenclature}
\item \verb"Dimension": \\
 This defines the dimension of the \(SU(N)\) gauge group: \verb"U[1]" for an
Abelian gauge group or \verb"SU[N]" with integer N for a non-Abelian gauge
group.
\item \verb"Name of Gauge Group": \\
This is the name of the gauge group, e.g. hypercharge, color or left. This
choice is import because all matter particles charged under a non-Abelian gauge
group carry an corresponding index. The name of the index consists of the first
three letter of the name plus a number. Hence, it must be taken care that the
first three letters of different gauge group names are not identical. Also the name
for the indices in the adjoint representation are derived from this entry. 
\item \verb"Coupling": The name of the coupling constant, e.g. \verb"g1" 
\item \verb"Expand": Values can be \verb"True" or \verb"False". If it is set to
\verb"True", all sums over the corresponding indices are evaluated during the
calculation of the Lagrangian. This is normally done non-Abelian gauge groups
which get broken like the \(SU(2)_L\) in the MSSM. 
\end{enumerate}
\SARAH adds for every vector superfield a soft-breaking gaugino mass
\begin{verbatim}
Mass<>"Superfield Name"
\end{verbatim}

\paragraph*{Example: Standard model color group}
\begin{verbatim}
Gauge[[3]] = {G,  SU[3],  color,  g3, False};
\end{verbatim}
The consequence of this entry is
\begin{enumerate}
\item Gluon and gluino are named \verb"VG" respectively \verb"fG"
\item The \(SU(3)\) generators, the Gell-Mann matrices, are used
\item The color index is abbreviated \verb"colX" (for \verb"X" = 1,2, ...)
\item The adjoint color index is abbreviated \verb"acolX" (for \verb"X" = 1,2, ...)
\item The strong coupling constant is named \verb"g3"
\item The sums over the color indices are not evaluated
\end{enumerate}

\paragraph*{Models with several $U(1)$ gauge groups}
The general, gauge invariant  Lagrangian for a theory including two $U(1)$ gauge
groups reads
\begin{equation}
 L = - \frac{1}{4} \hat{F}^{a, \mu \nu} \hat{F}^a_{\mu \nu} - \frac{1}{4}
\hat{F}^{b, \mu \nu} \hat{F}^b_{\mu \nu} - \frac{\sin \chi}{2}  \hat{F}^{a, \mu
\nu} \hat{F}^b_{\mu \nu} | \hat{D}_\mu \phi|^2 + i \bar{\Psi} \gamma^\mu
\hat{D}_\mu \Psi 
\end{equation}
with the covariant derivative
\begin{equation}
\hat{D} = \partial_\mu - i g_a^0 Q_a \hat{A}_\mu^a - i g_b^0 Q_b \hat{A}_\mu^b .
\end{equation}
It is possible to perform a field rotation to remove the mixing term between the
field strength tensors. In that case, the covariant derivate changes to
\begin{equation}
\label{eq:off_cov}
 D_\mu = \partial_\mu - i (\bar{g}_a Q_a + \bar{g}_{ba}) \bar{A}^a_\mu - i
(\bar{g}_{ab} Q_a + \bar{g}_b Q_b) \bar{A}^b_\mu .
\end{equation}
\SARAH uses eq.~(\ref{eq:off_cov}) to write the Lagrangian for models containing
several Abelian gauge groups. For that purposes, it generates new gauge
couplings
\begin{verbatim}
gA<>gB 
\end{verbatim}
for the off-diagonal couplings. Here \verb"gA" and \verb"gB" are the names for
the diagonal gauge couplings defined in \verb"Gauge". In addition, the gaugino
mass terms are written as
\begin{equation}
 \sum_i \sum_j M_{ij} \lambda_i \lambda_j + \mbox{h.c.} .
\end{equation}
The sum \(i\) and \(j\) runs over all Abelian gauge groups. The names for the
off-diagonal gaugino mass are
\begin{verbatim}
Mass<>A<>B 
\end{verbatim}
Here, \verb"A" and \verb"B" are the names of the vector superfields defined in
\verb"Gauge". \\
Note, to include these mixing effects in the RGEs, there have been done some
changes with respect to \cite{Martin:1993zk}. See section \ref{sec:U1RGE}.

\paragraph*{Example}
In the case of a gauge sector containing
\begin{verbatim}
Gauge[[1]] = {R,  U[1],  right,  gR, False};
Gauge[[2]] = {BL,  U[1],  bminusl,  gBL, False};
\end{verbatim}
the off-diagonal gauge couplings are called
\begin{verbatim}
gRgBL
gBLgR
\end{verbatim}
and the off-diagonal gaugino masses are
\begin{verbatim}
MassRBL
MassBLR
\end{verbatim}

\subsection{Chiral Superfields}
\label{chiralsuperfields}
Chiral superfields are defined as follows:
\begin{verbatim}
Field[[i]] = {Components, Generations, Superfield Name, Transformation 1,
Transformation 2... };
\end{verbatim}

\begin{enumerate}
\item \verb"Components": The basis of the name for the components. Two cases are
possible:
\begin{enumerate}
 \item The field transforms only trivially under the gauge groups with expanded
indices. In this case, the entry is one dimensional.
 \item The field transforms non-trivially under gauge groups with expanded
indices. In this case, the entry is a vector or higher dimensional tensor
fitting to the dimension of the field. Note, representations larger than the fundamental one 
are written as tensor products
\end{enumerate}
\item \verb"Generations": The number of generations
\item \verb"Superfield Name": The name for the superfield
\item \verb"Transformation 1", \verb"Transformation 2",..: Transformation under
the different gauge groups defined before. For \(U(1)\) this is the charge, for
non-Abelian gauge groups the dimensions is given as integer respectively
negative integer. The dimension D of an irreducible representation is not
necessarily unique. Therefore, to make sure, \SARAH uses the demanded
representation, also the corresponding Dynkin labels have to be added.
\end{enumerate}
\SARAH adds automatically for all chiral superfields soft-breaking squared
masses named
\begin{verbatim}
m <> "Name of Superfield" <> 2
\end{verbatim}
In addition, there are mixed soft-breaking parameters added which are allowed by
gauge invariance and R-parity. The last check can be disabled by setting
\begin{verbatim}
RParityConservation = False;
\end{verbatim}
in the model file. 
\paragraph*{Examples}
\begin{enumerate}
\item {\bf Fields with expanded indices} The definition of the left quark
superfield in the MSSM is
\begin{verbatim}
Field[[1]] = {{uL,  dL},  3, q,   1/6, 2, 3};
\end{verbatim}
The consequence of this definition is
\begin{enumerate}
\item Left up-squarks and quarks are called \verb"SuL" / \verb"FuL"
\item Left down-squarks and quarks are called \verb"SdL" / \verb"FdL"
\item There are three generations 
\item The superfield is named \verb"q"
\item The soft-breaking mass is named \verb"mq2"
\item The hypercharge is \(\frac{1}{6}\)
\item The superfield transforms as {\bf 2} under \(SU(2)\)
\item The superfield transforms as {\bf 3} under \(SU(3)\)
\end{enumerate}

\item {\bf Fields with no expanded indices} The right down-quark superfield is
defined in the MSSM as
\begin{verbatim}
Field[[3]] = {{conj[dR]}, 3, d, 1/3, 1, -3};
\end{verbatim}
The meaning is
\begin{enumerate}
\item The right squarks and quarks are called \verb"SdR" and \verb"FdR"
\item There are three generations
\item The Superfield name is \verb"d"
\item The soft-breaking mass is named \verb"md2"
\item The hypercharge is \(\frac{1}{3}\)
\item It does not transform under \(SU(2)\)
\item It does transform as \({\bf \bar{3}}\) under \(SU(3)\) 
\end{enumerate}

\item {\bf Specification of representation } Since the {\bf 10} under \(SU(5)\)
is not unique, it is necessary to add  the appropriate Dynkin labels, i.e. 
\begin{verbatim}
Field[[1]] = {Ten, 1, t, {10,{0,1,0,0}},...}; 
\end{verbatim}
or
\begin{verbatim}
Field[[1]] = {Ten, 1, t, {10,{0,0,1,0}},...}; 
\end{verbatim}

\item {\bf Mixed soft-breaking terms} In models which contain fields with the
same quantum numbers and same R-parity mixed soft-breaking terms are added. For
instance, in models with heavy squarks
\begin{verbatim}
Field[[3]] = {{conj[dR]}, 3, d, 1/3, 1, -3};
...
Field[[10]] = {{conj[dRH]}, 3, DH, 1/3, 1, -3};
\end{verbatim}
the term of the form
\begin{verbatim}
mdDH (conj[SdR] SdRH + SdR conj[SdRH])
\end{verbatim}
is automatically added. For the MSSM with $R$-parity violation, also the term
\begin{verbatim}
mlHd (conj[Sl] SHd + Sl conj[SHd])
\end{verbatim}
is created, when the flag {\tt RParityConservation = False} is used in the model file.
\end{enumerate}

\section{Superpotential}
\label{superpotenialModel}
The definition of the superpotential is straight forward
\begin{verbatim}
SuperPotential = {{{Coefficient,Parameter,(Contraction)},
                      {Particle 1, Particle 2, ...} }, ...}
\end{verbatim}
Each term of the superpotential is defined due to two list:
\begin{enumerate}
 \item The first list is two dimensional and defines a numerical coefficient
(\verb"Coefficient") and the name of the parameter (\verb"Parameter")
 \item The second list gives the involved superfields
 \item The contraction of the indices can be given optionally.
\end{enumerate}
Up to four superfields are possible, i.e. also effective operators can be
considered. 

\paragraph*{Contraction of indices}
The indices of the involved particles are automatically contracted by \SARAH.
Sometimes, there are more possibilities to contract all indices. Therefore, it
is also possible to fix the contraction of each term. The contraction used by
\SARAH can be seen by  
\begin{verbatim}
ShowSuperpotentialContractions; 
\end{verbatim}

\paragraph*{Properties of couplings and soft-breaking terms}
If the particles involved in the different interactions have more than one
generation, the couplings are in general complex tensors carrying up to three
generation indices. Assumptions like diagonality or no CP violation can added by
using the parameter file (see sec.~\ref{parameterFile}). \\
The corresponding soft-breaking term to each superfield coupling is
automatically added to the Lagrangian. The soft-breaking couplings carry the
same indices as the superpotential coupling. They are named as 
\begin{enumerate}
 \item Quartic terms: {\tt Q["Name of Coupling"]}
 \item Trilinear terms: {\tt T["Name of Coupling"]}
 \item Bilinear terms: {\tt B["Name of Coupling"]}
 \item Linear terms:  {\tt L["Name of Coupling"]}
\end{enumerate}
Simplifying assumptions for the soft-breaking terms can be made independently of
the assumptions for the superpotential parameters in the parameter file. \\

\paragraph*{Example} 
The term involving the up Yukawa coupling is
\begin{equation}
Y^u_{n_1,n_2} \hat{q}^i_{\alpha,n_1} \epsilon^{ij} \hat{H}_u^j
\hat\bar{{u}}_{\beta,n_2} \delta_{\alpha,\beta} 
\end{equation}
This can be defined in \SARAH with
\begin{verbatim}
 {{1, Yu},{q,Hu,u}}
\end{verbatim}
The explicit contraction of the indices would read as
\begin{verbatim}
 {{1, Yu,Delta[col1,col3] Eps[lef1,lef2] },{q,Hu,u}}
\end{verbatim}
There is a soft-breaking term automatically added to Lagrangian, which has the
same meaning as
\begin{equation}
T(Y^u)_{n_1,n_2} \tilde{q}^i_{\alpha,n_1} \epsilon^{ij} H_u^j
\tilde{u}^*_{\beta,n_2} \delta_{\alpha,\beta} 
\end{equation}

\section{Excluding terms or adding Dirac gaugino mass terms}
As shown in app.~\ref{sec:Lagrangian} \SARAH uses the information given so far about the vector and chiral superfields to calculate the entire Lagrangian for the 
gauge eigenstates. This includes in general the kinetic terms, the matter interactions from the superpotential, D- and F-terms
as well as soft-breaking terms. However, it is also possible to exclude parts by the following statements:
\begin{itemize}
 \item \verb"AddTterms = True/False;", default: \verb"True", includes/excludes trilinear softbreaking couplings
 \item \verb"AddBterms = True/False;", default: \verb"True", includes/excludes bilinear softbreaking couplings
 \item \verb"AddLterms = True/False;", default: \verb"True", includes/excludes linear softbreaking couplings
 \item \verb"AddSoftScalarMasses = True/False;", default: \verb"True", includes/excludes soft-breaking scalar masses
 \item \verb"AddSoftGauginoMasses = True/False;", default: \verb"True", includes/excludes Majorana masses for gauginos
 \item \verb"AddSoftTerms = True/False;", default: \verb"True", includes/excludes all soft-breaking terms
 \item \verb"AddDterms = True/False;", default: \verb"True", includes/excludes all D-terms
 \item \verb"AddFterms = True/False;", default: \verb"True", includes/excludes all F-terms
\end{itemize}
On the other hand, Dirac mass terms for gauginos are not written automatically if chiral superfields in the adjoint representation are present. The reason for this is just that models without these terms are still more common, e.g. the NMSSM is usually studied without a bino-singlino Dirac mass term. Therefore, to include Dirac mass terms for gauginos, one has to add explicitly
\begin{verbatim}
AddDiracGauginos = True;
\end{verbatim}
to the model file. In this case \SARAH writes down all possible mass terms between chiral and vector superfields and the corresponding D-terms for the model. In this context, it uses \verb"MD<>v<>c" as name for the new mass parameters where \verb"v" is the name of the vector and \verb"c" the name of the chiral superfield.  If several fields in the adjoint representation of one gauge group are present, \SARAH will generate the corresponding terms for all of them. To remove some of them, the parameters can be put to zero in the parameters file of the model definition. Furthermore, if several Abelian gauge groups are present, the impact of kinetic mixing is also respected. 
\paragraph*{Example}
One can add to the particle content of the MSSM, fields in the adjoint representations of the different gauge groups:
\begin{verbatim}
Fields[[8]] = {s, 1, S, 0, 1, 1};
Fields[[9]] = {{{T0/Sqrt[2],Tp},{Tm, -T0/Sqrt[2]}}, 1, T, 0, 3, 1};
Fields[[10]] = {Oc, 1, oc, 0, 1, 8}; 
\end{verbatim}
Here, the triplet superfield is defined as
\begin{equation}
\hat{T} = \left(\begin{array}{cc} \hat{T}^0/\sqrt{2} & \hat{T}^+ \\ \hat{T}^- & - \hat{T}^0/\sqrt{2} \end{array} \right)
\end{equation}
and to include the Dirac mass terms, use
\begin{verbatim}
AddDiracGauginos = True;
\end{verbatim}
This leads to the corresponding mass term and the D-terms. \SARAH names the new parameters using the corresponding superfield names as \verb"MDBS" (bino-singlet mass term), \verb"MDWBT" (wino-triplet mass term), \verb"MDGoc" (gluino-octet mass term).

\section{Properties of different eigenstates}
\label{NameOfStates}
For defining the properties of the different sets of eigenstates, the
\verb"DEFINITION" statement is used:
\begin{verbatim}
DEFINITION[$EIGENSTATES]["Property"]= {...};
\end{verbatim}
The possible properties are \verb"VEVs", \verb"GaugeSector",
\verb"MatterSector", \verb"Phase", \verb"Flavors" and
\verb"Additional".\\
But, before defining this properties, the names for all eigenstates must be
fixed in the correct order. This is done due to 
\begin{verbatim}
NameOfStates={List of Name};
\end{verbatim}
This list can, in principle, be arbitrary long. Common entries are, e.g. 
\begin{verbatim}
NameOfStates={GaugeES, EWSB}
NameOfStates={GaugeES, SCKM, EWSB}
\end{verbatim}

\subsection{Vacuum expectation values}
\subsubsection{Introduction}
\label{VEV}
The particles responsible for breaking a gauge symmetry receive a VEV. After the
symmetry breaking, these particles are parametrized by a scalar \(\phi\) and a
pseudo scalar \(\sigma\) part and the VEV \(v\): 
\begin{equation}
S = \frac{1}{\sqrt{2}} \left( \phi_S + i \sigma_S + v_S \right)
\end{equation}

\subsubsection{Implementation in \SARAH}
This is in \SARAH done by
\begin{verbatim}
DEFINITION[$EIGENSTATES][VEVs] = 
{Particle Name, {{VEV, Coefficient 1}, 
      {Pseudoscalar, Coefficient 2},{Scalar, Coefficient 3},({Phase})};
\end{verbatim}
\begin{enumerate}
\item \verb"Name": The name of the particle receiving a VEV
\item \verb"VEV": Name of the VEV
\item \verb"Scalar": Name of the scalar component
\item \verb"Pseudoscalar": Name of the pseudo scalar component
\item \verb"Coefficient 1,2,3": The different (numerical) coefficients. 
\item \verb"Phase": Optional phase 
\end{enumerate}
All indices carried by the particle receiving the VEV are automatically added to
the scalar and pseudo scalar part. The scalar, pseudo scalar and the VEV are
handled as real parameters in \SARAH. The phase is only an optional argument and
can be skipped for Higgs sectors without CP violation. \\

\paragraph*{Example}
In the MSSM, the Higgs \(H_d^0\) and $H_u^0$ get  VEVs \(v_d\) and $v_u$:
\begin{equation}
H_u^0 = \frac{1}{\sqrt{2}} \left(v_u + i \sigma_u +\phi_u \right) \, , \hspace{1cm}
H_d^0 =  \frac{1}{\sqrt{2}} \left(v_d + i \sigma_d +\phi_d \right)
\end{equation}
This is done in \SARAH by using
\begin{verbatim}
DEFINITION[EWSB][VEVs]=
{{SHd0, {vd, 1/Sqrt[2]}, {sigmad, I/Sqrt[2]},{phid,1/Sqrt[2]}},
 {SHu0, {vu, 1/Sqrt[2]}, {sigmau, I/Sqrt[2]},{phiu,1/Sqrt[2]}},
};
\end{verbatim}
To add a relative phase, use
\begin{verbatim}
DEFINITION[EWSB][VEVs]=
{{SHd0, {vd, 1/Sqrt[2]}, {sigmad, I/Sqrt[2]},{phid,1/Sqrt[2]}},
 {SHu0, {vu, 1/Sqrt[2]}, {sigmau, I/Sqrt[2]},{phiu,1/Sqrt[2]},{eta}},
};
\end{verbatim}
This is interpreted as
\begin{equation}
H_u^0 = \frac{e^{i \eta}}{\sqrt{2}} \left(v_u + i \sigma_u +\phi_u \right) \, , \hspace{1cm}
H_d^0 = \frac{1}{\sqrt{2}} \left(v_d + i \sigma_d +\phi_d \right)
\end{equation}

\subsection{Decomposition of Flavors}
\label{flavorExp}
\subsubsection{Introduction}
If a model without flavor violation is considered, it might be demanded to give
each generation of a family an own name, e.g. use different symbols for
electron, muon and tau.

\subsubsection{Implementation in \SARAH}
The expansion of flavors is done by
\begin{verbatim}
DEFINITION[EWSB][Flavors]= { ...
 {Field, {Name 1, Name 2, ... }},
 ... }
\end{verbatim}
There must be as many names as generations of \verb"Field" exist. 
\paragraph*{Example}
The down-type quarks can be expanded as follows:
\begin{verbatim}
DEFINITION[EWSB][Flavors]= {
 {Fd0L, {FdL,FsL,FbL}},
 {Fd0R, {FdR,FsR,FbR}} }
\end{verbatim}

\subsection{Mixings in the Gauge Sector}
\label{mixGauge}
\subsubsection{Introduction}
After breaking a gauge symmetry, the vector bosons mix among each other as well
as the gauginos do. In general, the different generations of a vector boson or
gaugino rotate to different mass eigenstates. Hence, it is not possible to use
the same parametrization as in the matter sector, shown in sec.~\ref{mixPart}.
\\
\subsubsection{Implementation in \SARAH}
It is possible to define the mixing in the gauge sector in the following way:
 \begin{verbatim}
DEFINITION[$EIGENSTATES][GaugeSector]=
{ {{Old 1a, Old 2a,...},{New 1a, New 2b,..},MixingMatrix 1},
  {{Old 1b, Old 2b,...},{New 1b, New 2b,..},MixingMatrix 2},
  ...
\end{verbatim} 
Here, \verb"Old Nx" is the name of the old and \verb"New Nx" of rotated eigenstates. \verb"MixingMatrix X"
is the rotatin matrix relating the old and new basis. \SARAH interprets this
definition as matrix multiplication:
\begin{equation}
(N_1, N_2,\dots)^T = M (O_1, O_2,\dots)^2
\end{equation}

\paragraph*{Example}
We consider the electroweak symmetry breaking of the MSSM
\begin{eqnarray}
W_1 &=& \frac{1}{\sqrt{2}} \left(W^- + {W^-}^* \right) \thickspace ,\\
W_2 &=& i \frac{1}{\sqrt{2}} \left({W^-}^* - W^- \right) \thickspace , \\
W_3 &=& \sin\Theta_W \gamma + \cos\Theta_W Z \thickspace , \\
B &=& \cos\Theta_W \gamma - \sin\Theta_W Z \thickspace ,
\end{eqnarray}
% This is done by
% \begin{verbatim}
% {VWB, {1,{VWm,1/Sqrt[2]},{conj[VWm],1/Sqrt[2]}},
%       {2,{VWm,-\[ImaginaryI]/Sqrt[2]},{conj[VWm],\[ImaginaryI]/Sqrt[2]}},
%       {3,{VP, Sin[ThetaW]},{VZ, Cos[ThetaW]}}},
% {B,   {1,{VP, Cos[ThetaW]},{VZ, -Sin[ThetaW]}}}
% \end{verbatim}
% in \SARAH. 
This is given in \SARAH by
\begin{verbatim}
DEFINITION[EWSB][GaugeSector]= 
{ {{VB,VWB[3]},{VP,VZ},ZZ},
  {{VWB[1],VWB[2]},{VWm,conj[VWm]},ZW},
  {{fWB[1],fWB[2],fWB[3]},{fWm,fWp,fW0},ZfW}
};    
\end{verbatim}
The rotation matrices $Z^{\gamma Z}$ (\verb"ZZ"), $Z^W$ (\verb"ZW") and $Z^{\tilde{W}}$ (\verb"ZfW") are defined in the parameter file of the corresponding model as
\begin{equation}
Z^{\gamma Z} = \left(\begin{array}{cc} \cos\Theta_W & - \sin\Theta_W \\ \sin\Theta_W & \cos\Theta_W \end{array}\right)\,, \hspace{0.5cm} Z^W = Z^{\tilde{W}} = \frac{1}{\sqrt{2}}\left(\begin{array}{cc} 1 & 1 \\ -i & i \end{array}\right) 
\end{equation}
\begin{center}  
\fbox{
\parbox{12cm}{
In the definition of EWSB \(W^+\) ({\tt VWp}) can not be used! This would
introduce a new complex field not related to \(W^-\) ({\tt VWm}) and therefore
change the degrees of freedom of the theory. Of course, the winos
\(\tilde{W}^+\) and \(\tilde{W}^-\) are not related by complex conjugation, so
two new fields {\tt fWm} and {\tt fWp} are used after mixing. 
}}
\end{center}

\SARAH always handles the vector bosons for unbroken gauge theories as real
particles. To define the mass eigenstates as real or complex \SARAH checks 
to number of degrees of freedom.

\subsection{Particle Mixings}
\label{mixPart}
Symmetry breaking or also bilinear terms in the superpotential lead to a
rotation of the former (gauge) eigenstates to new mass eigenstates. The
definition of these rotations depends on the fact, if the corresponding mass
matrix is hermitian or not. 

\subsubsection{Introduction}
\paragraph*{Properties of hermitian mass matrices}
In the hermitian case, the Lagrangian has the form
\begin{equation}
\La_{Mass} = \phi^\dagger M \phi .
\end{equation}
The matrix \(M\) can be diagonalized by a matrix \(U\):
\begin{equation}
M_{Dia} = U^{-1} M U .
\end{equation}
The eigenvalues of the mass matrix \(M\) are the masses of the new mass
eigenstates \(\psi\), which are related to the former eigenstates by
\begin{equation}
\psi^i = U^{ij} \phi_j 
\end{equation}  

\paragraph*{Properties of non-hermitian mass matrices}
If the mass term in the Lagrangian is built by two vectors \(\phi_1, \phi_2\),
i.e.
\begin{equation}
\La = \phi_1^T M \phi_2 \thickspace ,
\end{equation}
two mixing matrices are needed to diagonalize the mass matrix:
\begin{equation}
M_{Dia} = V^{-1} M U .
\end{equation}
The two mixing matrices \(V\) and \(U\) diagonalize the matrices \(M M^T\) and
\(M^T M\)
\begin{equation}
M^2_{Dia} = V^{-1} M M^T V, \hspace{1cm} M^2_{Dia} = U^{-1} M^T M U
\end{equation}
and connect the new eigenstates \(\vec{\psi}_1\) and \(\vec{\psi}_2\) with the
old ones by
\begin{equation}
\psi_1^i = V_{ij} \phi_1^j, \hspace{1cm} \psi_2^i = U_{ij} \phi_2^j 
\end{equation}

\subsubsection{Implementation in \SARAH}
Both type of mixings are defined due to 
\begin{verbatim}
DEFINITION[$EIGENSTATES][MatterSector] = { ... }
\end{verbatim}
\SARAH differs between hermitian and non-hermitian mixings by the form used for
the definition. Note, that first all generations of one particle in the basis
are inserted, before the next particle follows.

\paragraph{Hermitian mixings in \SARAH}
The form for defining hermitian rotations is 
\begin{verbatim}
{{List of Old Eigenstates},{Name of New Eigenstates, Name of Mixing Matrix}}
\end{verbatim} 
\begin{enumerate}
\item First, a list of the names of old eigenstates is given. 
\item The  name of the new eigenstates must be given. 
\item The name of the mixing matrix must be given. 
\end{enumerate}

\paragraph*{Examples}
The mixing in the down-squark sector is given by
\begin{verbatim}
{{SdL, SdR}, {Sd, ZD}}
\end{verbatim}
This means, the three left d-squarks \verb"SdL" and the three right d-squarks
\verb"SdR" mix to new eigenstates called \verb"Sd" with generation index running
from 1 to 6. The mixing matrix is called \verb"MD". The above statement leads to
the following relation between the states
\begin{eqnarray}
\label{eq:mix1} \tilde{d}^i &=& \sum_{j=1}^3 Z^D_{i,j} \tilde{d}_L^j \\ 
\label{eq:mix2} \tilde{d}^{i+3} &=& \sum_{j=1}^3 Z^D_{i+3,j} \tilde{d}_R^j
\end{eqnarray} 

In the flavor conserving case, this matrix is reducible. Therefore, \SARAH
checks all matrices for reducibility and sets the non-block elements
automatically to zero.

\paragraph{Non-hermitian mixings in \SARAH}
Non-hermitian rotations are defined in \SARAH by 
\begin{verbatim}
{{{First Basis},{Second Basis}},{{First States,First Matrix},{Second
States,Second Matrix}}}
\end{verbatim} 
Let us clarify this convention by an example.
\paragraph*{Example}
We consider the chargino sector in the MSSM. This mixing is specified by 
\begin{verbatim}
{{{fWm, FHdm}, {fWp, FHup}}, {{Lm,Um}, {Lp,Up}}}
\end{verbatim}
This means that the gauge eigenstates \(\tilde{W}^-\) (\verb"fWm") and
\(\tilde{H}_d^-\) (\verb"FHdm") mix to the negative charged mass eigenstates
\(\lambda^-\) (\verb"Lm"), while \(\tilde{W}^+\) (\verb"fWp") and
\(\tilde{H}_u^+\) (\verb"FHup") form the new eigenstates \(\lambda^+\)
(\verb"Lp"). The new and old eigenstates are connected by the mixing matrices
\(U^-\) (\verb"Um") and \(U^+\) (\verb"Up").
\begin{equation}
\left( \begin{array}{c} \lambda^-_1 \\ \lambda^-_2 \end{array} \right) = U^-
\left( \begin{array}{c} \tilde{W}^- \\ \tilde{H}_d^- \end{array} \right),
\hspace{1cm}
\left( \begin{array}{c} \lambda^+_1 \\ \lambda^+_2 \end{array} \right) = U^+
\left( \begin{array}{c} \tilde{W}^+ \\ \tilde{H}_u^+ \end{array} \right) 
\end{equation} 

\subsection{Gauge fixing terms and ghost interactions}
\label{sec:gaugefixing}
\subsubsection{Introduction}
As explained in app.~\ref{GaugeFixing}, the general form of a gauge fixing term
in \(R_\xi\)-gauge is
\begin{equation}
\label{gfstructure}
\La_{GF} = - \frac{1}{2 R_\xi} \sum_a |f^a|^2 
\end{equation}
with some gauge fixing functions \(f^a\).

\subsubsection{Implementation in \SARAH}
If ghost vertices were to be calculated by
   \SARAH 3.1 or earlier versions, it has been necessary
  to define the gauge fixing terms in $R_\xi$ gauge. However, since
 version 3.2 \SARAH derives these terms automatically using the calculated
 kinetic terms of the Lagrangian. To this end, the condition is applied
 that the mixing between scalar particles and vector bosons vanishes.
 Afterwards, the derived gauge fixing terms are used to calculate the ghost
 interactions. \\
Since it can happen in models with an extended gauge sector that several
 Goldstone bosons are a mixture of the same gauge eigenstate, for each massive
 vector boson, the corresponding Goldstone boson has to be defined
 \begin{verbatim}
 {{   Description -> "Z-Boson",
          ...
          Goldstone -> Ah[{1}]}}, 
 ...
 {{   Description -> "Z'-Boson",
          ...
          Goldstone -> Ah[{2}]}}, 
 \end{verbatim}
 The user can check the gauge fixing terms derived by \SARAH be looking at
 \begin{verbatim}
DEFINITION[$EIGENSTATES][GeneratedGaugeFixing]
 \end{verbatim}
 The general form of the gauge fixing term is:
\begin{verbatim}
DEFINITION[$EIGENSTATES][GeneratedGaugeFixing] = {{Function, Prefactor}, ... };
\end{verbatim}
Here, \verb"Function" is the \(f\) of eq.~(\ref{gfstructure}), and the corresponding
factor is \verb"Prefactor". If the gauge fixing functions involve derivatives of
gauge bosons,
\begin{verbatim}
Der["Gauge Boson"]
\end{verbatim} 
is used. 

% \paragraph*{Examples}
% \begin{enumerate}
% \item The gauge fixing term for the color group in \(R_\xi\) gauge is:
% \begin{equation}
% \La_{GF} = - \frac{1}{2 \xi_g} |\partial_\mu g |^2
% \end{equation}
% This is given in the model file by
% \begin{verbatim}
% DEFINITION[GaugeES][GaugeFixing]=
% {{Der[VG], -1/(2 RXi[G])},...};
% \end{verbatim} 
% \item The gauge fixing term corresponding to the Z-Boson after EWSB is (see
% app.~\ref{GFewsb}).
% \begin{equation}
% \La_{GF} = - \frac{1}{2 \xi_Z} \left( \partial^\mu Z_\mu + \xi_Z M_Z G^0
% \right)^2  
% \end{equation}
% The corresponding Goldstone boson is in \SARAH the first generation of the
% CP-Odd Higgs (see app.~\ref{sec:MSSM}), i.e. \verb"Ah[{1}]". Therefore, we can
% write the gauge fixing term as 
% \begin{verbatim}
% DEFINITION[EWSB][GaugeFixing]=
% {{Der[VZ] + Mass[VZ] RXi[Z] Ah[{1}],  - 1/(2 RXi[Z])},...};
% \end{verbatim}
% \end{enumerate}

\paragraph*{Examples}
\begin{enumerate}
\item The gauge fixing term for the color group in \(R_\xi\) gauge is:
\begin{equation}
\La_{GF} = - \frac{1}{2 \xi_g} |\partial_\mu g |^2
\end{equation}
The corresponding expression in \SARAH reads
\begin{verbatim}
DEFINITION[GaugeES][GeneratedGaugeFixing]=
{{Der[VG], -1/(2 RXi[G])},...};
\end{verbatim} 
\item The gauge fixing term corresponding to the Z-Boson after EWSB is (see
app.~\ref{GFewsb}).
\begin{equation}
\La_{GF} = - \frac{1}{2 \xi_Z} \left( \partial^\mu Z_\mu + \xi_Z M_Z G^0
\right)^2  
\end{equation}
The corresponding Goldstone boson is in \SARAH the first generation of the
CP-Odd Higgs (see app.~\ref{sec:MSSM}). Therefore, the gauge fixing term obtained by \SARAH are
\begin{verbatim}
DEFINITION[EWSB][GeneratedGaugeFixing]=
{(2*Der[VZ] - (sigmad*vd - sigmau*vu)*RXi[Z]*
    (g1*ZZ[1, 2] - g2*ZZ[2, 2]))/2, -1/(2*RXi[Z])}
\end{verbatim}
Here, {\tt ZZ} is the $\gamma - Z$ mixing matrices which can be expressed by the Weinberg angle. Therefore, this expression
is equivalent to the input used in older version of \SARAH
\begin{verbatim}
DEFINITION[EWSB][GaugeFixing]=
{{Der[VZ] + Mass[VZ] RXi[Z] Ah[{1}],  - 1/(2 RXi[Z])},...};
\end{verbatim}
\end{enumerate}

\subsection{Additional couplings or redefinition of existing couplings}
\subsubsection{Introduction}
Sometimes, it might be necessary to define interactions in the Lagrangian, which
can not be derived from the superpotential or the kinetic interaction.
Furthermore, it might be necessary to change the properties of some vertices by
hand. For example, integrating out particles might most likely spoil SUSY.
Therefore, supersymmetric relations are not longer valid: the standard model
Higgs self couplings are free parameters, while they are fixed in SUSY by the
gauge and Yukawa couplings. 
\subsubsection{Implementation in \SARAH}
\label{ReDef}
Both, defining new interactions and changing existing ones, is done in \SARAH
with one statement. For each set of eigenstates, new or changed terms can be
defined separately by declaring them in the model file via
\begin{verbatim}
DEFINITION[$EIGENSTATES][Additonal] = {
{Lag, {Options}}, 
  ... };
\end{verbatim} 
\verb"Lag" is a Lagrangian which is added to the complete Lagrangian, so it must
has mass dimension 4. A rather short notation can be used, only some points have
to be considered
\begin{itemize}
\item Fields are separated by dots
\item Weyl fermions are used
\item All indices are automatically added and contracted
\end{itemize}
The new couplings are handled in a similar way as the couplings of the
superpotential: tensor indices are added automatically and they are assumed to
be complex. Further assumptions about the coupling can be made in
\verb"parameters.m". \\
The manually defined terms will be handled like every other term of the
Lagrangian, i.e. they are affected by rotations and replacements if the
eigenstates are changed. \\ 
The two possible options are:
\begin{enumerate}
\item \verb"AddHC -> True/False": Defines if the hermitian conjugated of this
term is not added to the Lagrangian
\item \verb"Overwrite -> True/False": Defines if existing couplings involving
the same fields are overwritten. 
\end{enumerate}

\paragraph*{Example}
\begin{enumerate}
\item {\bf Define new terms} A mixed soft-breaking term of the form
\begin{equation}
m_{l H_d}^2 (\tilde{l}^* H_d + \tilde{l} H_d^*)
\end{equation}
is added to the Lagrangian of the gauge eigenstates by
\begin{verbatim}
DEFINITION[GaugeES][Additional] = 
{{mlHd2 conj[Sl].SHd, {Overwrite->False, AddHC->True}}};
\end{verbatim}
\item {\bf Adding additional terms to existing couplings} With
\begin{verbatim}
DEFINITION[EWSB][Additional] = 
{{1/24 Kappa hh.hh.hh.hh, {Overwrite->False, AddHC->False}}};
\end{verbatim}
the Higgs self couplings receive an additional contribution:
\begin{equation}
\Gamma_{h^4} ~ (a_i g_i^2 +  b_i Y_i^2) \hspace{0.5cm} \rightarrow
\hspace{0.5cm} \Gamma_{h^4} ~ (a_i g_i^2 +  b_i Y_i^2 + \kappa)
\end{equation}
\item {\bf Overwriting existing terms} 
To overwrite the former expressions for the Higgs self interactions, 
\begin{verbatim}
DEFINITION[EWSB][Additional] = 
{{LagNew, {Overwrite->True, AddHC->False}}};
LagNew = 1/24 Kappa hh.hh.hh.hh + 1/24 Lambda hh.hh.Ah.Ah;
\end{verbatim}
is used. This has the following effect:
\begin{equation}
\Gamma_{h^4} ~ (a_i g_i^2 +  b_i Y_i^2) \hspace{0.5cm} \rightarrow
\hspace{0.5cm} \Gamma_{h^4} ~ \kappa
\end{equation}
\item {\bf Interactions involving derivatives} For interactions involving
derivatives, \verb"Der" is used
\begin{verbatim}
DEFINITION[EWSB][Additional] = 
{{Kappa Der[SHd,lor3].conj[SHd].VB,{Overwrite->False, AddHC->True}}};
\end{verbatim}
\item {\bf Matter interactions of the standard model} This method can be used to
define the matter interactions of the standard model
\begin{verbatim}
SuperPotential = {};
DEFINITION[GaugeES][Additional]= {
        {LagHC, {Overwrite->True, AddHC->True}},
        {LagNoHC,{Overwrite->True, AddHC->False}}
};
LagNoHC = Mu conj[SH].SH + 1/24 Lambda1 conj[SH].SH.conj[SH].SH;
LagHC = - Yd conj[SH].Fq.conj[FdR] - Ye conj[SH].Fl.conj[FeR] + Yu SH.Fq.conj[FuR];
\end{verbatim}
\end{enumerate}

\section{Effective or non-supersymmetric theories}
\label{IntOutorDel}
It is easy in \SARAH to integrate particles out of the spectrum to get an
effective theory, or just to delete this particle to get a non-supersymmetric
limit of the model.

\subsection{Integrating out particles}
\subsubsection{General}
If in a theory very heavy particles \(\Phi\) exist, they are no physical degree
of freedom if the energy scale is below the mass of the particle. In this case,
an effective theory for the lighter fields \(\phi\) is derived by integrating
out the heavy states
\begin{equation}
L(\phi)_{eff} = \int L(\phi,\Phi) d\Phi 
\end{equation}
This procedure will lead to higher dimensional operators like the four fermion
interaction in Fermi's theory. 

\subsubsection{Implementation in \SARAH}
Integrating out particles in \SARAH is easy: the heavy particles, which should
be integrated by, are added to {\tt IntegrateOut}:
\begin{verbatim}
IntegrateOut = {Particle 1, Particle 2,...}
\end{verbatim} 
Here, \verb"Particle 1" can be the component name of a particle, e.g \verb"SdR"
for all right d-squarks. Moreover, if the superfield name is used, all
corresponding component fields are integrated out. If only specific generations
of one field should be integrated out, the assignment is
\begin{verbatim}
IntegrateOut = {Particle[{first,last}],...}
\end{verbatim} 
Here, \verb"first" is the first generation which should be integrated out and
\verb"last" is the last one. 

\begin{center}  
\fbox{
\parbox{12cm}{
The list {\tt IntegrateOut} can consist of particles belonging to different
eigenstates. All particles are integrated out at the first time they do appear. 
}}
\end{center}

\paragraph*{Example}
\begin{enumerate}
\item To get an effective theory by integrating out the gluino, use
\begin{verbatim}
IntegrateOut = { fG };
\end{verbatim} 
\item RGE running of soft-breaking masses leads to the effect that the first two
generations of sleptons are heavier than the third one. Therefore, building an
effective theory with only the third generation of squarks is done by
\begin{verbatim}
IntegrateOut = {SdR[{1,2}],SuR[{1,2}],SdL[{1,2}],SuL[{1,2}]};
\end{verbatim} 
\end{enumerate}

\subsection{Deleting particles}
Deleting particles can be done in the same way as integrating them out. The
difference is that there are no effective operators are calculated. Deleting is
therefore faster and should be used, if the higher dimensional operators are not
needed. \\
Deleting particles is done by
\begin{verbatim}
DeleteParticle={List Of Particles}; 
\end{verbatim}
The usage of \verb"DeleteParticles" Is the same as of \verb"IntegrateOut".
By deleting particles it is easy get a non supersymmetric limit of a model: it
is possible, i.e. one can get the standard model by deleting all SUSY particles and the
non-standard Higgs field. Since \SARAH supports are also a second \(SU(3)\) it might
also be possible to implement Technicolor (-like) theories with this procedure.\\
{\bf Caveat!} Of course, one must be careful: Some remaining interactions are still based on
SUSY relation, which are not longer valid for non-supersymmetric theories. For
example, the four point self-interaction of the light Higgs is in SUSY defined
by the gauge interactions, but a free parameter in the standard model. To delete
such relations the option for  redefining couplings can be used, see
sec.~\ref{ReDef}.   

\section{Definition of Dirac spinors}
\label{diracSpinors}
Event generators and programs for calculating Feynman Diagrams are
normally written for Dirac spinors, but \SARAH does all internal calculations
for Weyl spinors. Therefore, it is necessary to define, how the Weyl spinors
combine to Dirac spinors. That can be done for each all eigenstates separately
\begin{verbatim}
DEFINITION[$EIGENSTATES][DiracSpinors]={
 Dirac Spinor -> {Weyl 1, Weyl 2},
 ...
 }
\end{verbatim}

\verb"Dirac Spinor" is the new name of the Dirac spinor, while \verb"Weyl 1"
and \verb"Weyl 2" the names of the Weyl spinors building the left respectively
right component of the Dirac spinor. 

\paragraph*{Example}
The electron \verb"Fe" is built from the components \verb"FeL" and \verb"FeR" by
\begin{verbatim}
DEFINITION[EWSB][DiracSpinors]={
 Fe -> {FeL, FeR},
 ...
  }
\end{verbatim} 
while the neutralinos \verb"Chi", which are  Majorana particles, consists only
of the Weyl spinor \verb"L0"
\begin{verbatim}
Chi -> {L0, conj[L0]}
\end{verbatim} 

\section{Checking the model files}
After the initialization of a model via {\tt Start["MODEL"]} it can be checked for (self-) consistency using the command
\begin{verbatim}
CheckModel;
\end{verbatim}
The following checks are performed:
\begin{itemize}
 \item Leads the particle content to gauge anomalies?
 \item Leads the particle content to the Witten anomaly?
 \item Are all terms in the superpotential in agreement with charge, and if defined, $R$-parity conservation?
 \item Are there other terms allowed in the superpotential by gauge invariance and possibly $R$-parity conservation beyond those defined. 
 \item Are all mixings consistent with unbroken gauge groups?
 \item Are all definitions of Dirac spinors consistent with unbroken gauge groups?
 \item Are there terms in the Lagrangian of the mass eigenstates which can cause additional mixing between fields?
 \item Are all mass matrices irreducible?
 \item Are the properties of all particles and parameters defined correctly?
\end{itemize}
Note, the check for anomalies works so far only for supersymmetric models. 

\chapter{\LaTeX output}
\section{Writing a \LaTeX{} file}
It is possible to write a \LaTeX{} file with all information about the mode by
using
\begin{verbatim}
ModelOutput[Eigenstates, WriteTeX->True];
\end{verbatim}
This calculates first all interactions for the eigenstates. If this was already
done before, it is also possible to use
\begin{verbatim}
MakeTeX[Options];
\end{verbatim}
There are different Tex-files produced containing the following information:
\begin{enumerate}
\item List of the fields
\item Important parts of the Lagrangian (soft-breaking terms, gauge fixing terms)
\item Mass Matrices and tadpole equations
\item Renormalization Group Equations
\item One-loop self energies and tadpole equations
\item All interactions
\item Details about the conventions used in \SARAH
\end{enumerate} 
\section{Options}
The options are
\begin{enumerate}
\item \verb"FeynmanDiagrams", Values: \verb"True" or \verb"False", Default:
\verb"True" \\
Defines, if the Feynman diagrams for all interactions should be drawn. 
\item \verb"effectiveOperators", Values: \verb"True" or \verb"False", Default:
\verb"True"\\
Defines, if the higher dimensional operators should be included in the \LaTeX
file. By default, there are only the vertices involving up to four particles. For
switching on six particle interactions \newline \verb"SixParticleInteractions" is used.
\item \verb"SixParticleInteractions", Values: \verb"True" or \verb"False",
Default: \verb"False" \\
Defines, if also the six-particle interactions should be added to the \LaTeX{}
output
\item \verb"ShortForm", Values: \verb"True" or \verb"False", Default:
\verb"False" \\
Defines, if a shorter notation for the vertices should be used
\item \verb"WriteSARAH", Values: \verb"True" or \verb"False", Default:
\verb"False" \\
Defines, if the names and parameters used in \SARAH should be written   
\end{enumerate}

\section{Creating the pdf File}
The \LaTeX\ files are saved in the directory
\begin{verbatim}
../Output/$MODEL/$EIGENSTATES/TeX
\end{verbatim}
and the main file is {\tt \$MODEL <> \$EIGENSTATES <>.Tex}. All
other files are included in this file by using the \verb"input"-command of
\LaTeX. If \verb"Diagrams->True" is used, the following steps must be
done for generating an pdf document including the diagrams:
\begin{enumerate}
\item First, compile the Tex file, e.g. \verb"pdflatex model.tex"
\item Go to the directory \verb"Diagrams" and compile every \verb".mp" file with
\verb"mpost". This is done under Linux and under Windows with
\begin{verbatim}
mpost FeynmanDiaX.mp
\end{verbatim} 
It is also possible to apply the \verb"mpost" command on all \verb".mp"-files at
once by using
\begin{verbatim}
find . -name "*.mp" -exec mpost {} \;
\end{verbatim}
\item After generating all diagrams, go back and compile the \verb".tex"-file
again by using \verb"pdflatex". 
\end{enumerate}
To simplify this procedure, \SARAH will write a shell script in the Tex-output
directory which does exactly these three steps. It can be started under Linux
with
\begin{verbatim}
./MakePDF.sh
\end{verbatim}
or under Windows with
\begin{verbatim}
MakePDF.bat
\end{verbatim}
It is possible that the script must be first declared is executable in Linux via
\begin{verbatim}
chmod 755 MakePDF.sh
\end{verbatim}

\chapter{Output for \FeynArts}
\section{Generate model files for \FeynArts}
A model file for "\FeynArts" is created by 
\begin{verbatim}
ModelOutput[$EIGENSTATES, WriteFeynArts->True]
\end{verbatim}
or, if \verb"ModelOutput" was used before and the vertices are already
calculated, by 
\begin{verbatim}
MakeFeynArts[Options]
\end{verbatim}
As options can {\tt AddCounterTerms -> True} be used to add counter terms to
each parameter: in that case all appearing parameters are replaced by \(x
\rightarrow x + \delta x\). Note, this option has to be used carefully because 
the routines are not yet very sophisticate. For instance, the shift is applied for
mixing angles like the Weinberg angle itself and not for trigonometric functions as it is usually done. This will
be improved in the future. 

The following things are done:
\begin{enumerate}
\item A list of all particles present in the models is generated. The particles in
\FeynArts are named 
\begin{enumerate}
\item \verb"S[X]": For scalars, with some integer \verb"X"  
\item \verb"F[X]": For fermions, with some integer \verb"X"
\item \verb"V[X]": For vector bosons, with some integer \verb"X"
\item \verb"U[X]": For ghosts, with some integer \verb"X"
\end{enumerate} 
\verb"X" can be defined in the particle definitions file of \SARAH or is chosen
automatically. \verb"FeynArts" also supports labels for particles which are
easier to read for humans eyes by using a TeX-like output. The label for each
particle is generated from the defined \LaTeX\ name by \SARAH. 
\item A list with all appearing indices is written. 
\item The list with interactions is written. If the theory contains several
non-Abelian, unbroken gauge groups, the generators of these gauge groups will
appear in the vertices. By default, the generators of \(SU(3)\) are associated
to \verb"SUNT" and  automatically simplified when using \verb"FormCalc"
\cite{formcalc}. 
\end{enumerate}

\section{Dependences, numerical values and special abbreviations for
\FormCalc}
A second file is generated by \SARAH together with the model file for \FeynArts. 
This file is called
{\tt Substitutions-<> \$EIGENSTATES <>.m } and contains
additional information which might be useful for calculating diagrams:
\begin{enumerate}
\item Replacement rules with the defined dependences in \SARAH, see
sec.~\ref{sec:ParameterFile}:
 \begin{enumerate}
  \item \verb"Dependences"
  \item \verb"DependencesOptional"
  \item \verb"DependencesNum"
 \end{enumerate}
\item The definitions of the masses: \verb"Masses" in \SARAH
\item The numerical values for the parameters in \SARAH: \verb"NumericalValues" 
\item Special abbreviations for \FormCalc like those are also defined for the
MSSM and SM in the \FormCalc package: 
  \begin{enumerate}
   \item A complex conjugation is replaced by \verb"C": \verb"Conjugate[X] ->XC"
   \item A square is replaced by \verb"2": \verb"X^2 -> X2"
   \item The names of soft-breaking couplings are merged: \verb"T[X] -> TX"
  \end{enumerate}
 It is recommended to use this definitions to speed up the calculations with
\FormCalc. 
\end{enumerate}

\chapter{Output for \CalcHep/\CompHep}
\section{Generate model files for \CalcHep and \CompHep}
To generate model files for \CalcHep and \CompHep,
\begin{verbatim}
ModelOutput[$EIGENSTATES, WriteCHep->True]
\end{verbatim}
or
\begin{verbatim}
MakeCHep[options]
\end{verbatim}
is used. The second command offers more options to control the output:
\begin{enumerate}
\item \verb"FeynmanGauge", Values: \verb"True" or \verb"False", Default:
\verb"True" \\
By setting to \verb"True", the interactions of the Goldstone bosons are written
in the interaction file. 
\item \verb"CPViolation", Values \verb"True" or \verb"False", Default:
\verb"False" \\
By setting to \verb"True", the possibility of CP violation is included in the
model files, see sec.~\ref{CPCH}.
\item \verb"ModelNr", Values Integer, Default: \verb"1" \\
The number added to names of the files, see next section.
\item \verb"CompHep", Values: \verb"True" or \verb"False", Default: \verb"False"
\\
By setting to \verb"True" the model files are written in the \CompHep format.
\item \verb"NoSplittingWith", Values: Particles List, Default: \verb"{}"\\
If one of the given particles appears in a four-point interaction, the
interaction is not split using auxiliary fields
\item \verb"NoSplittingOnly", Values: Particles List, Default: \verb"{}"\\
If all particles of a four-point interaction appear in the given list, the
interaction is not split using auxiliary fields
\item \verb"UseRunningCoupling", Values: \verb"True" or \verb"False", Default:
\verb"False" \\
Defines, if the standard running of the strong coupling should be included in
the model file.
\item \verb"SLHAinput", Values: \verb"True" or \verb"False", Default:
\verb"False" \\
Defines, if parameter values should be read from a LesHouches input file, see
sec.~\ref{SLHAplus}.
\item \verb"CalculateMasses",  Values: \verb"True" or \verb"False", Default:
\verb"False"\\
The tree level mass matrices are diagonalized by \CalcHep to calculate the masses and
rotation matrices. Note, \verb"SLHAinput -> False" has to be used in addition
\item \verb"RunSPhenoViaCalcHep", Values: \verb"True" or \verb"False", Default:
\verb"False"\\
Writes C code to run \SPheno from the graphical interface of \CalcHep to calculate the input 
parameters on the fly. Note, the path to the \SPheno executable for the considered model
has to be set in the file {\tt SPhenoViaCalcHep.c} written by \SARAH
\item \verb"IncludeEffectiveHiggsVertices", Values: \verb"True" or \verb"False", Default:
\verb"False"\\
Includes the loop induced vertices of the CP even and odd Higgs to two photons or gluons.
The numerical values for these couplings can be calculated with \SPheno and they are given 
in the spectrum file and also included by \verb"SLHAinput->True".
\end{enumerate}

\subsection{Model Files}
The \CalcHep/\CompHep output of \SARAH generates the following four files
\begin{enumerate}
\item \verb"prtclX.mdl": Contains all particles
\item \verb"lgrngX.mdl": Contains the interactions
\item \verb"varsX.mdl": Contains the numerical values of the variables
\item \verb"funcX.mdl": Contains dependences between the parameters
\end{enumerate}
\verb"X" is a number, which can be chosen by the option \verb"ModelNr".

\subsubsection{Particles}
First, there are stringent constraints on the naming of particles in \CalcHep:
only names up to four letters are allowed and also indices aren't supported.
Therefore, it is not possible to use the \SARAH internal definitions of
particles. Thus, the names used for the model files are based on the defined
\verb"OutputName" of each particle in the following way
\begin{enumerate}
\item The basis of each name is the entry in \verb"OutputName" in the particle
file, see sec.~\ref{particleFile}
\item If the considered particle is not self-conjugated, for the anti particle
the first letter is changed from upper to lower case or vice versa. 
\item If there are more generations for one particle, the number of the
generation is appended at the end of the name
\item If the defined R-parity is -1, a \verb"~" is added to the beginning of the
name to assign SUSY particles. In this way, it is possible to use the model
files in {\tt MicrOmegas} without the need of an additional list of all SUSY
particles 
\end{enumerate} 
The steps above are the standard procedure for all vector bosons, fermions and
most scalars. Ghosts, Goldstone bosons and auxiliary fields handled in a
different way. There are three different kind of ghosts. These are not written
in the particle file, but appear in the Lagrangian file:
\begin{enumerate}
\item Faddeev-Popov Ghost: these are the well known Ghost derived from the gauge
transformations of the gauge fixing term. The name in the model file is 
\begin{verbatim}
"Name of Vector Boson".C
\end{verbatim}
\item Goldstone Ghosts: these are just the Goldstone bosons 'eaten up' by the
gauge particles. Their name is
\begin{verbatim}
"Name of Vector Boson".f
\end{verbatim}
\item Tensor Ghosts: Is needed to express the four gluon interaction. The name
is 
\begin{verbatim}
"Name of Gluon".t
\end{verbatim}
\end{enumerate}  
\SARAH derives the name of Goldstone and Faddeev-Popov ghost automatically from
the underlying vector boson, 
but the tensor ghost and its one interaction with two gluons is hard-coded. \\

The last kind of fields known by \CalcHep/\CompHep are auxiliary fields. Their
purpose is explained in the next section, but their names are as follows
\begin{verbatim}
~OX
\end{verbatim}
Here, \verb"X" is an integer. The antiparticle, if it is not the same, is
counted as X+1. \\

\subsection{Auxiliary fields in \CalcHep/\CompHep}
\label{auxCalc}
We mentioned in the last section that \CalcHep and \CompHep needs special
auxiliary fields. The reason is that the color structure is implicit. Hence,
interactions of four colored particles or two colored and two gluons suffer from
an ambiguity. Therefore, these interactions are split  in two three particle
interactions by inserting auxiliary fields. \\
\SARAH does a similar splitting for all interactions between four scalars  by
inserting  auxiliary fields when calculating the F- and D-Terms. Also the
interactions between two squarks and two gluons are split in two three particle
interactions. The splitting can be suppressed for specific vertices by using
\verb"NoSplittingWith" or \verb"NoSplittingOnly".

\subsection{Vertex functions}
All interactions are parametrized by a variable in the Lagrange file. The values
of these variables are defined by using the results of \SARAH for the
corresponding vertices. The following renaming had to be done:
\begin{enumerate}
\item Tensor indices are just added to the name, therefore all sums in the
vertices had do be evaluated:
\begin{verbatim}
sum[i1,1,3,MD[1,i1]]      ->       MD11 + MD12 + MD13
\end{verbatim}
Some values, which are known to be zero like in the flavor conserving case, are
set to zero at this point.
\item Variables names, which are longer then six letters, are truncated.
\item All parameters are assumed to be real, i.e. complex conjugation is removed
(see sec.~\ref{CPCH}).
\item The generators and structure constants of the strong interaction are
removed because they are defined implicitly.
\end{enumerate}

% Since, there is no index allowed, all interactions must be written for all
possible combinations of generation indices. This can lead for some models to a
very long time for writing the model file, e.g. in the \(\mu \nu SSM\) with 8
charged Higgs: All \(8^3\) combinations of the self interactions must be
written. 

\subsection{CP Violation}
\label{CPCH}
\CalcHep/\CompHep can't handle complex values in the function or vars file, but
only in the Lagrange file. Therefore, all variables are by default assumed to be
real, when \SARAH writes the model file with default options. This can be
changed by setting \verb"CPViolation" to \verb"True". In that case, \SARAH
splits all parameters, which are not explicitly defined as real, in real and
imaginary part:
\begin{verbatim}
X -> RX + i*IX
\end{verbatim}
The real and imaginary part for every interaction is calculated using that
splitting, and both parts are written separately in the Lagrange file:
\begin{verbatim}
v0001 -> Rv0001 + i*Iv0001
\end{verbatim} 

% \subsection{Variables and Dependences}
% The numerical values for parameters and masses are given in  \verb"vars1.mdl".
\SARAH writes in this file the output name of every needed variable. Since
\CalcHep/\CompHep does only support variable names with a length of maximal 6
letters, \SARAH cuts the name of all variables, which are longer then that. If a
numerical value for the variable is available, e.g. if it is defined in the
parameter file or it was read in from LesHouches file, it will also be added in
the \verb"vars" file, otherwise \verb"NaN" appears. \\
% The dependences defined in \SARAH are written as functions in
\verb"func1.mdl".

\subsection{SLHA input}
\label{SLHAplus}
\CalcHep supports the possibility to read LesHouches input files
\cite{Belanger:2010st}. \SARAH can write the corresponding definitions in the
functions file of \CalcHep. In this context, it is assumed that a LesHouches
file called \verb"SPheno.spc.[MODEL]" is located in the same directory. However, this
can easily be adjusted manually.

\subsection{What can be a problem...}
We have made the following experiences by testing model files with
\CalcHep/\CompHep:
\begin{enumerate}
\item A PDG number of \verb"0" is not allowed for other particles than auxiliary
particles. 
\item In the vars file is no discrimination between small and capital letters.
This must be taken into account by naming the mixing matrices and couplings in
\SARAH.
\item Higher dimensional operators are not supported
\item The color structure is implicit and indices are not supported in
\CalcHep/\CompHep. Therefore, it is difficult to implement models with other
unbroken non-Abelian gauge groups than the color group. 
\item  Writing the output for models with particles appearing in a large number
of generations and non reducible mixing matrices like in the flavor conserving
MSSM, last very long, because all possible combinations of indices have to be
written separately. 
\end{enumerate} 

\section{\micrOmegas}
\label{sec:MicrOmegas}
\micrOmegas \cite{Belanger:2006is} is a well known tool for the calculation of
the relic density of a dark matter candidate. \micrOmegas uses \CalcHep to
calculate the annihilation and co-annihilation
processes. Therefore, it is necessary to generate first a model file for
\CalcHep to implement new models in \micrOmegas as described in above.\\
\SARAH writes two files for \micrOmegas which can serve as so called main
files, i.e. they can be compiled with \micrOmegas and executed to perform
calculations. While {\tt CalcOmega.cpp} calculates only \(\Omega h^2\) and
writes the result to the file {\tt omg.out}. {\tt
CalcOmega\_with\_DDetection.cpp} computes also direct detection signals. In that
case, the numbers in the different lines in {\tt omg.out} correspond to
\begin{itemize}
 \item Relic density \(\Omega h^2\)
 \item Spin independent cross section with proton in pb
 \item Spin dependent cross section with proton in pb
 \item Spin independent cross section with neutron in pb
 \item Spin dependent cross section with neutron in pb
 \item Recoil at \({}^{73}\)Ge: number of events in 10 - 50 keV region
 \item Recoil at \({}^{131}\)Xe: number of events in 10 - 50 keV region
 \item Recoil at \({}^{23}\)Na: number of events in 10 - 50 keV region
 \item Recoil at \({}^{127}\)I: number of events in 10 - 50 keV region
\end{itemize}
Using the SLHA+ functionality of \CalcHep is also possible with \micrOmegas.
Therefore, it is sufficient to copy the spectrum file written by \SPheno to the
directory of \micrOmegas and start the calculation.

\chapter{Output for \WHIZARD}
{\bf \large (in collaboration with C. Speckner)} \vspace{1cm} \\
To generate model files for \WHIZARD
\begin{verbatim}
ModelOutput[$EIGENSTATES, WriteWHIZARD->True]
\end{verbatim}
is used.If the vertices have been calculated already by \verb"MakeVertexList",
\begin{verbatim}
MakeWHIZARD[options]
\end{verbatim}
can be used. The possible options are
\begin{enumerate}
\item \verb"WriteOmega", Values: \verb"True" or \verb"False", Default:
\verb"True" \\
Defines, if the model files for \OMEGA should be written
\item \verb"WriteWHIZARD", Values: \verb"True" or \verb"False", Default:
\verb"True" \\
Defines, if the model files for \WHIZARD should be written
\item \verb"Exclude", Values: list of generic type, Default: \verb"{SSSS}" \\
Defines, which generic diagrams are excluded when writing the model file
\item \verb"WOModelName", Values: string, Default: defined model name \\
Gives the possibility to change the model name
\item \verb"MaximalCouplingsPerFile", Values: Number, Default: 500\\
Defines the maximal number of couplings written in one file
\item \verb"Version", Values: Number, Default: \verb"2.0.3"\\
Defines the version of \WHIZARD for which the model file is generated
\item \verb"ReadLists", Values: \verb"True" or \verb"False", Default:
\verb"False" \\
Defines, if the information from a former evaluation should be used
\end{enumerate}

\paragraph*{Using the generated model files with \WHIZARD}\mbox{}\\
After the interface has completed, the generated files can be found in the
\verb"WHIZARD_Omega" subdirectory of {\SARAH}s output directory. In order to use
the model with \WHIZARD 2.x, the generated code must be compiled and installed.
For most applications, this is done by simply issuing (inside the output
directory)
\begin{lstlisting}[frame=shadowbox] 
./configure
make
make install
\end{lstlisting} 
By default, the third command installs the compiled model into \verb".whizard"
in current user's home directory where it is automatically picked up by
\WHIZARD. Alternative installation paths can be specified using the
\verb"--prefix" option to WHIZARD.
\begin{lstlisting}[frame=shadowbox] 
./configure --prefix=/path/to/installation/prefix
\end{lstlisting} 
If the files are installed into the \WHIZARD
installation prefix, the program will also pick them up automatically, while
{\WHIZARD}'s \verb"--localprefix" option must be used to communicate any other
choice to \WHIZARD. In case \WHIZARD is not available in the binary search
path, the \verb"WO_CONFIG" environment variable can be used to point
\verb"configure" to the binaries
\begin{lstlisting}[frame=shadowbox] 
./configure WO_CONFIG=/path/to/whizard/binaries
\end{lstlisting} 
More information on the available options and their syntax can be obtained with
the
\verb"--help" option.

In the case of \WHIZARD 1.x output, the generated files must be patched into the
\WHIZARD source tree. To this end, the interface creates a script called
\verb"inject". In most cases, it is sufficient to simply call the script as
\begin{lstlisting}[frame=shadowbox] 
./inject /path/to/whizard
\end{lstlisting} 
(from within the output directory). Issuing \verb"./inject --help" will display
a list of options which can be used to adapt the script to more complicated
usage scenarios.

\chapter{Output in \UFO format}
To generate model files in the \UFO \cite{Degrande:2011ua} format which can be used e.g. with \Madgraph,
\begin{verbatim}
ModelOutput[$EIGENSTATES, WriteUFO->True]
\end{verbatim}
is used.If the vertices have been calculated already by \verb"MakeVertexList" or \verb"ModelOutput",
\begin{verbatim}
MakeUFO[options]
\end{verbatim}
can be used. The possible options are
\begin{enumerate}
\item \verb"Exclude", Values: list of generic type, Default: \verb"{SSSS,GGS,GGV}" \\
Defines, which generic diagrams are excluded when writing the model file
\end{enumerate}
The output written by \SARAH consists of the files
\begin{itemize}
 \item {\tt particles.py}: contains the particles  present in the model
 \item {\tt parameters.py}: contains all parameters present in the model 
 \item {\tt lorentz.py}: defines the Lorentz structures needed for the vertices
\item {\tt vertices.py}: defines the vertices 
\item {\tt couplings.py}: expressions to calculate the couplings
\item {\tt coupling\_orders.py}: defines hierarchies for the coupling orders
\end{itemize}
These files are saved in the directory
\begin{verbatim}
 $SARAH/Output/$MODEL/$EIGENSTATES/UFO/
\end{verbatim}
This directory contains also additional files which are model independent and
 can therefore be used with all models:  {\tt function\_library.py},
 {\tt object\_library.py}, {\tt \_\_init\_\_.py} and
 {\tt write\_param\_card.py}. These files were kindly provided by Olivier
 Mattelaer. \\
 
 \subsubsection*{Using the UFO model files of \SARAH with \Madgraph} 
To use the model files with \Madgraph, it is sufficient to copy all files
 to {\tt \$MADGRAPH/models/\$NAME}. Here, {\tt \$MADGRAPH} is the
 directory containing the local \Madgraph installation and {\tt \$NAME} is a
 freely-chosen name of a new subdirectory. This
 directory name is used afterwards to load the model in \Madgraph via 
\begin{verbatim}
> import model $NAME
\end{verbatim}
\Madgraph has a list of pre-defined names for the particles of the SM and MSSM which
 are used by default. However, it could be that there are conflicts between
 these names and the names used by \SARAH in an extension of the MSSM. For
 instance, {\tt h3} is defined in \Madgraph as the
pseudoscalar Higgs in the MSSM, but \SARAH uses it in
 the NMSSM for the third-heaviest scalar Higgs.
 In order to avoid such clashes, a model can be loaded
 using only the names defined by
 the UFO files via
\begin{verbatim}
> import model $NAME -modelname
\end{verbatim}

To use these model files for the calculation of cross sections, it is, of
 course, necessary to provide the numerical values of all masses and parameters.
 The necessary input can be obtained by using a \SPheno module created by \SARAH
 for  the given model. The SLHA
 spectrum files written by this \SPheno module can directly be used with \Madgraph. 
\chapter{Output for \SPheno}
\section{Introduction}
\SARAH is based on \Mathematica and therefore it is usually not sensible to do
exhaustive
numerical calculations in {\tt SARAH's} native environment. As opposed to that,
there is \SPheno \cite{Porod:2003um,Porod:2011nf}, a well tested spectrum
calculator written in Fortran. \SPheno
provides fast numerically routines for the evaluation of the RGEs, calculating
the phase space of 2- and 3-body decays as well as Passarino Veltman integrals
and much more. Since these routines are model independent, they can be used in
principle for all SUSY models implemented in \SARAH.

The generation of the source code for \SPheno is started via
\begin{verbatim}
MakeSPheno[Options]
\end{verbatim}

The different options are:
\begin{itemize}
\item \verb"ReadLists->True" can be used if all vertices and RGEs have already
been calculated for the model and the former results should be used to save
time. 
\item \verb"InputFile". The name of the \SPheno input file. If not defined,
\verb"SPheno.m" is used.
\item  \verb"StandardCompiler->Compile". The compiler which should be set as standard in the created Makefile. 
      Default is {\tt gfortran}.
\end{itemize}

The generated source code is located in
\begin{verbatim}
/Directory of SARAH/Output/$MODEL/$EIGENSTATES/SPheno/ 
\end{verbatim}
To compile the code, a new sub-directory called {\tt \$MODEL} in local installation of \SPheno has to be created and the 
code has to be copied into that directory. Afterwards, 
\begin{verbatim}
make Model=$MODEL
\end{verbatim}
has the be executed in the root directory of \SPheno. This compiles the code and a new binary 
\begin{verbatim}
SPheno <> $MODEL
\end{verbatim}
is generated in \verb"bin/".

\section{Input file}
\label{sec:sphenoinputfile}
For the \SPheno output, a new file {\tt SPheno.m} is needed to define the
properties of the generated \SPheno version. The file {\tt SPheno.m} must be
located in the same directory as the other input files of the current model.
The content of this can be the following:
\lstset{frame=none}
  \begin{enumerate}
\item {\tt MINPAR}: A list of parameters which should be read from a LesHouches
 file by SPheno. First, the number in the block is defined, afterwards the
variable. For example:
\begin{verbatim}
 MINPAR = {{1,m0},
           {2,m12},
           {3,TanBeta},
           {4,SignMu},
           {5,Azero}};
\end{verbatim}
Now, all information about these parameters can later on given to \SPheno by
using an input file containing the part
\begin{verbatim}
 Block MINPAR #
    1  7.000000E+01 #  m_0
    2  2.500000E+02 #  M_1/2
    3  1.000000E+01 #  Tan(beta)
    4. 1.000000E+00 #  Sign(mu)
    5. 0.000000E+00 #  A_0 
\end{verbatim}
In case that several sets of parameters are demanded, e.g. to support mSugra- and GMSB-like boundary conditions, 
{\tt MINPAR} can also be a nested like, e.g.
\begin{verbatim}
MINPAR=Table[{},{2}];

MINPAR[[1]]={{1,m0},
             {2,m12},
             {3,TanBeta},
             {4,SignumMu},
             {5,Azero}};
             
(* GMSB input parameters *)             
             
MINPAR[[2]]={{1,LambdaInput},   
             {2,MessengerScale},   
             {3,TanBeta},
             {4,SignumMu},
             {6,cGrav},
             {7,n5plets},
             {8,n10plets}};
\end{verbatim}
In that case {\tt MINPAR[[X]]} is associated with the boundary conditions {\tt BoundarySUSYScale[[x]]}, {\tt BoundaryHighScale[[x]]},
{\tt BoundaryEWSBScale[[x]]} described below. 
\item {\tt EXTPAR}: It is also possible to define additional parameters for the
block {\tt EXTPAR} of the LesHouches input file by
\begin{verbatim}
EXTPAR = {{Nr1,  Var1}, 
          {Nr2,  Var2},
          ...}; 
\end{verbatim}
For instance, in order to give three additional VEVs as input, we can use
\begin{verbatim}
EXTPAR = {{100, v1},
          {101, v2},
          {102, v3}};
\end{verbatim}
and set the values later on in the input file by
\begin{verbatim}
 Block EXTPAR #
    100  1.000000E-04 #  v_1
    101  1.500000E-04 #  v_2
    103  2.000000E-04 #  v_3
\end{verbatim}
Note, there are no hardcoded entries for {\tt MINPAR} or {\tt EXTPAR}. That
makes it necessary to define these blocks also for models for which in principle
SLHA conventions exist. However, that provides also more freedom in varying the
model and the free parameters.
\item {\tt ParametersToSolveTadpoles}: \SARAH derives for each VEV the
corresponding minimum condition for the vacuum. These equations give
constraints to the same number of parameters as VEV are in the models.  {\tt
ParametersToSolveTadpoles} is used to set the parameter which fixed by the
tadpole equations. \\
For example, to use the standard choice in the MSSM \(\mu, B_\mu\), the entry
reads:
\begin{verbatim}
 ParametersToSolveTadpoles = {\[mu], B[\mu]};
\end{verbatim}
\SARAH uses the {\tt Solve} command of \Mathematica to solve the tadpole
equations for the given set of parameters. If the solution is not unique because
a parameter \(X\) appears squared, \SARAH solves the equations for the absolute
squared. The phase is defined by the automatically generated variable  {\tt
SignumX}, which is expected to be given as input.\\
The expressions for the constrained parameters are automatically used during the
numerical analysis at the SUSY as well at the electroweak scale. That's
possible, because also the RGE evaluation of all VEVs is included in the
generated \SPheno version. \\
If models with CP violation in the Higgs sector are studied, it is often necessary to solve
the tadpole equations for complex parameters. This can be done by demanding that Mathematica
should solve the tadpole equations for the real and imaginary part of the corresponding parameter, e.g.
\begin{verbatim}
 ParametersToSolveTadpoles = {\[mu], re[B[\mu]], im[B[\mu]]};
\end{verbatim}
\item {\tt AssumptionsTadpoleEquations}
It is possible to define a list with replacements which are by \SARAH when it tries to solve
the tadpole equations. For instance, to approximate some matrices as diagonal and assume that
all parameters are real, use
\begin{verbatim}
AssumptionsTadpoleEquations = {Ye[a__]->Delta[a] Ye[a], 
    T[Ye][a__]->Delta[a] T[Ye][a], conj[x_]->x}; 
\end{verbatim}
\item {\tt UseGivenTapdoleSolution}: In cases, in which \Mathematica won't find
an analytical solution for the tadpole equations for the given set of
parameters, this variable has to be set to {\tt True} and an approximated
solution can be given. These solutions are defined by
\begin{itemize}
\item {\tt SubSolutionsTadpolesTree}: For the solution at tree level
\begin{verbatim}
 SubSolutionsTadpolesTree = {x1 -> sol1, x2 -> sol2,...};
\end{verbatim}
Here, {\tt x1}, {\tt x2} are the names of the parameters which are fixed by the
tadpole equations and {\tt sol1}, {\tt sol2} are the approximated expressions
for them. 
\item {\tt SubSolutionsTadpolesLoop}: The solutions of the one loop corrected
tadpole equations. The one loop corrections to the different VEVs have to be
named {\tt Tad1Loop[i]}.
\end{itemize}
\item {\tt RenormalizationScale}: It is possible to use a dynamical adjusted renormalization scale, e.g. a function of the stop masses
\begin{verbatim}
RenormalizationScale = MSu[1]*MSu[6]; 
\end{verbatim}
\item  {\tt RenormalizationScaleFirstGuess}: For a first evaluation of the RGEs,
before any mass has been calculated and \SPheno hasn't any glue about the renormalization scale, 
an approximated scale can be used as 'first guess'. For example, for a mSugra scenario the common choice is
\begin{verbatim}
  RenormalizationScaleFirstGuess = m0^2 + 4 m12^2;
\end{verbatim}
This affects the running only if the SUSY scale is not fixed and SPA conventions
are disabled in the LesHouches input file.
\item {\bf Two loop contributions to the Higgs masses}: if the Higgs sector of the
model is the same as for the MSSM, the original SPheno routines for calculating
the two loop tadpole equations and two loop self energies to the the scalar and
pseudo scalar Higgs can be activated by setting
\begin{verbatim}
 UseHiggs2LoopMSSM = True;
\end{verbatim}
\item {\bf Boundary Conditions}: It is possible to define boundary conditions at three
different scales:
\begin{itemize}
\item Electroweak scale: {\tt BoundaryEWSBScale}
\item SUSY scale: {\tt BoundarySUSYScale}
\item GUT scale: {\tt BoundaryHighScale} 
\end{itemize}
All these conditions are applied when running up and down with RGEs. In contrast, there is
also the possibility to define a boundary condition at the EW scale which is only applied
when running down from the SUSY scale:
\begin{verbatim}
BoundaryEWSBScaleRunningDown = ...
\end{verbatim}
In addition, if thresholds are involved, boundary conditions can be set at the
threshold scale. See section \ref{sec:SPhenoThresholds}. It is also possible to
use a low scale input without any RGE running. In that case special boundary
conditions can be defined by  the array {\tt BoundaryLowScaleInput}.\\
All boundaries are defined by a two dimensional array. The first entry is the
name of the parameter, the second entry is the used condition at the considered
scale. The condition can be \dots
\begin{itemize}
\item \dots an input parameter from {\tt MODSEL} or {\tt EXTPAR}, e.g.
\begin{verbatim}
 {MassB, m12};
\end{verbatim}
\item \dots a block in the SLHA input file, e.g.
\begin{verbatim}
 {Yv, LHInput[Yv]};
\end{verbatim}
\item \dots a function of different parameters, e.g. 
\begin{verbatim}
 {TYd, Azero*Yd};
\end{verbatim}
\item \dots a diagonal matrix, e.g.
\begin{verbatim}
 {md2, DIAGONAL m0^2};
\end{verbatim}
\item \dots matrix  multiplications or the inverse of a matrix, e.g.
 \begin{verbatim}
 {X, MatMul2[A,InverseMatrix[B], FortranFalse]};  
 \end{verbatim}
\item \dots a self defined function
\begin{verbatim}
 {X, Func[A,B,C]};  
 \end{verbatim}
It is also possible to use some self defined function. The Fortran code of that
function has to included in the array {\tt SelfDefinedFunctions} in {\tt
SPheno.m}. It will later on be written to {\tt Model\_Data.f90}. Note, that the
standard functions needed for GMSB are already included \cite{Giudice:1998bp}:
\begin{itemize}
\item {\tt fGMSB[X]}:
\begin{eqnarray}
\nonumber f(x) &=& \frac{1+x}{x^2} \left(\ln(1+x) - 2 \text{Li}_2(\frac{x}{1+x})
+ \frac{1}{2} \text{Li}_2(2 \frac{x}{1+x}) \right) + \\
 && \frac{1-x}{x^2} \left(\ln(1-x) - 2 \text{Li}_2(\frac{x}{x-1}) + \frac{1}{2}
\text{Li}_2 (2 \frac{x}{x-1})\right)
\end{eqnarray}
\item {\tt gGMSB[X]}:
\begin{equation}
 g(x) = \frac{1+x}{x^2} \ln(1+x) + \frac{1-x}{x^2} \ln(1-x)
\end{equation}
\end{itemize}
\end{itemize}
For the matrix multiplication \verb"MatMul2" has to be used. The third argument
controls whether if only diagonal elements (\verb"FortranTrue") should be
considered or
not ( \verb"FortranFalse").

Boundary conditions can be overridden by assigning a value to a parameter in the
LesHouches input file. For example, the Higgs soft breaking masses at
the GUT scale can be forced to have specific values instead of \(m_0^2\) by
declaring
\begin{verbatim}
 Block MSOFTIN    #
  21  10000.000   # mHd2
  22  20000.00    # mHu2
\end{verbatim}
in the SLHA file.

\paragraph*{Several sets of boundary conditions} In order to implement different
versions of a single model which differ only by the used boundary conditions,
{\tt BoundaryEWSBScale}, {\tt BoundarySUSYScale}, {\tt BoundaryHighScale} can be
also a nested list, e.g.
\begin{verbatim}
BoundarySUSYScale = Table[{},{2}];
BoundaryGUTScale = Table[{},{2}];

BoundarySUSYScale[[1]] = {{KappaNMSSM, KappaInput},
                          {LambdaNMSSM, LambdaInput}};
BoundaryGUTScale[[1]]  = {};

BoundarySUSYScale[[2]] = {};
BoundaryGUTScale[[2]]  = {{KappaNMSSM, KappaInput},
                          {LambdaNMSSM, LambdaInput}};
\end{verbatim}
In the first case, the input values for \(\lambda\) and \(\kappa\) are
used at the SUSY scale, in the second on at the GUT scale. To
communicate to \SPheno which set of boundary conditions should be used
for a run, flag 2 in {\tt MODSEL} is used:
\begin{verbatim}
Block MODSEL #
 2  X  # This uses the X. set of boundary conditions. 
\end{verbatim}
The default value is 1. 
\item {\bf Lists for calculating decay widths}:
\begin{itemize}
\item {\tt ListDecayParticles}: List of particles for which the two-body decays
are to be calculated. This can be a list of particles using the names inside
\SARAH, e.g.
\begin{verbatim}
ListDecayParticles = {Sd,Su,Se,hh,Ah,Hpm,Chi};
\end{verbatim}
or just {\tt Automatic}. If {\tt Automatic} is used, the widths of all particles
not defined as standard model particles as well as for the top quark are
calculated.
\item {\tt ListDecayParticles3B}; Three body decays of fermions. This can be a
list with the names of the particles and the corresponding files names, e.g.
\begin{verbatim}
ListDecayParticles3B =  {{Chi,"Neutralino.f90"},
                         {Cha,"Chargino.f90"},
                         {Glu,"Gluino.f90"}}; 
\end{verbatim}
or just {\tt Automatic}. If {\tt Automatic} is used, the widths of all fermions
not defined as standard model particles are calculated. The auto generated
file names are {\tt ParticleName.f90}.
\end{itemize}
%
% \item {\bf Ordering of mass eigenstates}: normally, all particles of one kind are
% ordered in \SPheno by their mass. However, it might be desirable to override
% this behavior and instead define another ordering scheme, for example
% if several massless CP odd particles at tree level exist which can be assigned
% to the Goldstone boson. For this purpose, a condition can be defined by using
% %
% 
% \begin{verbatim}
%  ConditionForMassOrdering = { {Particle, Condition}, ... };
% \end{verbatim}
% %
% The condition has to be Fortran source code and is added to the corresponding
% routine. For instance, a condition for the NMSSM would read
% %
% \begin{verbatim}
% ConditionForMassOrdering={
% {Ah,
% "If ((Abs(ZA(1,3)).gt.Abs(ZA(2,3))).And.(MAh2(1).lt.1.E-5_dp) &
%         &.And.(MAh2(2).lt.1.E-5_dp)) Then \n
%    MAh2temp = MAh2 \n
%    ZAtemp = ZA \n
%    ZA(1,:) = ZAtemp(2,:) \n
%    ZA(2,:) = ZAtemp(1,:) \n
%    MAh2(1) = MAh2temp(2) \n
%    MAh2(2) = MAh2temp(1) \n
% End If \n \n"}
% };
% \end{verbatim}
%
This checks whether two massless pseudo scalars are present in the spectrum and,
if this is the case, it uses as Goldstone boson the not singlet-like particle.
\item {\bf Flag to switch off loop-corrections}: if
\begin{verbatim}
FlagLoopContributions = True;  
\end{verbatim}
is used, flags are created which can be used to disable loops which contain specific fields. 
This applies to the 1-loop mass corrections as well as the calculation of low-energy observables.
The necessary entries to are added to the block {\tt SPhenoInput} in the LesHouches input file
and also listed in the template created by \SARAH. 
\item Low scale input: it is also possible to define the free parameters of the
model at a specific scale without RGE running. These parameters are afterward
used to calculate the loop corrected mass spectrum and the decays. The
corresponding flag is
\begin{verbatim}
Block MODSEL #
 1  0     #  Low scale input
12  1000. #  Renormalization scale
\end{verbatim}
which in this example would declare the parameters to be renormalized at 1~TeV
if no explicit renormalization scale is defined.

\end{enumerate}
 
\subsection*{Information about particles and parameters}
\SARAH needs some information about the physical meaning of some particles and
parameters. These information are used to calculate the gauge and Yukawa
couplings at the electroweak scale, calculate the CKM matrix, use the correct
on-shell masses, etc. All definitions are done by the description statement in
the parameters and particles file. 
\begin{enumerate}
\item The following particles are needed: 
\begin{itemize}
\item {\tt Leptons}
\item {\tt Down-Quarks}
\item {\tt Up-Quarks}
\item {\tt Photon}
\item {\tt Gluon}
\item {\tt W-Boson}
\item {\tt Z-Boson}
\item {\tt Up-Squarks}
\item {\tt Higgs}
\item {\tt Pseudo-Scalar Higgs}
\item {\tt Charged Higgs}
\item {\tt Neutrinos}
\item {\tt Sneutrinos}
\end{itemize}

\item The following parameters have to be defined:
\begin{itemize}
\item {\tt Up-Yukawa-Coupling}
\item {\tt Down-Yukawa-Coupling}
\item {\tt Lepton-Yukawa-Coupling}
\item {\tt Hypercharge-Coupling}
\item {\tt Left-Coupling}
\item {\tt Strong-Coupling}
\item {\tt Up-Squark-Mixing-Matrix}
\item {\tt Down-Squark-Mixing-Matrix}
\item {\tt Left-Lepton-Mixing-Matrix}
\item {\tt Right-Lepton-Mixing-Matrix}
\item {\tt Left-Down-Mixing-Matrix}
\item {\tt Right-Down-Mixing-Matrix}
\item {\tt Left-Up-Mixing-Matrix}
\item {\tt Right-Up-Mixing-Matrix}
\item {\tt Weinberg-Angle}
\item {\tt Down-VEV}
\item {\tt Up-VEV}
\item {\tt Scalar-Mixing-Matrix}
\item {\tt Pseudo-Scalar-Mixing-Matrix}
\item {\tt Softbreaking right Down-Squark Mass} 
\item {\tt Softbreaking right Up-Squark Mass}
\item {\tt Softbreaking left Slepton Mass}
\item {\tt Softbreaking right Slepton Mass}
\item {\tt Softbreaking left Squark Mass}
\item {\tt Trilinear-Up-Coupling}
\item {\tt Trilinear-Down-Coupling}
\item {\tt Trilinear-Lepton-Coupling}
\end{itemize}
\end{enumerate}

\subsection*{Mixing angles in gauge sector}
The mixings in the gauge sector are normally expressed by some angles, e.g,
\(\Theta_W\) in SM/MSSM. \SPheno diagonalizes the mass matrix for the gauge 
bosons and calculates also the values of the rotation matrices. Those can be
used to calculate the corresponding mixing angle. For this purpose, the relation
between the mixing angle and the mixing matrix have to be defined in the {\tt 
parameter.m} file using an additional dependence called {\tt DependenceSPheno} For instance,
\begin{verbatim}
 {{ Description -> "Weinberg-Angle",
             ...
             DependenceSPheno -> ArcCos[Abs[ZZ[1,1]]] }},
\end{verbatim}

\subsection{Including Thresholds}
\label{sec:SPhenoThresholds}
Using \SARAH it is possible to include thresholds in the RGE evaluation.  
\subsubsection{Thresholds without gauge symmetry breaking}
If all scales have the same gauge structure, it is possible  for \SARAH to
derive the RGEs for all scales from the RGEs for the highest scale by performing
the following steps:
\begin{itemize}
\item For those fields which should be integrated out during the run new
variables \(n_{gen}(\Phi_i)\) are introduced, which define the number of
generation of the heavy field \(\Phi_i\). All gauge group constants like the the
Dynkin index summed over chiral superfields, \(S(R)\), are expressed as function
of \(n_{gen}(\Phi_i)\). These \(n_{gen}(\Phi_i)\) are dynamically adjusted, when
the energy scale crosses a threshold.
\item It is also necessary to set the couplings which involve heavy fields  to
zero when a threshold is crossed. For example, the Yukawa type coupling of the
form \(Y^{ij} \Phi_i \phi_j H\) involves three generations of the heavy field
\(\Phi\). At the threshold of \(\Phi_k\), the \(k\)-th row of
\(Y\) is set to zero. That happens similarly for all other superpotential and
soft-breaking parameters. 
\item It is assumed, that the masses of the scalar and fermionic component of a
heavy superfield are the same, i.e. the masses are much larger than the soft-breaking
masses. Furthermore, it is assumed that the masses are given by a bilinear superpotential term.
\end{itemize}

To include thresholds without gauge symmetry breaking, the following
steps have to be performed:
\begin{enumerate}
\item The heavy fields must be deleted in the \SARAH model definition:
\begin{verbatim}
 DeleteFields = {...};
\end{verbatim}
This ensures, that the these particle are not take into account for the 
calculation of decays or loop corrections  at the SUSY scale.
\item The thresholds have to be defined in \verb"SPheno.m" :
\begin{verbatim}
 Thresholds = {{Scale1, {HeavyFields1}},
               {Scale2, ... }};
\end{verbatim}
For all scales an entry in the array \verb"Thresholds" has to be added. Each
entry defines first the threshold scale, at second position a list with the
heavy superfields is given. Also specific generations for a superfield  can be
given.
\end{enumerate}

It is possible to define boundary conditions at each threshold scale for running
up and down separately:
\begin{verbatim}
 BoundaryConditionsUp[[x]] = { ...};
 BoundaryConditionsDown[[x]] = { ...};
\end{verbatim}

\paragraph*{Threshold corrections}
Using 2-loop RGEs demands 1-loop boundary condition. Therefore, at each
threshold scale the one loop threshold corrections to gauge couplings and
gaugino masses  are calculated. The general
expressions are \cite{Hall:1980kf}
\begin{eqnarray}
\label{eq:shift1}
 g_i & \rightarrow & g_i \left( 1\pm \frac{1}{16 \pi^2} g_i^2 I^i_2(r) 
\ln\left(\frac{M^2}{M_T^2}\right)\right)  \thickspace ,\\
\label{eq:shift2}
 M_i & \rightarrow & M_i \left( 1\pm \frac{1}{16 \pi^2} g_i^2 I^i_2(r) 
\ln\left(\frac{M^2}{M_T^2}\right)\right) \thickspace .
\end{eqnarray}
\(I^i_2(r)\) is the Dynkin index of a field transforming as
representation \(r\) with respect to the gauge group belonging to the
gauge coupling \(g_i\), \(M\) is the mass of this particle and \(M_T\)
is the threshold scale. When evaluating the RGEs from the low to the
high scale the contribution is positive, when running down, it is
negative. 
\paragraph*{Example}
As an example, a version of \SPheno implementing the seesaw type~II and
type~III models can be generated by adding the following entries to
\verb"Spheno.m"
\begin{enumerate}
 \item Seesaw II: 
\begin{verbatim}
Thresholds={
  {Abs[MTMIN],{s,sb,t,tb,z,zb}}
};
\end{verbatim}
 \item Seesaw III:
\begin{verbatim}
Thresholds={
  {Abs[MWM3IN[1,1]],{Hx3[1],Hxb3[1],Hg3[1],Hb3[1],Hw3[1]}},
  {Abs[MWM3IN[2,2]],{Hx3[2],Hxb3[2],Hg3[2],Hb3[2],Hw3[2]}},
  {Abs[MWM3IN[3,3]],{Hx3[3],Hxb3[3],Hg3[3],Hb3[3],Hw3[3]}}
};
 \end{verbatim}
\end{enumerate}

\subsubsection{Thresholds with gauge symmetry breaking}

If the gauge structure at the different scales are different, each set of RGEs
is calculated separately and this information is then combined
into one consistent version of \SPheno. This code includes
routines for calculating finite shifts in the gauge couplings and gaugino 
mass parameters. As an example, the implementation of a left-right
supersymmetric model with two symmetry breaking scales is shown in
app.~\ref{app:Omega}.
In order to implement such a model, the following steps are necessary:
\begin{enumerate}
\item For each regime, a separate model file for \SARAH has to be created.
These model file have to be saved in the subdirectories \verb"Regime-1",
\verb"Regime-2", \dots of  
\begin{verbatim}
[SARAH Directory]/Models/[Model]/
\end{verbatim}
They are numbered beginning with the highest scale.
\item The \SPheno input file for the higher scales must provide the following
information:
\begin{itemize}
\item {\tt IntermediateScale = True}
\item {\tt RegimeNr = X}
%\item A list of the heavy fields, which should be integrated out: 
\item A list of the heavy fields, which should be integrated out, the gauge
sector below the threshold as well as the corresponding quantum numbers of the
fields integrated. That's needed to calculate the finite shifts at the threshold
scale.  For instance, the entries might read
\begin{verbatim}
HeavyFields = {Field_1, Field_2,..};
NextGauge =   {U[1],   SU[2],   SU[2],    SU[3]};
NextQN = {
       {Field_1, 0,       2,       1,         1},
       {Field_2, 1/3,     1,       2,         4},
 ...
};
\end{verbatim}
\end{itemize}
\item All necessary information for combining the regimes to one \SPheno is
given in \verb"SPheno.m" of the lowest scale.
\begin{itemize}
\item {\tt IntermediateScale = False}
\item {\tt RegimeNr = X}
\item The threshold scales: {\tt ThresholdScales = {...} }
\item The boundary conditions for running up and down at each threshold scale:
\begin{verbatim}
 BoundaryConditionsUp[[x]] = {...};
 BoundaryConditionsDown[[x]] = {...};
\end{verbatim}
In the boundary conditions \verb"index1", \verb"index2", \dots can be used for
defining sums over indices.
\item The usual information for \SPheno, defined in the sec.
\ref{sec:sphenoinputfile}.
\end{itemize}
\end{enumerate}
When starting the \SPheno output of the lowest scale, automatically all other
scales are evolved. Note, to calculate the RGEs of the different regimes
requires \SARAH to start one additional \Mathematica kernel.
For passing the information between the different \Mathematica kernels
 a directory \verb"Meta" in the model directory is created by \SARAH. Also the
screen output of \Mathematica during the evaluation of the higher regimes
is written to that directory (\verb"Output-Regime-X.m"). So, the user can
supervise the progress and see potential error messages. The necessary
information of each regime for writing the combined source code for \SPheno at
the end is saved by \SARAH in the files \verb"Regime-X.m".

\subsection{Supported models and known issues}
\label{sec:SPheno_restrictions}
While \SARAH can create valid \SPheno code for many different
models, there are some requirements on the model and some minor restrictions
on the functionality  of the resulting \SPheno module. At the moment, those are

\begin{itemize}
 \item {\bf Fit to low energy data}: in order to perform a fit to low energy data
(e.g.\ for fermion masses, $m_Z$, $G_F$ and $\alpha_{em}$) as
starting point of the RGE evaluation, the following parameters must be present
in the model: Yukawa couplings for lepton and quarks, two Higgs VEVs and, of
course, the three SM gauge couplings and the SM particle content. However, it
is still possible to use at least some features of the \SPheno output of \SARAH
by manually supplying model parameters for \SPheno. In that way, the RGE evaluation and
the fit the electroweak data is skipped, but the one-loop corrected masses as well as the decay widths
and branching ratios are calculated.
\item {\bf Flavor decomposition}: with \SARAH it is possible to assign a unique
name to each generation of a particular field and this way treat the individual generations as
independent fields. That is not yet supported in the \SPheno output. Furthermore, mixing
matrices generated with the option {\tt NoFlavorMixing} can not yet be handled
by the numerical code. 
\end{itemize}
\lstset{frame=none}

\subsection{Low energy \SPheno version}
\label{sec:LowScaleSPheno}
It is also possible to create a \SPheno version with much less features which
only accepts low energy input. That means, the RGEs are not written out and
also the fit to the electroweak data is not performed in the numerical
evaluation of one point. It just solves the tadpole equations, calculates the
tree- and one-loop masses as well as the decay widths and branching ratios. 
The advantage of such a \SPheno version is that it
works with a larger set of models, e.g. also non-SUSY models or other models
not supported by a full evaluation as explained in
sec.~\ref{sec:SPheno_restrictions}. To get a \SPheno version without
RGE evolution, insert
\begin{verbatim}
OnlyLowEnergySPheno = True;
\end{verbatim}
in {\tt SPheno.m}. The remaining information needed by \SARAH is only a small
subset of the settings discussed above and consists of
\begin{itemize}
 \item {\tt MINPAR}
 \item {\tt ParametersToSolveTadpoles}
 \item {\tt BoundaryLowScaleInput}
 \item {\tt ListDecayParticles} and {\tt ListDecayParticles3B}. Note that the {\tt
Automatic} statement for automatically deriving the decays of all non-SM particles
does not work in
this case as \SARAH doesn't differ between SUSY or Non-SUSY
particle in order to make the output as generic as possible. Therefore, the lists
of the decaying particles have to be supplied manually. 
\end{itemize}
If all SM couplings (gauge and Yukawa couplings) as well as VEVs for up- and down Higgs  are present
in the considered model, \SARAH calculates the running gauge and Yukawa couplings at the SUSY scale. 
For this purpose, it uses the $\overline{\text{DR}}$ masses from \cite{Xing:2007fb} and 2-loop SM
RGEs. 

\subsection{Models with another gauge group at the SUSY scale}
Some SUSY models have the distinct feature that they gauge group at the SUSY scale doesn't consist of
$SU(3)_C \times SU(2)_L \times U(1)_Y$. This is for instance the case in left-right models in which 
$U(1)_R \times U(1)_{B-L}$ is just broken around the SUSY scale (see e.g. Ref.~\cite{Hirsch:2012kv} and references therein). 
This special feature has to be taken 
into account in some calculations. For instance, the calculation of the running couplings at the EW scale assumes that the hypercharge is present.\\
Therefore, \SARAH has to create in this kind of models an auxiliary variable for the hypercharge coupling called {\tt gYaux}. This is done by
adding 
\begin{verbatim}
AuxiliaryHyperchargeCoupling = True;
\end{verbatim}
in {\tt SPheno.m}. In addition, the user has to define a relation between the existing gauge couplings and the hypercharge coupling. 
For instance, in the model mentioned above, this relation reads
\begin{verbatim}
ExpressionAuxHypercharge =Sqrt[(gBL*gR - gBLgR*gRgBL)^2/((gBLgR - gR)^2 + (gBL - gRgBL)^2)];
\end{verbatim}
Note, if kinetic mixing is neglected, this reduces to the more familiar form of $\sqrt{g^2_{BL} g_R/(g_R^2 + g_{BL}^2)}$. When setting the boundary conditions
to relate the gauge couplings, one has to make sure that always the relations for the not GUT-normalized values are used. For instance,
\begin{verbatim}
BoundaryEWSBScale = {
  {gYauxt, Sqrt[5/3]*gYaux},
  {gR, g1RBLFactor*gYauxt}, 
  {gRgBLt, 1*gRgBL},
  {gBLgRt, Sqrt[2/3]*gBLgR},
  {gBLt, (5 gBLgRt gR gRgBLt - Sqrt[6] gRgBLt gYauxt^2 
      + Sqrt[(3 gBLgRt^2 - 2 Sqrt[6] gBLgRt gR + 2 gR^2) *
        (5 (gR^2 + gRgBLt^2) - 3 gYauxt^2) gYauxt^2])/(5 gR^2 - 3 gYauxt^2)},
  {gBL, Sqrt[3/2]*gBLt},
  {TanBetaR, TanBetaRinput},
  {vChiR, vR*TanBetaR/Sqrt[1 + TanBetaR^2]},
  {vChiRb, vR*1/Sqrt[1 + TanBetaR^2]}};
\end{verbatim}
Here, {\tt g1RBLFactor} is the ratio of $g_R/g_Y^{aux}$ which has been calculated in the iteration before
\begin{verbatim}
BoundaryEWSBScaleRunningDown = {
  {gBLt, gBL*Sqrt[2/3]},
  {gRgBLt, 1*gRgBL},
  {gBLgRt, Sqrt[2/3]*gBLgR},
  {gYaux, Sqrt[5*(gBLt*gR - gBLgRt*gRgBLt)^2/(3*(gBLt^2 + gBLgRt^2) 
     + 2*(gR^2 + gRgBLt^2) - 2*Sqrt[6]*(gR*gBLgRt + gBLt*gRgBLt))]},
  {g1RBLFactor, gR/gYaux},
  {gYaux, Sqrt[3/5]*gYaux}
 };
\end{verbatim}

\section{Writing input files for \HB with \SPheno}
\label{sec:HB}
\HB \cite{Bechtle:2008jh,Bechtle:2011sb} is a tool to test the neutral and
charged Higgs 
sectors against the current exclusion bounds from the Higgs
searches at the LEP, Tevatron and LHC experiments. The required input consists
of the masses, width and branching ratios of the Higgs fields. In addition, it
is either possible to provide full information about production cross sections
in \(e^+ e^-\) and \(p p\) collisions, or to work with a set of effective
couplings.

\HB can be downloaded from
\begin{verbatim}
http://projects.hepforge.org/higgsbound
\end{verbatim}

Although \HB supports a LesHouches interface, this functionality is restricted
so far to at most 5 neutral Higgs fields, and therefore, we don't use it.
Instead, a \SPheno module generated by \SARAH can create all necessary input
files needed for a run of \HB with effective couplings (option {\tt
whichinput=effC}). To write these file, the flag 76 in the block {\tt
SPhenoInfo} in the LesHouches input file has to be set to {\tt 1}.
\begin{verbatim}[frame=shadowbox]
 Block SPhenoInput #
 76 1 # Write files for HiggsBounds
\end{verbatim}
Unfortunately, we can not provide all information which can be used by \HB to
check the constraints. So far, the effective couplings \(H \rightarrow \gamma
Z\) and \(H \rightarrow g g Z\) are not calculated by \SPheno and therefore they are
set to zero in the output. In addition, as already mentioned, the \SPheno
version created by \SARAH does not include the calculations of \(e^+ e^-\) cross
sections. For this reason, also the LEP production cross section of charged
Higgs fields is not available for \SPheno and it sets this value also to 0 in the output.
However, it is of course possible to calculate this cross section as well as all
other cross sections needed for the options ({\tt whichinput=hadr} or {\tt
whichinput=part}) of \HB using \CalcHep, \Madgraph or \WHIZARD.

The following files are created by \SPheno
\begin{itemize}
 \item {\tt MH\_GammaTot.dat}: \\
 Masses and widths of all neutral Higgs fields
 \item {\tt MHplus\_GammaTot.dat}: \\
Masses and widths of all charged Higgs fields
 \item {\tt BR\_H\_NP.dat}: \\
Branching ratios for neutral Higgs fields into invisible and other neutral
Higgs fields.
 \item {\tt BR\_Hplus.dat}: \\
 Branching ratios of charged Higgs fields into \(c \bar{s}\), \(c \bar{b}\) and
\(\tau \bar{\nu}\) final states
 \item {\tt BR\_t.dat}: \\
 Branching ratios of top quark into bottom quark and a \(W\) boson or charged
Higgs
 \item {\tt effC.dat}:
effective couplings of neutral Higgs fields to
\(s\bar{s}\), \(c\bar{c}\), \(b\bar{b}\), \(t\bar{t}\), \(\mu\bar{\mu}\), 
\(\tau\bar{\tau}\), \(\gamma \gamma\), \(g g\), \(\gamma Z\), \(g g Z\) as well
as to all other neutral Higgs fields. 
 \item {\tt LEP\_HpHm\_CS\_ratios.dat}: \\
LEP production cross section of charged Higgs. (set to zero, see above). 
\end{itemize}
To run \HB, use in the \HB directory
\begin{verbatim}[frame=none]
> ./HiggsBounds LandH effC [NN] [NC] '[SPheno Directory]'
\end{verbatim}
where {\tt [NN]} has to be replaced by the number of neutral Higgs particle,
and {\tt [NC]} by the number of the charged ones. For more information, see
also the \HB manual. The results of the check are written to the file {\tt
HiggsBounds\_results.dat} in the directory from which \HB has been called.

\chapter{Other output}
\section{LHPC spectrum plotter}
The LHPC spectrum plotter is a small but handy tool to produce plots of the SUSY mass
 spectrum based on the information given in a SLHA output file \cite{LHPC}. \\
For the output it is necessary to provide a second control file in addition to
 the SLHA spectrum file. The control file includes information about the paths
 to the necessary shell tools ({\tt gnuplot}, {\tt latex}, {\tt dvips},
 {\tt ps2eps}, {\tt rm}, {\tt mv}) and the \LaTeX{} name associated with a PDG
 number. In addition, the color and column used for the different particles are
 defined in that file. \SARAH can provide such a file which works nicely
 together with the spectrum file written by a \SPheno module also created by
 \SARAH. By default it assumes the standard paths under Linux, while the color
 and column of each particle can be defined in {\tt particles.m} using the new
 option {\tt LHPC}. For instance, to put the gluino in the fourth column and to
 use purple for the lines, the entry reads
\begin{verbatim}
 {{   Description -> "Gluino",
            ...
            LaTeX -> "\\tilde{g}",
            LHPC -> {4, "purple"},
            ... }},
\end{verbatim}
As name for the colors all available colors in {\tt gnuplot} can be used. The
 control file for a given set of eigenstates of the initialized model is written
 via
\begin{verbatim}
MakeLHPCstyle[$EIGENSTATES]; 
\end{verbatim}
and saved in the directory
\begin{verbatim}
$SARAH/Output/$MODEL/$EIGENSTATES/LHPC/ 
\end{verbatim}
It is used together with a spectrum file to create the figure by the shell
 command 
\begin{verbatim}
./LhpcSpectrumPlotter.exe SPheno.spc.$MODEL LHPC_$MODEL_Control.txt
\end{verbatim}

\section{All at once}
To generate the entire output for \SPheno, \CalcHep, \WHIZARD, \FeynArts as well as model files in the UFO format and the \LaTeX{} file, use 
\begin{verbatim}
MakeAll[Options];
\end{verbatim}
The options are
\begin{itemize}
 \item {\tt ReadLists}, Values: {\tt True/False}, Default: {\tt False}:
  Should results for earlier runs are used. 
 \item {\tt  IncludeSPheno}, Values: {\tt True/False}, Default: {\tt True}:
 Includes/excludes the \SPheno output
 \item {\tt  IncludeFeynArts}, Values: {\tt True/False}, Default: {\tt True}:
 Includes/excludes the \FeynArts output
  \item {\tt  IncludeCalcHep}, Values: {\tt True/False}, Default: {\tt True}:
 Includes/excludes the \CalcHep output
  \item {\tt  IncludeWHIZARD}, Values: {\tt True/False}, Default: {\tt True}:
 Includes/excludes the \WHIZARD output
   \item {\tt  IncludeUFO}, Values: {\tt True/False}, Default: {\tt True}:
 Includes/excludes the UFO output
   \item {\tt  IncludeTeX}, Values: {\tt True/False}, Default: {\tt True}:
 Includes/excludes the \LaTeX{} output
\end{itemize}

\chapter{Numerical values}
\SARAH offers some basic routines for working with numerical values of parameters
and calculating mixing matrices and masses. 

\section{Adding numerical values}
\label{addNumValues}
There are different ways to define numerical values for a parameter: they can
just be added in the parameter file as described in sec.~\ref{parameterFile} or
the values from a LesHouches spectrum file can be used. To read automatically a
LesHouches file when evaluating the command \verb"Start", add 
\begin{verbatim}
SpectrumFile="Name of File"
\end{verbatim}
to the model file. The spectrum file must be in the same directory as the model
file. \\  
It is also possible to read a spectrum file afterwards via the command
\begin{verbatim}
ReadSpectrum["Spectrum File"]
\end{verbatim}

An additional possibility to add a numerical value during the work with \SARAH
is to use
\begin{verbatim}
SetParameterValue[parameter,value];
\end{verbatim}
\verb"parameter" is the name of the parameter and \verb"value" the numerical
Value. Of course, it is also possible to remove easily a value. For this purpose
 the command
\begin{verbatim}
DeleteParameterValue[parameter]; 
\end{verbatim}
is used. 

\section{Calculate Mixing Matrices}
After numerical values for all free  parameters are defined, the mass
eigenstates and entries of mixing matrices can be calculated by 
\begin{verbatim}
CalcMatrices;
\end{verbatim}

The eigenvalues and mixing matrices of each mass matrix are saved in two
additional variables:
\begin{enumerate}
\item Eigenvalues: the eigenvalues of the mass matrix are saved in 
\begin{verbatim}
Mass <> $PARTICLE
\end{verbatim}
\item The numerical value of the mixing matrix are saved in
\begin{verbatim}
"Name of Mixing Matrix" <> Num
\end{verbatim}
\end{enumerate}

\paragraph*{Negative Squared Masses}
It could happen that for some parameter points some eigenvalues of the mass
matrices for scalars are getting negative. If this happens, there will appear a
warning in the output and the variable \verb"warning" is set to \verb"True". 

\paragraph*{Example}
After using \verb"CalcMatrices", the numerical value for the down-squark mixing
matrix are saved in \verb"MDNum" and the corresponding squark masses are saved
in \verb"MassSd".

\section{Calculate Numerical Values}
To get the numerical value for a term
\begin{verbatim}
NumericalValue[x];
\end{verbatim}
is used. \verb"x" can be e.g. the entry of a mass matrix or a vertex 

\paragraph*{Examples}
\begin{enumerate}
\item The numerical value for the Higgs mass matrix is calculated with
\begin{verbatim}
NumericalValue[MassMatricesFullEWSB[[4]]]; 
\end{verbatim}

\item The numerical value for the interaction between the photon and down
squarks is calculated with 
\begin{verbatim}
NumericalValue[Vertex[{VP,Sd,conj[Sd]}][[2,1]]];
\end{verbatim}
\end{enumerate}

\begin{appendix}
\chapter{Calculation of Group Factors}
\label{group}
SARAH supports not only chiral superfields in the fundamental representation but
in any irreducible representation of \(SU(N)\). In most cases, it is possible to
fix the transformation properties of the chiral superfield by stating the
corresponding dimension \(D\). If the dimension is not unique, also the Dynkin
labels are needed. For calculating kinetic terms and D-terms, it is necessary to
derive from representation the corresponding generators. Furthermore,  the
eigenvalues \(C_2\) of the quadratic Casimir for any irreducible representation
\(r\) 
\begin{equation}
T^a T^a \phi(r) = C_2(r) \phi(r)
\end{equation}
as well as the Dynkin index \(I\)
\begin{equation}
Tr(T^a T^b) \phi(r) = I \delta_{a b} \phi(r)
\end{equation}
are needed for the calculation of the RGEs. All of that is derived by \SARAH due
to the technique of Young tableaux. The following steps are evolved:
\begin{enumerate}
\item The corresponding Young tableaux fitting to the dimension \(D\) is
calculated using the hook formula:
\begin{equation}
D = \Pi_i \frac{N + d_i}{h_i}
\end{equation}
\(d_i\) is the distance of the \(i.\) box to the left upper corner and \(h_i\)
is the hook of that box. \\
\item The vector for the highest weight \(\Lambda\) in Dynkin basis is extracted
from the tableaux.
\item The fundamental weights for the given gauge group are calculated.
\item The value of \(C_2(r)\) is calculated using the Weyl formula
\begin{equation}
C_2(r) = ( \Lambda, \Lambda + \rho) \thickspace .
\end{equation}
\(\rho\) is the Weyl vector. 
\item The Dynkin index \(I(r)\) is calculated from \(C_2(r)\). For this step,
the value for the fundamental representation is normalized to \(\frac{1}{2}\). 
\begin{equation}
I(r) = C_2(r) \frac{D(r)}{D(G)}
\end{equation}
With \(D(G)\) as dimension of the adjoint representation.
\item The number of co- and contra-variant indices is extracted from the Young
tableaux. With this information, the generators are written as tensor product.
\end{enumerate}
The user can calculate this information independently from the model using the
new command
\begin{verbatim}
CheckIrrepSUN[Dim,N]
\end{verbatim}
\verb"Dim" is the dimension of the irreducible representation and \verb"N" is
the dimension of the \(SU(N)\) gauge group. 
The result is a vector containing the following information: (i) repeating the
dimension of the field, (ii) number of covariant indices, (iii) number of
contra-variant indices, (iv) value of the quadratic Casimir \(C_2(r)\), (v) value
of the Dynkin index \(I(r)\), (vi) Dynkin labels for the highest weight.
\paragraph*{Examples}
\begin{enumerate}
\item {\bf Fundamental representation} The properties of a particle,
transforming under the fundamental representation of \(SU(3)\) are calculated
via {\tt CheckIrrepSUN[3,3]}. The output is the well known result
\begin{verbatim}
{3, 1, 0, 4/3, 1/2, {1, 0}}
\end{verbatim}
\item {\bf Adjoint representation } The properties of a field transforming as
{\bf 24} of \(SU(5)\) are calculated by
{\tt  CheckIrrepSUN[24,5] }. The output will be
\begin{verbatim}
{24, 1, 1, 5, 5, {1, 0, 0, 1}}
\end{verbatim}
\item {\bf  Different representations with the same dimension }
The {\bf{70}} under \(SU(5)\) is not unique. Therefore, {\tt CheckIrrepSUN[\{70,
\{0, 0, 0, 4\}\}, 5] }  returns
\begin{verbatim}
{70, 0, 4, 72/5, 42, {0, 0, 0, 4}} 
\end{verbatim}
while {\tt CheckIrrepSUN[\{70, \{2, 0, 0, 1\}\}, 5] } leads to
\begin{verbatim}
{70, 2, 1, 42/5, 49/2, {2, 0, 0, 1}} 
\end{verbatim}
\end{enumerate}

\chapter{Calculation of the Lagrangian of supersymmetric models and deriving
the vertices}
\section{The supersymmetric Lagrangian}
\label{sec:Lagrangian}
We describe in this section the calculation of the complete Lagrangian for a
supersymmetric model based on the superpotential and the gauge structure. 
\paragraph*{Interactions of chiral superfields}
If we call the superpotential for a given theory \(W\) and use \(\phi_i\) for
the scalar and \(\psi_i\) for the fermionic component of a chiral
supermultiplet, the matter interactions can by derived by
\begin{equation}
\label{eq:MatterInteractions}
\La_Y = - \frac{1}{2} W^{ij} \psi_i \psi_j + \mbox{h.c.} \thickspace,
\hspace{1cm} \La_F = F^{* i} F_i + \mbox{h.c.}
\end{equation}
with 
\begin{equation}
\label{eq:FTerms}
W^{ij} = \frac{\delta^2}{\delta \phi_i \delta \phi_j} W \hspace{1cm} \mbox{and}
\hspace{1cm} F^i = - W^{* i} = \frac{\delta W}{\delta \phi_i} \thickspace .
\end{equation}
The first term of eq. (\ref{eq:MatterInteractions}) describes  the interaction
of two fermions with one scalar, while the second term forms the so called
F-terms which describe  four-scalar interactions.  

\paragraph*{Interactions of vector superfields}
We name  the spin-\(\frac{1}{2}\) component of a vector supermultiplet
\(\lambda\) and  the spin-1 component \(A^\mu\). The most general Lagrangian
only involving these fields is
\begin{equation}
\label{eq:LagVS}
\La = - \frac{1}{4} F^a_{\mu\nu} F^{\mu\nu a} - i \lambda^{\dagger a }
\bar{\sigma}^\mu D_\mu \lambda^a
\end{equation}
with the field strength
\begin{equation}
\label{eq:FieldStrength}
F_{\mu\nu}^a = \partial_\mu A_\nu^a - \partial_\nu A_\mu^a + g f^{abc} A_\mu^b
A_\nu^c \thickspace ,
\end{equation}
and the covariant derivative
\begin{equation}
D_\mu \lambda^a = \partial_\mu \lambda^a + g f^{abc} A_\mu^b \lambda^c
\thickspace .
\end{equation}
Here, \(f^{abc}\) is the structure constant of the gauge group. Plugging
eq.~(\ref{eq:FieldStrength}) in the first term of eq.~(\ref{eq:LagVS}) leads to
self-interactions of three and four gauge bosons
\begin{equation}
\La_{V} = - \frac{1}{4} (\partial_\mu A_\nu^a - \partial_\nu A_\mu^a)  g f^{abc}
A^{\mu,b} A^{\nu,c}  - \frac{1}{4} g^2 (f_{abc} A_\mu^b A_\nu^c) (f^{ade}
A^{\mu,e} A^{\nu,e})  \thickspace .
\end{equation}
The second term of eq.~(\ref{eq:LagVS})  describes the  interactions between
vector bosons and gauginos.
\paragraph*{Supersymmetric gauge interactions}
The parts of the Lagrangian with both chiral and vector superfields are the
kinetic terms for the fermions and scalars
\begin{equation}
\La_{kin} = - D^\mu \phi^{*i} D_\mu \phi_i  - i \psi^{\dagger i }
\bar{\sigma}^\mu D_\mu \psi_i 
\end{equation}
as well as the interaction between a gaugino and a matter fermion and scalar
\begin{equation}
\La_{GFS} = - \sqrt{2} g (\phi^* T^a \psi) \lambda^a + \mbox{h.c.} \thickspace .
\end{equation} 
Here, \(T^a\) are the fundamental generators of the gauge group. Furthermore,
the covariant derivatives are 
\begin{eqnarray}
D_\mu \phi_i &=& \partial_\mu \phi_i - ig A^a_\mu (T^a \phi)_i  \thickspace ,\\
D_\mu \phi^{*i} &=& \partial_\mu \phi^{*i} + ig A^a_\mu (\phi^* T^a)^i
\thickspace , \\
D_\mu \psi_i &=& \partial_\mu \psi_i - i g A^a_\mu (T^a \psi)_i \thickspace,
\end{eqnarray}
In addition, the D-Terms are defined by
\begin{equation}
\La_D =  \frac{1}{2} D^a D^a \thickspace .
\end{equation}
The solution of the equations of motion for the auxiliary fields leads to 
\begin{equation}
\label{eq:DTerms}
D^a = - g (\phi^* T^a \phi) \thickspace .
\end{equation}
\paragraph*{Soft-breaking terms} SUSY must be a broken. This can be parametrized
by adding soft-breaking terms to the Lagrangian. The possible terms are the mass
terms for all scalar matter fields and gauginos
\begin{equation}
\La_{SB} = - m_{\phi_i}^2 \phi_i \phi_i^* - \frac{1}{2} M_{\lambda_i} \lambda_i
\lambda_i
\end{equation}
as well as soft-breaking interaction corresponding to the superpotential terms
\begin{equation}
\La_{Soft,W} = T \phi_i \phi_j \phi_k + B \phi_i \phi_j + S \phi_i \thickspace .
\end{equation}
\paragraph*{Dirac gauginos} Using the just described method and input it was possible to implement the MSSM and many extensions of it in \SARAH. However, in the last years another structure had been become popular in SUSY model building: Dirac mass terms for gauginos. If one demands for instance a continuous R-symmetry this forbids to write down Majorana mass terms of the gauginos. However, Dirac mass terms between a gaugino and a chiral superfields in the adjoint representation might be allowed. Such a mass term between a vector and a chiral superfield expanded in component fields leads to two physical relevant terms in the Lagrangian  \cite{Benakli:2011vb}:
\begin{equation}
\La_{GF} = - m_D \lambda_a \Psi_a + \sqrt{2} m_D \phi_a D_a 
\end{equation}
The first term is the Dirac mass term of the gauginos, the second term is an additional D-term contribution. \\
Models with Dirac mass terms are fully supported since version 3.2.0 of \SARAH. Also the corresponding RGEs are calculated at two-loop level. \\
\paragraph*{Gauge fixing terms and ghost interactions}
\label{GaugeFixing}
The Lagrangian of a theory without further restrictions is invariant under a
general gauge transformation. This invariance leads to severe problems in the
quantization of the theory as can be seen in the divergence of functional
integrals. Therefore, it is in necessary to add gauge fixing terms to break this
gauge invariance. \\
The general form of the gauge fixing Lagrangian is
\begin{equation}
\La_{GF} = - \frac{1}{2} \sum_a|f(x)^a|^2 \thickspace .
\end{equation} 
\(f_a\) can be a function of partial derivatives of a gauge boson and a
Goldstone boson. The corresponding ghost terms of the ghost fields
\(\bar{\eta}\) and \(\eta\)  are
\begin{equation}
\La_{Ghost} = - \bar{\eta}_a (\delta f^a) \thickspace. 
\end{equation}
Here, \(\delta\) assigns the operator for a BRST transformation. For an unbroken
gauge symmetry, the gauge fixing terms in the often chosen \(R_\xi\)-gauge  are
\begin{equation}
\La_{GF} = - \frac{1}{2 R_\xi} \sum_a \left(\partial^\mu V_\mu^a \right)^2 
\thickspace.
\end{equation}
Here, \(V_\mu\) are the gauge boson of the unbroken gauge group. It is often
common to choose a distinct value for \(R_\xi\). The most popular gauges are the
unitary gauge \(R_\xi \rightarrow \infty\) and the Feynman-'t Hooft-gauge
\(R_\xi = 1\). For broken symmetries, the gauge fixings terms are chosen in a
way that the mixing terms between vector bosons and scalars disappear from the
Lagrangian. Therefore, the common choice for the gauge fixing Lagrangian for
theories with the standard model gauge sector after EWSB is  
\begin{equation}
\label{GFewsb}
\La_{GF, R_\xi} =  - \frac{1}{2 \xi_\gamma} \left( \partial^\mu
\gamma_\mu\right)^2 - \frac{1}{2 \xi_Z} \left( \partial^\mu Z_\mu + \xi_Z M_Z
G^0 \right)^2 + - \frac{1}{\xi_{W^+}} \left( \partial^\mu W^+_\mu + \xi_{W^+}
M_W G^+\right)^2 \thickspace . 
\end{equation}
Here, \(G^0\) and \(G^+\) are the Goldstone bosons, which build the longitudinal
component of the massive vector bosons. \\

\section{Deriving the vertices}
\SARAH calculates the vertices as partial derivatives with respect to the
external fields and applies afterwards the vacuum conditions
\begin{equation}
 c \bra \frac{\partial^n L}{\partial \phi_i \dots \partial \phi_j} \ket
\end{equation}
The numerical coefficient \(c\) depends on the generic type of the interactions
and takes the value \(i\) if one or more scalars are involved and is 1
in the other cases. Furthermore, we define a momentum flow by the replacement
\begin{equation}
 \partial_\mu \phi \rightarrow - i \p_\mu (\phi)
\end{equation}

\chapter{Parts of the Lagrangian in \SARAH}
\label{PartsLag}
By default, \SARAH writes down the most general Lagrangian derived by the rules shown in app.~\ref{sec:Lagrangian}. However, there are also flags to suppress specific terms in the Lagrangian. This might especially interesting for models with a $R$-symmetric which forbids Majorana mass terms for gauginos as well as trilinear soft-breaking terms, or it can be used for the implementation of non-SUSY models.  The following options exist:
\begin{itemize}
 \item \verb"AddTterms = True/False;", default: \verb"True", includes/excludes trilinear softbreaking couplings
 \item \verb"AddBterms = True/False;", default: \verb"True", includes/excludes bilinear softbreaking couplings
 \item \verb"AddLterms = True/False;", default: \verb"True", includes/excludes linear softbreaking couplings
 \item \verb"AddSoftScalarMasses = True/False;", default: \verb"True", includes/excludes soft-breaking scalar masses
 \item \verb"AddSoftGauginoMasses = True/False;", default: \verb"True", includes/excludes Majorana masses for gauginos
 \item \verb"AddDiracGauginos = True/False;", default: \verb"False", includes/excludes Dirac masses for gauginos
 \item \verb"AddSoftTerms = True/False;", default: \verb"True", includes/excludes all soft-breaking terms
 \item \verb"AddDterms = True/False;", default: \verb"True", includes/excludes all D-terms
 \item \verb"AddFterms = True/False;", default: \verb"True", includes/excludes all F-terms
\end{itemize}
\SARAH saves the different parts of the Lagrangian for the different eigenstates
in the following variables:
\begin{itemize}
\item \verb"LagSV[$EIGENSTATES]": Parts with scalars and vector bosons (=
kinetic terms for scalars)
\item \verb"LagFFV[$EIGENSTATES]": Parts with fermions and vector bosons (=
kinetic terms of scalars)
\item \verb"LagSSSS[$EIGENSTATES]": Parts with only scalars (= scalar
potential)
\item \verb"LagFFS[$EIGENSTATES]": Parts with fermions and scalars
\item \verb"LagVVV[$EIGENSTATES]": Parts with three vector bosons
\item \verb"LagVVVV[$EIGENSTATES]": Parts with four vector bosons
\item \verb"LagGGS[$EIGENSTATES]": Parts with ghosts and scalars  
\item \verb"LagGGV[$EIGENSTATES]": Parts with ghosts and vector bosons
\item \verb"LagSSA[$EIGENSTATES]": Parts with scalars and auxiliary fields
(only needed for \CalcHep output)
\end{itemize}
In addition, for an effective theory, there might exist
\begin{itemize}
\item \verb"LagSSSSSS[$EIGENSTATES]": Dimension 6 operators with only scalars
\item \verb"LagSSSVVV[$EIGENSTATES]": Dimension 6 operators with scalars and
fermions
\item \verb"LagFFFF[$EIGENSTATES]": Dimension 6 operators with only fermions
\item \verb"LagFFSS[$EIGENSTATES]": Dimension 5 operators with fermions and
scalars
\item \verb"LagFFVV[$EIGENSTATES]": Dimension 6 operators with fermions and
vector bosons
\end{itemize}
Moreover, the different results of the calculation of the Lagrangian in gauge
eigenstates are also saved separately. The variable names are:
\begin{itemize}
 \item Superpotential: {\tt Superpotential}
 \item Fermion - scalar interactions coming from the superpotential: {\tt Wij}
 \item F-Terms: {\tt FTerms}
 \item Scalar soft-breaking masses: {\tt SoftScalarMass }
 \item Gaugino masses: {\tt SoftGauginoMass }
 \item Soft-breaking couplings: {\tt SoftW}
 \item Kinetic terms for scalars: {\tt KinScalar}
 \item Kinetic terms for fermions: {\tt KinFermion}
 \item D-Terms: {\tt DTerms}
 \item Interactions between gauginos and a scalar and a fermion: {\tt FSGaugino}
 \item Trilinear self-interactions of gauge bosons: {\tt GaugeTri}
 \item Quartic self-interactions of gauge bosons: {\tt GaugeQuad}
 \item Interactions between vector bosons and gauginos: {\tt BosonGaugino}
\end{itemize}
Furthermore, the additional interactions and the redefinition of existing
interactions are saved in {\tt LagRedefinition}.

\chapter{Gauge anomalies}
\label{sec:GaugeAnomaly}
Before \SARAH starts the calculation of the Lagrangian it checks the model for the different triangle anomalies. These anomalies can involve diagrams with three external gauge bosons belonging to the same \(U(1)\) or \(SU(N)\) gauge group. To be anomaly free the sum over all internal fermions has to vanish
\begin{eqnarray}
 U(1)^3_i &:&  \sum_n {Y^i_n}^3 = 0 \thickspace , \\
 SU(N)^3_i &:&  \sum_n \mbox{Tr}(T^i_n T^i_n T^i_n) = 0 \thickspace .
 \end{eqnarray}
We label the different gauge groups with the indices \(i,j,k\). \(Y^i_n\) is the charge of particle \(n\) under the abelian gauge group \(i\) while \(T^i_n\) is the generator with respect to a non-abelian gauge group.\\
Combinations of two different gauge groups are possible, if one group is an \(U(1)\). Hence, another condition for the absence of anomalies is
\begin{equation}
 U(1)_i\times SU(N)^2_j  :  \sum_n Y^i_n\, \mbox{Tr}(T^j_n T^j_n) = 0  \thickspace .
\end{equation}
If more than one \(U(1)\) gauge group are present, anomalies can be generated by  two or three different \(U(1)\) gauge bosons as external fields, too. Therefore, it has to be checked, that 
\begin{eqnarray}
 U(1)_i\times U(1)_j^2 &:& \thinspace  \sum_n  Y^i_n {Y^j_n}^2 = 0 \thickspace , \\
 U(1)_i\times U(1)_j\times U(1)_k &:& \thinspace  \sum_n  Y^i_n Y^j_n  Y^k_n= 0
\end{eqnarray}
holds. In addition, it has to be checked that there is an even number of \(SU(2)\) doublets. This is the necessary for a model in order to be free of the Witten anomaly \cite{Witten:1982fp}. If one condition is not fulfilled, a warning is given by \SARAH but the model can be evaluated anyway.

\chapter{Conventions and generic expressions}
\section{Renormalization group equations}
\label{app:RGE_con}
\subsection{Generic form of $\beta$ functions}
We summarize in this section the used equations for the calculation of the one-
and two-loop RGEs in \SARAH. These equations are extensively discussed in
literature, see e.g. \cite{Martin:1993zk,Yamada:1993ga, Jack:1998iy,
Jack:1994rk, Jones:1984cx, West:1984dg, Jack:1997eh, Yamada:1993uh,
Yamada:1993ga, Yamada:1994id}.  \\
For a  general $N=1$ supersymmetric gauge theory with superpotential  
\begin{equation}
 W (\phi) = L_i \phi_i + \frac{1}{2}{\mu}^{ij}\phi_i\phi_j + \frac{1}{6}Y^{ijk}
\phi_i\phi_j\phi_k \thickspace ,
\end{equation}
the  soft SUSY-breaking scalar terms are given by
\begin{equation}
V_{\hbox{soft}} = \left(S^i \phi_i +   \frac{1}{2}b^{ij}\phi_i\phi_j
+ \frac{1}{6}h^{ijk}\phi_i\phi_j\phi_k +\hbox{c.c.}\right)
+(m^2)^i{}_j\phi_i\phi_j^* + \frac{1}{2} M_\lambda \lambda_a \lambda_a
\thickspace.
\end{equation}
The anomalous dimensions are given by 
\begin{align}
 \gamma_i^{(1)j} = & \frac{1}{2} Y_{ipq} Y^{jpq} - 2 \delta_i^j g^2 C(i)
\thickspace ,\\
 \gamma_i^{(2)j}  = &  -\frac{1}{2} Y_{imn} Y^{npq} Y_{pqr} Y^{mrj} + g^2
Y_{ipq} Y^{jpq} [2C(p)- C(i)] \nonumber \\
 & \; \;  + 2 \delta_i^j g^4 [ C(i) S(R)+ 2 C(i)^2 - 3 C(G) C(i)] \thickspace ,
\end{align}
and the \(\beta\)-functions for the gauge couplings are given by
\begin{align}
 \beta_g^{(1)}  =  & g^3 \left[S(R) - 3 C(G) \right] \thickspace , \\
 \beta_g^{(2)}  =  & g^5 \left\{ - 6[C(G)]^2 + 2 C(G) S(R) + 4 S(R) C(R)
\right\}
    - g^3 Y^{ijk} Y_{ijk}C(k)/D(G) \thickspace .
\end{align}
Here, \(C(i)\) is the quadratic Casimir for a specific superfield and
$C(R),C(G)$ are the quadratic Casimirs for the matter and adjoint 
representations, respectively. \(D(G)\) is the dimension of the adjoint
representation.  \\
The $\beta$-functions for the superpotential parameters can be obtained by using
superfield technique. The obtained expressions are 
\begin{eqnarray}
 \beta_W^{ijkl} &=& W^{ijkp} \left [
 \frac{1}{16\pi^2}\gamma_p^{(1)l} +
 \frac{1}{(16\pi^2)^2}  \gamma_p^{(2)l} \right ]
+ (l \leftrightarrow i) + (l\leftrightarrow j) + (l\leftrightarrow k)
\thickspace , \\
 \beta_Y^{ijk} &= & Y^{ijp} \left [
 \frac{1}{16\pi^2}\gamma_p^{(1)k} +
 \frac{1}{(16\pi^2)^2}  \gamma_p^{(2)k} \right ]
+ (k \leftrightarrow i) + (k\leftrightarrow j) \thickspace , \\
 \beta_{\mu}^{ij} &= & \mu^{ip} \left [
 \frac{1}{16\pi^2}\gamma_p^{(1)j} +
 \frac{1}{(16\pi^2)^2} \gamma_p^{(2)j} \right ]
+ (j \leftrightarrow i) \thickspace , \\
\beta_L^i & = & L^{p} \left [
 \frac{1}{16\pi^2}\gamma_p^{(1)i} +
 \frac{1}{(16\pi^2)^2}  \gamma_p^{(2)i} \right ] \thickspace .
\end{eqnarray}

The expressions for trilinear, soft-breaking terms are
\begin{align}
\frac{d}{dt} h^{ijk}  =  & \frac{1}{16\pi^2} \left [\beta^{(1)}_h\right ]^{ijk}
+  \frac{1}{(16\pi^2)^2} \left [\beta^{(2)}_h\right ]^{ijk} \thickspace ,
\end{align}
with
\begin{align}
\left [\beta^{(1)}_h\right ]^{ijk}   =
 & \frac{1}{2} h^{ijl} Y_{lmn} Y^{mnk}
+ Y^{ijl} Y_{lmn} h^{mnk} - 2 \left (h^{ijk} - 2 M Y^{ijk}  \right )
g^2  C(k) \nonumber \\
& + (k \leftrightarrow i) + (k \leftrightarrow j) \thickspace , \\
\left [\beta^{(2)}_h\right ]^{ijk}   =
 &
-\frac{1}{2} h^{ijl} Y_{lmn} Y^{npq} Y_{pqr} Y^{mrk} \nonumber \\
& - Y^{ijl} Y_{lmn} Y^{npq} Y_{pqr} h^{mrk}
- Y^{ijl} Y_{lmn} h^{npq} Y_{pqr} Y^{mrk} \nonumber \\
& + \left ( h^{ijl} Y_{lpq} Y^{pqk} +  2 Y^{ijl} Y_{lpq} h^{pqk}
- 2 M Y^{ijl} Y_{lpq} Y^{pqk} \right ) g^2\left[ 2 C(p) - C(k) \right ] 
\nonumber \\
& + \left (2h^{ijk} - 8 M Y^{ijk} \right )
g^4 \left [  C(k)S(R)+ 2 C(k)^2  - 3 C(G)C(k)\right ]\nonumber  \\
&+ (k \leftrightarrow i) + (k \leftrightarrow j)   \thickspace .
\end{align}
For the bilinear soft-breaking parameters, the expressions read
\begin{align}
\frac{d}{dt} b^{ij}  =  &
 \frac{1}{16\pi^2}\left [\beta^{(1)}_b \right ]^{ij} +
 \frac{1}{(16\pi^2)^2} \left [\beta^{(2)}_b \right ]^{ij} \thickspace , 
\end{align}
with
\begin{align}
\left [\beta^{(1)}_b \right ]^{ij}   = 
&\frac{1}{2} b^{il} Y_{lmn} Y^{mnj} +\frac{1}{2}Y^{ijl} Y_{lmn} b^{mn}
+ \mu^{il} Y_{lmn} h^{mnj}
- 2 \left (b^{ij} - 2 M \mu^{ij} \right )g^2 C(i)  \nonumber \\ 
& + (i \leftrightarrow j) \thickspace , \\
\left [\beta^{(2)}_b \right ]^{ij}   = 
& -\frac{1}{2} b^{il} Y_{lmn} Y^{pqn} Y_{pqr} Y^{mrj}
-\frac{1}{2} Y^{ijl} Y_{lmn} b^{mr} Y_{pqr} Y^{pqn} \nonumber \\
&-\frac{1}{2} Y^{ijl} Y_{lmn} \mu^{mr} Y_{pqr} h^{pqn}
- \mu^{il} Y_{lmn} h^{npq} Y_{pqr}  Y^{mrj} \nonumber \\
& - \mu^{il} Y_{lmn} Y^{npq}  Y_{pqr} h^{mrj}
+ 2 Y^{ijl} Y_{lpq} \left ( b^{pq} - \mu^{pq} M \right ) g^2 C(p)
 \nonumber \\
& + \left ( b^{il} Y_{lpq} Y^{pqj} + 2 \mu^{il} Y_{lpq} h^{pqj}
- 2 \mu^{il} Y_{lpq} Y^{pqj} M \right )
g^2 \left[ 2 C(p) - C(i) \right ]  \nonumber \\
& +  \left ( 2 b^{ij} - 8 \mu^{ij} M\right )
g^4 \left [ C(i)S(R)+ 2 C(i)^2- 3 C(G)C(i)   \right ] \nonumber \\
&+ (i \leftrightarrow j)  \thickspace , 
\end{align}
Finally, the RGEs for the linear soft-breaking parameters are
\begin{align}
\frac{d}{dt} S^{i}  =  &
 \frac{1}{16\pi^2}\left [\beta^{(1)}_S + \beta^{(1)}_{SD} \right ]^{i} +
 \frac{1}{(16\pi^2)^2} \left [\beta^{(2)}_S + \beta^{(2)}_{SD} \right ]^{i} \thickspace ,
\end{align}
with
\begin{align}
\left [\beta^{(1)}_S \right]^i  =&
\frac{1}{2}Y^{iln}Y_{pln}S^{p}
+L^{p}Y_{pln}h^{iln}
+\mu^{ik}Y_{kln}B^{ln}+2Y^{ikp}(m^2)_{p}^l\mu_{kl}
+h^{ikl}B_{kl} \thickspace ,\\
\left [\beta^{(2)}_S \right]^i  = &
2g^2C(l)Y^{ikl}Y_{pkl}S^{p} -\frac{1}{2}Y^{ikq}Y_{qst}Y^{lst}Y_{pkl}S^{p}
-4g^2C(l)(Y^{ikl}M-h^{ikl}) Y_{pkl}L^{p}  \nonumber \\
& -\big[Y^{ikq}Y_{qst}h^{lst}Y_{pkl} +h^{ikq}Y_{qst}Y^{lst}Y_{pkl}\big]L^{p}
-4g^2C(l)Y_{jnl} (\mu^{nl}M-B^{nl})\mu^{ij} 
\nonumber \\
& -\big[Y_{jnq}h^{qst}Y_{lst}\mu^{nl} +Y_{jnq}Y^{qst}Y_{lst}B^{nl}\big]\mu^{ij}
+4g^2C(l)(2Y^{ikl}\mu_{kl}|M|^2-Y^{ikl}B_{kl}M
\nonumber \\
&-h^{ikl}\mu_{kl} M^* +h^{ikl}B_{kl}+Y^{ipl}(m^2)_{p}^k\mu_{kl}
+Y^{ikp}(m^2)_{p}^l\mu_{kl}) 
\nonumber \\
 &-\Big[Y^{ikq}Y_{qst}h^{lst}B_{kl}+h^{ikq}Y_{qst}Y^{lst}B_{kl}
+Y^{ikq}h_{qst}h^{lst}\mu_{kl}  +h^{ikq}h_{qst}Y^{lst}\mu_{kl} \nonumber \\
\nonumber &
+Y^{ipq}(m^2)_{p}^kY_{qst}Y^{lst}\mu_{kl}
+Y^{ikq}Y_{qst}Y^{pst}(m^2)_{p}^l\mu_{kl}
+Y^{ikp}(m^2)_{p}^qY_{qst}Y^{lst}\mu_{kl}
\\ & +2Y^{ikq}Y_{qsp}(m^2)_t^{p}Y^{lst}\mu_{kl} \Big] \thickspace .
\end{align}
as well as additional contributions in the presence of Dirac gaugino mass terms \cite{Goodsell:2012fm}.
The new terms are  
\begin{align}
\beta^{(1)}_{SD} =&2 \sqrt{2} g_Y m_D^{aY} \mbox{Tr}(\mathcal{Y} m^2)  + \big[(m_D^2)_{ef} (A^{aef}  + M Y^{aef}) + Y_{efk} \mu^{ka} (m_D^2)^{ef} \big] \\
\beta^{(2)}_{SD} =&2 \sqrt{2} g_Y m_D^{aY} \mbox{Tr}( \mathcal{Y} m^2 (4 g^2 C_2 - Y_2)) + \nonumber \\
& + 4 (\beta_{m_D}^{(1)}/m_D)^f_g \big[(m_D^2)_{ef} (A^{aeg}  + M Y^{aeg}) + Y_{efk} \mu^{ka} (m_D^2)^{eg} \big] 
\end{align}
with
\begin{equation}
(\beta_{m_D}^{(1)}/m_D)^f_g = \frac{1}{2} (Y_2)_g^f + g^2 (S_2 - 5C_{2} (G)) \delta^f_g 
\end{equation}

With this results, the list of the \(\beta\)-functions for all couplings is
complete. Now, we turn to  the RGEs for the gaugino masses, squared masses of
scalars and vacuum expectation values. The result for the gaugino masses is 
\begin{align}
 \frac{d}{dt} M = & \frac{1}{16 \pi^2} \beta_M^{(1)}
+ \frac{1}{(16 \pi^2)^2 } \beta_M^{(2)} \thickspace ,
\end{align}
with
\begin{align}
\beta_M^{(1)} = & g^2 \left[ 2 S(R) - 6 C(G) \right] M \thickspace ,
\\
\beta_M^{(2)}
= & g^4\left\{ -24[C(G)]^2 + 8 C(G) S(R) + 16  S(R)C(R)\right\} M\cr
 &\hbox{\hskip 110pt} + 2 g^2 \left [h^{ijk}  - M Y^{ijk}\right] Y_{ijk}
C(k)/D(G) \thickspace .
\end{align}
The results for the Dirac masses can be expressed in the short from \cite{Goodsell:2012fm}
\begin{align}
\beta_{m_D^{iA}} =& \gamma^i_j m_D^{jA} + \frac{\beta_g}{g}  m_D^{iA}
\end{align}
using the anomalous dimension of the chiral superfield in the adjoint representation and 
the $\beta$ function of the corresponding gauge coupling. \\
The one- and two-loop RGEs for the scalar mass parameters read
\begin{align}
\frac{d}{dt} (m^2)_{i}^j  =  &
 \frac{1}{16\pi^2} \left [\beta^{(1)}_{m^2} \right ]_i^j +
 \frac{1}{(16\pi^2)^2} \left [\beta^{(2)}_{m^2} \right ]_i^j \thickspace , \\
\end{align}
with
\begin{align}
\left [\beta^{(1)}_{m^2} \right ]_i^j  =  &
 \frac{1}{2} Y_{ipq} Y^{pqn} {(m^2)}_n^j
+ \frac{1}{2} Y^{jpq} Y_{pqn} {(m^2)}_i^n + 2 Y_{ipq} Y^{jpr} {(m^2)}_r^q
\nonumber \\
& + h_{ipq} h^{jpq}  - 8 \delta_i^j M M^\dagger g^2 C(i) +
 2 g^2 {\bf t}^{Aj}_i {\rm Tr} [ {\bf t}^A m^2 ] \thickspace , \\
\left [\beta^{(2)}_{m^2} \right ]_i^j  =  &
 -\frac{1}{2} {(m^2)}_i^l Y_{lmn} Y^{mrj} Y_{pqr} Y^{pqn}
 -\frac{1}{2} {(m^2)}^j_l Y^{lmn} Y_{mri} Y^{pqr} Y_{pqn} \nonumber \\
& - Y_{ilm} Y^{jnm} {(m^2)}_r^l Y_{npq} Y^{rpq}
 - Y_{ilm} Y^{jnm} {(m^2)}_n^r Y_{rpq} Y^{lpq} \nonumber  \\
& - Y_{ilm} Y^{jnr} {(m^2)}_n^l Y_{pqr} Y^{pqm}
- 2 Y_{ilm} Y^{jln}  Y_{npq} Y^{mpr} {(m^2)}_r^q \nonumber \\
& - Y_{ilm} Y^{jln} h_{npq} h^{mpq} - h_{ilm} h^{jln} Y_{npq} Y^{mpq} 
\nonumber\\
& - h_{ilm} Y^{jln} Y_{npq} h^{mpq} - Y_{ilm} h^{jln} h_{npq} Y^{mpq}
\nonumber\\
& + \biggl [{(m^2)}_i^l Y_{lpq} Y^{jpq}
+ Y_{ipq} Y^{lpq} {(m^2)}_l^j + 4 Y_{ipq} Y^{jpl} {(m^2)}_l^q
+  2 h_{ipq} h^{jpq}
\nonumber \\
&  - 2 h_{ipq} Y^{jpq} M -2 Y_{ipq} h^{jpq} M^\dagger
+ 4Y_{ipq} Y^{jpq} M M^\dagger
\biggr ]
g^2 \left [C(p) + C(q)- C(i) \right ] \nonumber \\
& -2 g^2 {\bf t}^{Aj}_i ({\bf t}^A m^2)_r^l Y_{lpq} Y^{rpq}
+ 8 g^4 {\bf t}^{Aj}_i {\rm Tr} [ {\bf t}^A C(r) m^2 ]  \nonumber \\
& + \delta_i^j g^4 M M^\dagger \left [
24C(i) S(R) + 48 C(i)^2 - 72 C(G) C(i) \right ]
\nonumber \\ 
& + 8 \delta_i^j g^4 C(i) ( {\rm Tr} [S(r) m^2] - C(G) M M^\dagger ) \thickspace
.
\end{align}
The RGEs for a VEV \(v^i\) is proportional to the anomalous dimension of the
chiral superfield whose scalar component receives the VEV
\begin{equation}
 \frac{d}{dt }v^i =  v^{p} \left [
 \frac{1}{16\pi^2}\gamma_p^{(1)i} +
 \frac{1}{(16\pi^2)^2}  \gamma_p^{(2)i} \right ]
\end{equation}

\subsection{Direct product of gauge groups}
To calculate the RGEs for a direct product of gauge group, the following
substitution rules are needed \cite{Martin:1993zk}. Note, these replacements are
not sufficient in the case of several $U(1)$ gauge groups as discussed in app.~\ref{sec:U1RGE}.
For the \(\beta\) functions of gauge couplings and gauginos the rules are
\begin{align}
 g^3 C(G)  &\rightarrow g_a^3 C(G_a) \thickspace ,\\
 g^3 S(R)  &\rightarrow  g_a^3 S_a(R)  \thickspace ,\\
g^5 C(G)^2  &\rightarrow  g_a^5 C(G_a)^2 \thickspace ,\\
g^5 C(G) S(R)  &\rightarrow  g_a^5 C(G_a) S_a(R) \thickspace ,\\
g^5 S(R) C(R)  &\rightarrow  \sum_b g_a^3 g_b^2 S_a(R) C_b(R) \thickspace ,\\
16 g^4 S(R) C(R) M  &\rightarrow  8 \sum_b g_a^2 g_b^2 S_a(R) C_b(R) (M_a + M_b)
\thickspace ,\\
g^3 C(k)/d(G)  &\rightarrow  g_a^3 C_a(k) d(G_a) \thickspace .
\end{align}
For the other \(\beta\) functions, we need
\begin{align}
 g^2 C(r) &\rightarrow \sum_a g_a^2 C_a(r) \thickspace , \\
 g^4 C(r) S(R) &\rightarrow \sum_a g_a^4 C_a(r) S_a(R) \thickspace , \\
 g^4 C(r) C(G) &\rightarrow \sum_a g_a^4 C_a(r) C(G_a)\thickspace , \\
 g^4 C(r)^2 &\rightarrow \sum_a \sum_b g_a^2 g_b^2 C_a(r) C_b(r)\thickspace , \\
 48 g^4 M M^\dagger C(i)^2 &\rightarrow \sum_a \sum_b g_a^2 g_b^2 C_a(i) C_b(i)
(32 M_a M_a^\dagger + 8 M_a M_b^\dagger + 8 M_b M_a^\dagger) \thickspace ,\\
g^2 t_i^{A j} Tr(t^A m^2) &\rightarrow \sum_a g_a^2 (t_a^A)^j_i Tr(t_a^A
m^2)\thickspace , \\
g^2 t_i^{A j} (t^A m^2)^l_r Y_{lpq} Y^{rpq} &\rightarrow \sum_a g_a^2
(t_a^A)^j_i (t_a^A m^2)^l_r Y_{lpq} Y^{rpq}\thickspace , \\
g^4 t_i^{A j} Tr(t^4 C(r) m^2) &\rightarrow \sum_a \sum_b g_a^2 g_b^2
(t_a^A)^j_i Tr(t_a^A C_b(r) m^2)\thickspace , \\
g^4 C(i) Tr(S(r) m^2) &\rightarrow \sum_a g_a^4 C_a(i) Tr(S_a(r) m^2)
\thickspace .
\end{align}

\subsection{Several $U(1)$ gauge groups}
\label{sec:U1RGE}
In the case of several Abelian gauge groups, it is necessary to use a
generalized form of the some terms given in the last section. The complete set of
two-loop RGEs has been published in \cite{Fonseca:2011vn}, but we use here a
notation closer to \cite{Jack:2000jr}. First, we define a generalized charge
\(\mathscr{Y}\)
for each chiral superfield \(r\)
\begin{equation}
\mathscr{Y}(r)_\alpha = \sum_\beta Y(r)_\beta g_{\beta \alpha} .
\end{equation}
\(Y(r)_\beta\) is the charge of the chiral superfield \(r\) with respect to the
gauge group \(\beta\). 
The sum runs over all Abelian gauge groups. Here and in the following we use
Greek letters for $U(1)$ gauge groups and Latin ones for $SU(N)$. Using that
definition, we can define a generalized Dynkin index for Abelian gauge groups as
\begin{equation}
Q_{\alpha \beta} = \sum_r  \mathscr{Y}(r)_\alpha \mathscr{Y}(r)_\beta .
\end{equation}
In addition, we define
\begin{eqnarray}
 \tilde{t}^{\alpha i}_j &=& \delta_{i j} \mathscr{Y}(i)_\alpha \hspace{1cm}
\mbox{for} \hspace{0.5cm} U(1) \thickspace , \\
 \tilde{t}^{a i}_j &=& t^{a i}_j \hspace{1cm} \mbox{for} \hspace{0.5cm} SU(N)
\thickspace . 
\end{eqnarray}
as well as
\begin{eqnarray}
\tilde{C}_\alpha(r) &=& \Y(r)_\alpha \Y(r)_\alpha    \hspace{1cm} \mbox{for}
\hspace{0.5cm} U(1) \thickspace , \\
\tilde{C}_a(r) &=& g_a^2 C_a(r) \hspace{1cm} \mbox{for} \hspace{0.5cm} SU(N) 
\thickspace .
\end{eqnarray}
Now, to write the RGEs of the last section in the case of several \(U(1)\) gauge
groups, we have to perform the following replacements for gauge couplings and
gauginos. Note, we give only the parts involving Abelian gauge groups. The expressions
without Abelian gauge groups keep unchanged. 
\begin{align}
g^3 S(R) &\rightarrow \sum_\gamma g_{\alpha \gamma} Q_{\gamma \beta} \thickspace
,  \\
g^5 S(R) C(R) &\rightarrow \sum_\gamma g_{\alpha \gamma} \sum_r \sum_a
\tilde{C}_a(r) \Y(r)_\gamma \Y(r)_\beta \hspace{1cm} \mbox{for} \hspace{0.5cm}
U(1) \thickspace ,\\
g^3 C(k) &\rightarrow \sum_\gamma g_{\alpha \gamma} \Y(k)_\gamma \Y(k)_\beta
\thickspace ,\\
g^2 S(R) M &\rightarrow \sum_\gamma \frac{1}{2} \left(M_{\alpha \gamma}
Q_{\gamma \alpha} + Q_{\alpha \gamma} M_{\gamma \beta } \right) \thickspace ,\\
16 g^4 S(R) C(R) M &\rightarrow 8 \left(\sum_r \sum_\gamma \frac{1}{2}
\left( M_{\alpha \gamma} \Y(r)_\gamma \Y(r)_\beta + M_{\gamma \beta }
\Y(r)_\gamma \Y(r)_\alpha \right) \bar{C}(r) + \sum_r \Y(r)_\alpha \Y(r)_\beta
\bar{C}^M(r) \right) \thickspace ,\\
g^2 C(k) &\rightarrow \Y(k)_\alpha \Y(k)_\beta \thickspace ,\\
g^2 C(k) M &\rightarrow \sum_\gamma \frac{1}{2} \left( M_{\alpha \gamma}
\Y(k)_\gamma \Y(k)_\beta + \Y(k)_\gamma \Y(k)_\alpha M_{\gamma \beta }
\right) \thickspace .
\end{align}
For all other couplings we need
{\allowdisplaybreaks
\begin{align}
g^2 C(r) &\rightarrow \sum_\alpha \mathscr{Y}(r)_\alpha \mathscr{Y}(r)_\alpha +
\sum_a g_a^2 C_a(r) \equiv \bar{C}(r) \thickspace , \\
M g^2 C(r) &\rightarrow \sum_\alpha \sum_\beta M_{\alpha \beta} \Y(r)_\alpha
\Y(r)_\beta + \sum_a g_a^2 M_a C_a(r) \equiv \bar{C}^M(r) \thickspace ,  \\
M^* g^2 C(r) &\rightarrow \sum_\alpha \sum_\beta M^*_{\beta \alpha} \Y(r)_\alpha
\Y(r)_\beta + \sum_a g_a^2 M_a C_a(r) \thickspace , \\
M M^* g^2 C(r) &\rightarrow \sum_\alpha \sum_\beta (M M^*)_{\alpha \beta}
\Y(r)_\alpha \Y(r)_\beta + \sum_a g_a^2 M_a M_a^* C_a(r) \thickspace ,  \\
g^4 C(r) S(R) &\rightarrow \sum_\alpha \sum_\beta \Y(r)_\alpha Q_{\alpha \beta}
\Y(r)_\beta + \sum_a g_a^4 C_a(r) S_a(R)  \thickspace ,\\
M g^4 C(r) S(R) &\rightarrow \sum_\alpha \sum_\beta \sum_\gamma \Y(r)_\alpha
Q_{\alpha \beta} M_{\beta \gamma} \Y(r)_\gamma + \sum_a g_a^4 M_a C_a(r) S_a(R)
\thickspace , \\
g^4 C^2(r) &\rightarrow \bar{C}(r)^2 \thickspace ,  \\
M g^4 C^2(r) &\rightarrow \bar{C}^M(r) \bar{C}(r) \thickspace ,\\
g^2 t_i^{A j} \mbox{Tr}(t^4 m^2) &\rightarrow \tilde{t}^{A j}_i \sum_r (m^2)^r_r
\tilde{t}^{A r}_r \thickspace ,\\
g^2 t_i^{A j} (t^A m^2)^l_r &\rightarrow \tilde{t}^{A j}_i \sum_q \tilde{t}^{A
l}_q (m^2)^q_r  \thickspace ,\\
g^4 t_i^{A j} \mbox{Tr}(t^4 C(r) m^2) &\rightarrow   \tilde{t}^{A i}_j \sum_r
\sum_\beta \tilde{t}^{\alpha r}_r \Y(r)_\beta \Y(r)_\beta (m^2)^r_r \thickspace
,\\
g^4 C(i) \mbox{Tr}(S(r) m^2) &\rightarrow \sum_\alpha \sum_\beta \Y(i)_\alpha
\Y(i)_\beta  \sum_r \Y(r)_\alpha \Y(r)_\beta (m^2)^r_r + \sum_a g_a^4 C_a(i)
\sum_r S_a(r) (m^2)^r_r   \thickspace ,\\
\nonumber 24 g^4 M M^* C(r) S(R) &\rightarrow 24 \sum_a g_a^4 M_a M_a^2 C_a(r)
S_a(R) + 8\Big( \sum_{\alpha,\beta,\gamma,\delta} \Y(r)_\alpha M_{\alpha\beta}
M^*_{\delta\beta} Q_{\delta \gamma} \Y(r)_\gamma + \\
&\hspace{0.5cm} + \sum_{\alpha,\beta,\gamma,\delta} \Y(r)_\alpha M^*_{\beta
\alpha} M_{\beta \delta} Q_{\delta \gamma} \Y(r)_\gamma +
\sum_{\alpha,\beta,\gamma,\delta} \Y(r)_\alpha M_{\alpha\beta} Q_{\beta \gamma}
M^*_{\delta \gamma} \Y(r)_\alpha \Big) \thickspace ,\\
\nonumber 48 g^4 M M^* C(r)^2 &\rightarrow \sum_a \sum_b g_a^2 g_b^2 C_a(r)
C_b(r) (32 M_a M_a^* + 8 M_a M_b^* + 8 M_b M_a^*) \\
& \nonumber \hspace{0.5cm} + \sum_a g_a^2 C_a(r) \Big( 32 M_a M_a^* \sum_\alpha
\Y(r)_\alpha \Y(r)_\alpha + 16 M_a \sum_{\alpha,\beta} \Y(r)_\alpha M^*_{\beta
\alpha} \Y(r)_\beta + \\
& \nonumber \hspace{1.5cm}  16 M_a^* \sum_{\alpha,\beta} \Y(r)_\alpha M_{\alpha
\beta} \Y(r)_\beta + 32 \sum_{\alpha,\beta,\gamma} \Y(r)_\alpha M_{\alpha\beta}
M^*_{\gamma \beta} \Y(r)_\gamma \Big) \\
&  \nonumber \hspace{0.5cm} + 32 \sum_{\alpha,\beta,\gamma} \Y(r)_\alpha
M_{\alpha\beta} M^*_{\gamma\beta} \Y(r)_\gamma \sum_{\alpha} \Y(r)_\alpha
\Y(r)_\alpha + \\
& \hspace{1.5cm} 16 \sum_{\alpha,\beta} \Y(r)_\alpha M_{\alpha\beta} \Y(r)_\beta
\sum_{\alpha,\beta} \Y(r)_\alpha M^*_{\beta\alpha} \Y(r)_\beta \thickspace .
\end{align}
}

\section{One-loop amplitudes for one- and two-point functions}
\label{sec:Integrals}
We used for the calculation of the one-loop self energies and the one-loop
corrections to the tadpoles in \(\DR\)-scheme the scalar functions defined in
\cite{Pierce:1996zz}. The basic integrals are
\begin{eqnarray}
A_0(m) &=& 16\pi^2Q^{4-n}\int{\frac{d^nq}{ i\,(2\pi)^n}}{\frac{1}{
q^2-m^2+i\varepsilon}} \thickspace ,\\
B_0(p, m_1, m_2) &=& 16\pi^2Q^{4-n}\int{\frac{d^nq}{ i\,(2\pi)^n}}
{\frac{1}{\biggl[q^2-m^2_1+i\varepsilon\biggr]\biggl[
(q-p)^2-m_2^2+i\varepsilon\biggr]}} \thickspace ,
\label{B0 def}
\end{eqnarray}
with the renormalization scale \(Q\). The integrals are regularized by
integrating in $n=4-2\epsilon$ dimensions. The result for \(A_0\) is
\begin{equation}
A_0(m)\ =\ m^2\left({\frac{1}{\hat\epsilon}} + 1 -
\ln{\frac{m^2}{Q^2}}\right)~,\label{A}
\end{equation}
where $1/\hat\epsilon =1/\epsilon-\gamma_E+\ln 4\pi$. The function $B_0$  has
the analytic expression
\begin{equation}
 B_0(p, m_1, m_2) \ =\ {\frac{1}{\hat\epsilon}} -
\ln\left(\frac{p^2}{Q^2}\right) - f_B(x_+) - f_B(x_-)~,
\end{equation}
with
\begin{equation}
 x_{\pm}\ =\ \frac{s \pm \sqrt{s^2 - 4p^2(m_1^2-i\varepsilon)}}{2p^2}~,
\qquad f_B(x) \ =\ \ln(1-x) - x\ln(1-x^{-1})-1~,
\end{equation}
and $s=p^2-m_2^2+m_1^2$. All the other, necessary functions can be expressed by
$A_0$ and $B_0$. For instance,
\begin{equation}
 B_1(p, m_1,m_2) \ =\ {\frac{1}{2p^2}}\biggl[ A_0(m_2) - A_0(m_1) + (p^2
+m_1^2 -m_2^2) B_0(p, m_1, m_2)\biggr]~,
\end{equation}
and
\begin{eqnarray}
B_{22}(p, m_1,m_2) &=& \frac{1}{6}\ \Bigg\{\,
\frac{1}{2}\biggl(A_0(m_1)+A_0(m_2)\biggr)
+\left(m_1^2+m_2^2-\frac{1}{2}p^2\right)B_0(p,m_1,m_2)\nonumber \\ &&+
\frac{m_2^2-m_1^2}{2p^2}\ \biggl[\,A_0(m_2)-A_0(m_1)-(m_2^2-m_1^2)
B_0(p,m_1,m_2)\,\biggr] \nonumber\\ && +  m_1^2 + m_2^2
-\frac{1}{3}p^2\,\Bigg\}~.
\end{eqnarray}
Furthermore, for the vector boson self-energies it is useful to define
\begin{align}
F_0(p,m_1,m_2) =& A_0(m_1)-2A_0(m_2)- (2p^2+2m^2_1-m^2_2)B_0(p,m_1,m_2)
\ , \\ 
G_0(p,m_1,m_2) =&
(p^2-m_1^2-m_2^2)B_0(p,m_1,m_2)-A_0(m_1)-A_0(m_2)\ ,\\
H_0 (p,m_1,m_2) =& 4B_{22}(p,m_1,m_2) + G(p,m_1,m_2)\ ,\\
\tilde{B}_{22}(p,m_1,m_2) =& B_{22}(p,m_1,m_2) - \frac{1}{4}A_0(m_1) -
\frac{1}{4}A_0(m_2)
\end{align}
In all calculations, specific coefficient are involved:
\begin{itemize}
 \item \(c_S\) is the symmetry factor: if the particles in the loop are
indistinguishable, the weight of the contribution is only half of the weight in
the case of distinguishable particles. If two different charge flows are
possible in the loop, the weight of the diagram is doubled.
 \item \(c_C\) is a charge factor: for corrections due to vector bosons in the
adjoint representation this is the Casimir of the corresponding group. For
corrections due to matter fields this can be, for instance, a color factor for
quarks/squarks. For corrections of vector bosons in the adjoint representation
this is normally the Dynkin index of the gauge group.
 \item  \(c_R\) is 2 for real fields and Majorana fermions in the loop and 1
otherwise. \\
\end{itemize}
We use in the following \(\Gamma\) for non-chiral interactions and
\(\Gamma_L\)/\(\Gamma_R\) for chiral interactions. If two vertices are involved,
the interaction of the incoming particle has an upper index 1 and for the
outgoing field an upper index 2 is used. 
\subsection{One-loop tadpoles}
\begin{enumerate}
\item Fermion loop (generic name in \SARAH: \verb"FFS"):
\begin{equation}
T = 8 c_S c_C m_F \Gamma A_0(m_F^2) 
\end{equation}
\item Scalar loop (generic name in \SARAH: \verb"SSS"):
\begin{equation}
T = - 2 c_S c_C \Gamma A_0(m_S^2) 
\end{equation}
\item Vector boson loop (generic name in \SARAH: \verb"SVV"):
\begin{equation}
T = 6 c_S c_C \Gamma A_0(m_V^2) 
\end{equation}
\end{enumerate}

\subsection{One-loop self-energies}
\paragraph{Corrections to fermion}
\begin{enumerate}
\item Fermion-scalar loop (generic name in \SARAH: \verb"FFS"):
\begin{eqnarray*}
\Sigma^S(p^2) &=& m_F c_S c_C c_R \Gamma^1_R \Gamma^{2,*}_L B_0(p^2,m_F^2,m_S^2)
\\
\Sigma^R(p^2) &=& - c_S c_C c_R \frac{1}{2} \Gamma^1_R \Gamma^{2,*}_R
B_1(p^2,m_F^2,m_S^2) \\
\Sigma^L(p^2) &=& - c_S c_C c_R \frac{1}{2} \Gamma^1_L \Gamma^{2,*}_L
B_1(p^2,m_F^2,m_S^2) 
\end{eqnarray*}
\item Fermion-vector boson loop (generic name in \SARAH: \verb"FFV"):
\begin{eqnarray*}
\Sigma^S(p^2) &=& - 4 c_S c_C c_R m_F \Gamma^1_L \Gamma^{2,*}_R
B_0(p^2,m_F^2,m_S^2) \\
\Sigma^R(p^2) &=& - c_S c_C c_R \Gamma^1_L \Gamma^{2,*}_L B_1(p^2,m_F^2,m_S^2)
\\
\Sigma^L(p^2) &=& - c_S c_C c_R \Gamma^1_R \Gamma^{2,*}_R B_1(p^2,m_F^2,m_S^2) 
\end{eqnarray*}
\end{enumerate}

\paragraph{Corrections to scalar}
\begin{enumerate}
\item Fermion loop (generic name in \SARAH: \verb"FFS"): 
\begin{equation}
\label{eq:SE_FFS}
\Pi(p^2) = c_S c_C c_R \left((\Gamma^1_L \Gamma^{2,*}_L + \Gamma^1_R
\Gamma^{2,*}_R) G_0(p^2,m_F^2,m_S^2) +  (\Gamma^1_L \Gamma^{2,*}_R + \Gamma^1_R
\Gamma^{2,*}_L) B_0(p^2,m_F^2,m_S^2) \right)
\end{equation}
\item Scalar loop (two 3-point interactions, generic name in \SARAH:
\verb"SSS"):
\begin{equation}
\Pi(p^2) = c_S c_C c_R \Gamma^1 \Gamma^{2,*} B_0(p^2,m_F^2,m_S^2) 
\end{equation}
\item Scalar loop (4-point interaction, generic name in \SARAH: \verb"SSSS"):
\begin{equation}
\label{eq:SE_SSSS}
\Pi(p^2) =  - c_S c_C \Gamma A_0(m_S^2) 
\end{equation}
\item Vector boson-scalar loop (generic name in \SARAH: \verb"SSV"):
\begin{equation}
\label{eq:SE_SSV}
\Pi(p^2) = c_S c_C c_R \Gamma^1 \Gamma^{2,*} F_0(p^2,m_F^2,m_S^2) 
\end{equation}
\item Vector boson loop (two 3-point interactions, generic name in \SARAH:
\verb"SVV"):
\begin{equation}
\Pi(p^2) =  c_S c_C c_R \frac{7}{2} \Gamma^1 \Gamma^{2,*} B_0(p^2,m_F^2,m_S^2) 
\end{equation}
\item Vector boson loop (4-point interaction, generic name in \SARAH:
\verb"SSVV"):
\begin{equation}
\Pi(p^2) =   c_S c_C \Gamma A_0(m_V^2) 
\end{equation}
\end{enumerate}

\paragraph{Corrections to vector boson}
\begin{enumerate}
\item Fermion loop (generic name in \SARAH: \verb"FFV"):
\begin{equation}
\Pi^T(p^2) = c_S c_C c_R \left((|\Gamma^1_L|^2+|\Gamma^1_R|^2)
H_0(p^2,m_V^2,m_F^2)+ 4 Re(\Gamma^1_L \Gamma^2_R)B_0(p^2,m_V^2,m_F^2) \right)
\end{equation}
\item Scalar loop (generic name in \SARAH: \verb"SSV"):
\begin{equation}
\Pi^T(p^2) = -4 c_S c_C c_R |\Gamma|^2 B_{22}(p^2,m_{S_1}^2,m_{S_2}^2)
\end{equation}
\item Vector boson loop (generic name in \SARAH: \verb"VVV"):
\begin{equation}
\Pi^T(p^2) =  |\Gamma|^2  c_S c_C c_R \left(-(4 p^2 + m_{V_1}^2 + m_{V_2}^2 )
B_0(p^2,m_{V_1}^2,m_{V_1}^2) - 8 B_{22}(p^2,m_{S_1}^2,m_{S_2}^2) \right)
\end{equation}
\item Vector-Scalar-Loop (generic name in \SARAH: \verb"SVV"):
\begin{equation}
\Pi^T(p^2) =  |\Gamma|^2 c_S c_C c_R B_0(p^2,m_V^2,m_S^2)
\end{equation}
\end{enumerate}
We need here only the diagrams involving three point interactions because the
4-point interactions are related to them due to gauge invariance. 
\subsection{One-loop corrections to masses} 
\label{sec:OneLoopMass}
% The one-loop self-energies can be used to calculate the one-loop masses and mass
% matrices. For the one-loop corrections of scalars, the radiatively corrected
% mass matrix is 
% \begin{equation}
% 	m^{2,S}_{1L}(p^2_i) = m^{2,S}_{T} - \Pi_{S S}(p^2_i) , 
% \end{equation}
% while the one-loop mass of a vector boson \(V\) is given by
% \begin{equation}
%  m^{2,V}_{1L}(Q) =  m^{2,V}_T + \mathrm{Re}\big\{ \Pi^T_{VV}(m^{2,V}_T) \big\}.
% \end{equation}
% According to the conventions of the counter terms of \cite{Pierce:1996zz}, the
% one-loop mass matrices \(M^{\tilde\chi^0}_{1L}\) of Majorana fermions are
% connected to the one-loop self-energies and tree-level mass matrix
% \(M^{\tilde\chi^0}_T\)  by
% \begin{eqnarray}
% M^{\tilde\chi^0}_{1L} (p^2_i) &=& M^{\tilde\chi^0}_T - 
% \frac{1}{2} \bigg[ \Sigma^0_S(p^2_i) + \Sigma^{0,T}_S(p^2_i)
%  + \left(\Sigma^{0,T}_L(p^2_i)+   \Sigma^0_R(p^2_i)\right) M^{\tilde\chi^0}_T
%  \nonumber \\
% && \hspace{16mm}
% + M^{\tilde\chi^0}_T \left(\Sigma^{0,T}_R(p^2_i) +  \Sigma^0_L(p^2_i) \right)
% \bigg] .
% \end{eqnarray}
% In the case of Dirac fermions, the one-loop corrected mass matrix is 
% \begin{eqnarray}
% M^{\tilde\chi^+}_{1L}(p^2_i) =  M^{\tilde\chi^+}_T - \Sigma^+_S(p^2_i)
%  - \Sigma^+_R(p^2_i) M^{\tilde \chi^+}_T - M^{\tilde \chi^+}_T \Sigma^+_L(p^2_i)
% .
The one-loop self-energies can be used to calculate the one-loop masses and mass
matrices.
\begin{enumerate}
 \item {\bf Real scalars}:  for a real scalar $\phi$, the one-loop corrections  are included by calculating
 the real part of the poles of the corresponding propagator matrices \cite{Pierce:1996zz}
\begin{equation}
\mathrm{Det}\left[ p^2_i \mathbf{1} - m^2_{\phi,1L}(p^2) \right] = 0,
\label{eq:propagator}
\end{equation}
where
\begin{equation}
 m^2_{\phi,1L}(p^2) = \tilde{m}^{2}_{\phi,T} -  \Pi_{\phi}(p^2) .
\end{equation}
Equation (\ref{eq:propagator}) has to be solved for each
eigenvalue $p^2=m^2_i$ which can be achieved in an iterative
procedure. This has to be done also for charged scalars as well as the fermions. 
Note, $\tilde{m}^2_T$ is the tree-level mass matrix
but for the parameters fixed by the tadpole equations the one-loop
corrected values $X^{(1)}$ are used.

 \item {\bf Complex scalars}: for a complex scalar $\eta$ field we use at one-loop level
\begin{equation}
	m^{2,\eta}_{1L}(p^2_i) = \tilde{m}^{2,\eta}_{T} - \Pi_{\eta}(p^2_i) ,
\end{equation}
While in case of sfermions $\tilde{m}^{2,\eta}_{T}$ agrees exactly with the tree-level mass matrix, 
for charged Higgs bosons $\mu^{(1)}$ and $B^{(1)}_\mu$ or $m_{H_d}^{(1)}$ and $m_{H_d}^{(1)}$  has 
to be used depending on the set of parameters the tadpole equations are solved for.

 \item {\bf Majorana fermions}: the one-loop mass matrix of a Majorana $\chi$ fermion is related
to the tree-level mass matrix by
\begin{eqnarray}
M^{\chi}_{1L} (p^2_i) &=& M^{\chi}_T - 
\frac{1}{2} \bigg[ \Sigma^0_S(p^2_i) + \Sigma^{0,T}_S(p^2_i)
 + \left(\Sigma^{0,T}_L(p^2_i)+   \Sigma^0_R(p^2_i)\right) M^{\chi}_T
 \nonumber \\
&& \hspace{16mm}
+ M^{\tilde\chi^0}_T \left(\Sigma^{0,T}_R(p^2_i) +  \Sigma^0_L(p^2_i) \right)
 \bigg] ,
\end{eqnarray}
where we have denoted the wave-function corrections by $\Sigma^{0}_R$,
 $\Sigma^{0}_L$ and the direct one-loop contribution to the mass by 
$\Sigma^{0}_S$.
 
\item {\bf Dirac fermions}: for a Dirac fermion $\Psi$ one has to add the self-energies as
\begin{eqnarray}
M^{\Psi}_{1L}(p^2_i) =  M^{\Psi}_T - \Sigma^+_S(p^2_i)
 - \Sigma^+_R(p^2_i) M^{\Psi}_T - M^{\Psi}_T \Sigma^+_L(p^2_i) .
\end{eqnarray}
\end{enumerate}

\chapter{More information about the \SPheno output}
\section{Generated Files for SPheno}
All routines generated by \\SARAH are strongly inspired by the intrinsic routines
of \SPheno to ensure that they interact nicely with all other \SPheno
functions. We give here some details of the different files which are written
by \\SARAH as well as of the main functions for the case that the user wants to
do some changes.

\subsection{{\tt BranchingRatios\_[Model].f90}}
Contains the routines for calculating the branching ratios for all particles.
Checks are done, if three body decays for fermions are necessary or not. 
 %See also
%sec.~\ref{sec:SPheno_decays} for more information. %For the
%couplings to real and virtual vector boson the SM values are used. 

\subsection{{\tt Couplings\_[Model].f90}}
All vertices calculated by \\SARAH are written to this files. Also functions for
calculating different subsets of these couplings are generated:
\begin{itemize}
 \item {\tt AllCouplings}: all couplings for the mass eigenstates (used for the
calculation of decays)
 \item {\tt CouplingsForSMfermion}: couplings involved in the one loop self
energy of SM fermions
 \item {\tt CouplingsForVectorBosons}: couplings involved in the one loop self
energy of SM vector bosons
 \item {\tt CouplingsForTadpoles}: couplings involved in the calculation of the
one loop tadpoles
 \item {\tt CouplingsForLoop}: couplings needed for the one loop self energies
of all particles
\end{itemize}
For all couplings involved in loop calculation, the mixing matrices of external
particles are replaced by the identity matrix.

\subsection{{\tt CouplingsForDecays\_[Model].f90}}
Calculates the running couplings at the mass scale of each decaying particle. The SUSY parameters
are obtained by running the RGEs from the SUSY scale to the desired energy scale. 
The calculation of the SM gauge and Yukawa coupling is different for particles with masses above or
below the SUSY scale. For lighter particles, the SM RGEs are used to run from $M_Z$ to the mass scale, 
for heavier particles the SUSY RGEs are taken to run from $M_{SUSY}$. \\
In addition, the running tree-level masses are calculated. These are used to get 
effective couplings of the Higgs bosons to two gluons or photons. Furthermore, also the couplings
of a light Higgs to a real and virtual vector boson \((W,Z)\) are calculated.

\subsection{{\tt InputOutput\_[Model].f90}}
This file contains the routines for reading the {\tt LesHouches.in.[Model]}
input file and writing the output {\tt SPheno.spc.[Model]}. 
\paragraph*{Input}
 The following blocks are changed in comparison to the standard
LesHouches/\SPheno input, see also sec.~\ref{sec:SPHENOINPUT}.
 \begin{itemize}
 \item {\tt Block MODSEL}
 \item {\tt Block MINPAR}
 \item {\tt Block EXTPAR} 
 \item {\tt Block MODSEL}%: Since the generated \SPheno works just for one
%model, % the flag {\tt 1} is ignored
  \item {\tt Block SPHENOINPUT}%: New flags have been added:
 %\begin{itemize}
 %\item {\tt 50}: If set to {\tt 0}, majorana fermions are not rotated to make
%all % masses positive
% \item {\tt 51}: If set to {\tt 0}, the parameters
% \(Y_u,Y_d,T_u,T_d,m_q^2,m_d^2,m_u^2\) are not rotated in SCKM basis in
% \verb"SPheno.spc"
% \item {\tt 52}: Negative mass squares of particles are ignored and those
%masses % are set back to zero. 
% \end{itemize}
 \end{itemize}

\paragraph*{Output}
Routines for writing the following information to the LesHouches spectrum file
are generated:
\begin{itemize}
\item Parameters at the SUSY scale: \\
As block names for the output, the entries in {\tt LesHouches} in the parameter
file of \\SARAH are used. If this entry is missing, the name for the block is
generated automatically.
\item Masses at the SUSY scale: \\
The PDG of the particles file of \SARAH is used. If this is missing or 0, the
mass is not written in order to exclude unphysical states.
\item Optionally, decay widths and branching ratios for the particle in {\tt
ListDecayParticles} and {\tt ListDecayParticles3B}
\item Optionally, results for low energy constraints 
\item Optionally, effective couplings of the Higgs to SM particles which can be used for instance by \HB
\item Optionally, the GUT values of all parameters
\item Optionally, the loop contributions of all particles to the effective coupling of the Higgs to two photons or gluons
\end{itemize}
In addition, that file contains a routine to write the parameters to a  \WHIZARD
specific output file

\subsection{{\tt LoopCouplings\_[Model].f90}}
Routines for calculating the following couplings:
\begin{itemize}
 \item Corrections to SM gauge couplings due to all particles heavier than
\(M_Z\), i.e. normally SUSY particles and top quark ({\tt AlphaSDR, AlphaEwDR})
 \item Effective couplings of scalar/pseudoscalar Higgs to SM fermions (e.g.
\verb"CouphhtoFe"). 
 \item \verb"DeltaVB": one-loop correction to \(G_F\) calculated from \(\mu
\rightarrow e \sum_{i j} \nu_i \bar{\nu}_j\) 
\end{itemize}

\subsection{{\tt LoopMasses\_[Model].f90}}
This file contains the routines for calculating the one loop contributions to
the tadpoles and the self energies:
\begin{itemize}
 \item {\tt OneLoopMasses}: main routine for calling all other functions
 \item {\tt TadpolesOneLoop}: calculates the one loop tadpoles. 
 \item {\tt Pi1LoopX}: one loop self energy for particle {\tt X} (scalars or
vector bosons (transverse part))
 \item {\tt Sigma1LoopX}: calculates \(\Sigma_L, \Sigma_R,\Sigma_S\) for
fermions
 \item {\tt OneLoopX}: calculates the one loop corrected mass for particle {\tt
X} in an iterative way in order to solve
  \begin{equation}
 \mathrm{Det}\left[ p^2_i \mathbf{1} - m^{2,X}_{1L}(p^2) \right] = 0,
 \end{equation}
for external masses on-shell, i.e. \(p^2 = m^{2,X}_{Tree}\).
\end{itemize}
All these calculation are done in 't Hooft gauge. Normally, the results are
one-loop \DRbar\ masses and mixing matrices. For further calculations, the mixing
matrices corresponding to
an external momentum equal to the heaviest mass eigenstate is taken. However, for SM 
particles the pole masses are used in the decays as well as in the output.

\subsection{{\tt LowEnergy\_[Model].f90}}
Contains the routines for the calculation of precision observables:
\begin{itemize}
 \item $b \to s \gamma$ \cite{Bobeth:2001jm,Baek:2001kh,Huber:2005ig}  ({\tt BToQGamma})
 \item $l_i \to  l_j \gamma$ \cite{Arganda:2005ji}  ({\tt BrLgammaLp})
 \item $l_i \to 3 l_j$ (with $l = (e,\mu,\tau)$),  \cite{Arganda:2005ji} ({\tt BR1LeptonTo3Leptons})
 \item the anomalous magnetic moment of leptons \cite{Ibrahim:1999hh}  ({\tt Gminus2})
 \item electric dipole moments of the charged leptons \cite{Bartl:1999bc,Bartl:2003ju}  ({\tt LeptonEDM})
 \item  $\delta\rho=1-\rho=\frac{\Pi_{WW}(0)}{m_W^2}-\frac{\Pi_{ZZ}(0)}{m_Z^2}$ ($\Pi_{ZZ}$, $\Pi_{WW}$ are the self-energies of the massive vector bosons). \cite{Drees:1990dx}  ({\tt DeltaRho})
 \item $\mu - e$ conversion in nuclei (Al, Ti, Sr, Sb, Au, Pb) based on the
 results of \cite{Arganda:2007jw}  ({\tt BrLLpHadron})
 \item $\tau \to l P^0$ with a pseudoscalar meson $P^0$   ({\tt BrLLpHadron})
 ($\pi^0$, $\eta$, $\eta'$) based on the results of \cite{Arganda:2008jj}
 \item $Z \to l_i l_j$ calculated and implemented by Kilian Nickel (based on generic results obtained by \FeynArts/\FormCalc)  \cite{Dreiner:2012mx}  ({\tt BrZLLp})
 \item $B_{s,d} \to l_i l_j$ calculated and implemented by Kilian Nickel (based on generic results obtained by \FeynArts/\FormCalc)  ({\tt BrB0LLp})
\end{itemize}
Note, the calculation is based on the approach and conventions given in the references. However, these expressions are generalized and \\SARAH includes new contributions possible in an extended SUSY model. This is done be calculating and implementing the amplitudes for each generic possible diagram for a given process in a general form. The necessity of this is for instance pronounced in Ref.~\cite{Hirsch:2012ax} where it has been shown that the $Z$-penguins to $l_i \to 3 l_j$ often small in the MSSM can dominate in other models like inverse seesaw. 
% \end{itemize}

\subsection{{\tt ModelData\_[Model].f90}}
Contains the declaration of global variables and a function for initializing all
variables with 0. Only masses of particles, which are integrated out, are
initialized with \(10^{16}\)~GeV. That's used for the threshold corrections in
the first iteration. \\
In addition, functions for the calculation of GMSB boundary conditions as well
as the content of {\tt SelfDefinedFunctions} is written to that file, see also
sec.~\ref{sec:sphenoinputfile}.

\subsection{{\tt RGEs\_[Model].f90}}
That file includes all information about the RGEs. In principle, \\SARAH
generates three different sets of RGEs for \dots
\begin{itemize}
 \item \dots running of all parameters (gauge couplings, superpotential
parameter, soft breaking parameters) from the GUT scale to the SUSY scale
 \item \dots running of all parameters and VEVs from SUSY scale to electroweak
scale and back
 \item \dots running of a minimal set of parameters from electroweak scale to
GUT scale: that includes all SM gauge and Yukawa couplings as well as the other
parameter the running depends on. Above a threshold scale, the set of
parameters consists of all parameter related to the SM gauge and Yukawa
couplings.
 \end{itemize}

For each set of RGEs the following function are generated:
\begin{itemize}
 \item {\tt ParametersToGXX}: saves the parameters in a vector of length XX
 \item {\tt GToParametersXX}: extracts the parameters from a vector of length XX
 \item {\tt rgeXX}: definition of all \(\beta\)-functions. 
\end{itemize}
\SARAH does the following simplification/modifications of the \(\beta\)-functions
before exporting it to Fortran code, in order to increase the speed of the
numerical calculation
\begin{itemize}
 \item All matrix multiplications are replaced by constants, which are
calculated at the beginning of each {\tt rgeXX} routine
 \item All powers of numbers/parameters are replaced by a constants, which are
also calculated first
\end{itemize}

\subsection{{\tt Shifts\_[Model].f90}}
That file is only created for models including at least one threshold scale
with gauge symmetry breaking. It contains the necessary routines to calculate
the finite shifts for gauge couplings and gaugino masses at the threshold.

\subsection{{\tt SPheno\_[Model].f90}}
The main program. Calls routines for reading the input, calculating the
spectrum, decay widths and low energy observables and writing the output.

\subsection{{\tt SugraRuns\_[Model].f90}}
Main routines for calculating the (s)particle spectrum:
\begin{itemize}
 \item {\tt FirstGuess}: calculates a first, approximate spectrum as starting
point
 \item {\tt SugraRuns}: calculates in an iterative way the spectrum
 \item {\tt BoundaryEW}: calculates the starting point of the running for gauge
and Yukawa couplings from electroweak data
 \item {\tt BoundaryHS, BoundarySUSY}: sets the boundary conditions at the GUT
and SUSY scale.  
 \item {\tt BoundaryConditionsUp,BoundaryConditionsDown}: applies the boundary
conditions at the threshold scales when running up/down 
\end{itemize}

\subsection{{\tt SusyDecays\_[Model].f90}}
Calculates all two body decays by using the phase space functions of \SPheno. If
scalar and pseudo scalar Higgs as well as the quarks are defined in the
particles file by the corresponding {\tt Description} statement, the one loop
corrections to the decays \(H,A \rightarrow q\bar{q}\) from gluons are
added, see app.~\ref{app:SPhenoDecays} for more information. \\

\subsection{{\tt SusyMasses\_[Model].f90}}
This file contains functions to calculate all running masses at tree level. The mass
matrices involving Goldstone bosons are taken in 't Hooft gauge. Note, that works only, if the
Goldstone bosons have been associated with the corresponding vector boson in the {\tt particles.m}
file, see sec.~\ref{sec:gaugefixing}. 
The main routines in that file are
\begin{itemize}
 \item {\tt TreeMasses}: main function to call the other subroutines 
 \item {\tt CalculateX}: derives the tree level mass and mixing matrix of
particle {\tt X} by solving the eigensystem of the corresponding mass matrix.
\end{itemize}

\subsection{Three Body Decays, e.g. {\tt Glu\_[Model].f90} }
Contains all necessary routines to calculate the branching ratios and decay
width of a fermion to three other fermions. All possible processes and diagrams
are generated by \\SARAH and mapped to the phase space function of \SPheno.

\subsection{Additional Files}
\begin{itemize}
 \item A model specific {\tt Makefile}
 \item Template for an input file of the current model: {\tt
LesHouches.in.[Model]}, see app.~\ref{sec:LHinput:file}.
\item Template for an input file assuming a SUSY scale input: {\tt
LesHouches.in.[Model]\_low}, see also app.~\ref{sec:LHinput:file}.
\end{itemize}

\section{LesHouches input file}
\label{sec:LHinput:file}
The LesHouches input file for each model is by default named
\lstset{frame=none}
\begin{lstlisting}
 LesHouches.in.[Model]
\end{lstlisting}
However, it is also possible to give the name of the in- and output file as
option when running \SPheno, see also Ref.~\cite{Porod:2011nf}:
\begin{lstlisting}
./[SPheno Directory]/bin/SPheno[Model] [Input file] [Output file]
\end{lstlisting}
While the entries of the common block {\tt MINPAR} and {\tt EXTPAR} are
completely  defined by the \SPheno input file for \\SARAH, some other blocks have
distinct values.
\lstset{frame=shadowbox}
\subsection*{{\tt SMINPUTS}}
The entries of this block accordingly to the SLHA conventions are: \\
\numentry{1}{$\alpha_\mrm{em}^{-1}(m_{\Z})^{\MSbar}$. Inverse 
 electromagnetic coupling at the $\Z$ pole in the $\MSbar$ scheme  (with 5
 active flavours).} 
\numentry{2}{$G_F$. Fermi constant (in units of $\GeV^{-2}$).}
\numentry{3}{$\alpha_s(m_{\Z})^{\MSbar}$. Strong coupling at
  the $\Z$ pole in the $\MSbar$ scheme (with 5 active flavours).}
\numentry{4}{$m_\Z$, pole mass.}
\numentry{5}{$m_b(m_b)^{\MSbar}$. $b$ quark running mass in the $\MSbar$
  scheme.}
\numentry{6}{$m_t$, pole mass.}
\numentry{7}{$m_\tau$, pole mass.}
\numentry{8}{$m_{\nu_3}$, pole mass.}
\numentry{11}{$m_\e$, pole mass. }
\numentry{12}{$m_{\nu_1}$, pole mass.} 
\numentry{13}{$m_\mu$, pole mass. }
\numentry{14}{$m_{\nu_2}$, pole mass.} 
\numentry{21}{$m_d(2\ \GeV)^{\MSbar}$. $d$ quark running mass in the $\MSbar$
  scheme.}
\numentry{22}{$m_u(2\ \GeV)^{\MSbar}$. $u$ quark running mass in the $\MSbar$
  scheme.}
\numentry{23}{$m_s(2\ \GeV)^{\MSbar}$. $s$ quark running mass in the $\MSbar$
  scheme.}
\numentry{24}{$m_c(m_c)^{\MSbar}$. $c$ quark running mass in the $\MSbar$
  scheme.}

\subsection*{{\tt MODSEL}}
The block {\tt MODSEL} is not totally equivalent to the SLHA conventions: since
each \SPheno module generated by \\SARAH handles one specify model, the flags
{\tt 3} and {\tt 4} have no effect. In addition, we added the flag {\tt 2} to
choose between the different boundary conditions, and flag {\tt 1} is used to choose
between a GUT scale and SUSY scale input. \\
\numentry{1}{(Default=\ttt{1}) Choice of input type\\
\snumentry{0}{low scale input}
\snumentry{1}{Gut scale input}
}
\numentry{2}{(Default=\ttt{1}) Choice of boundary conditions. \\
\snumentry{X}{GUT scale input. {\tt X} gives the number of the set of boundary
conditions}
}
\numentry{3}{No effect}
\numentry{4}{No effect} 
\numentry{5}{(Default=\ttt{0}) CP violation. Switches defined are:\\
\snumentry{0}{CP is conserved. No information even on the CKM phase
is used. This corresponds to the SLHA1.}
\snumentry{1}{CP is violated, but only by the standard CKM
phase. All other phases are assumed zero.}
\snumentry{2}{CP is violated. Completely general CP phases
allowed.}
}
\numentry{6}{(Default=\ttt{0}) Flavour violation. Switches defined are:\\
\snumentry{0}{No (SUSY) flavour violation. }
\snumentry{1-3}{Flavour is violated. }
}

\subsection*{{\tt SPHENOINPUT}}
\label{sec:SPHENOINPUT}
The block with \SPheno specific commands is similar to the conventions
used by the standard \SPheno version 3.1 and above. However, some switches have
no effect and there are also some new entries which we highlight with in bold.
\\
\numentry{1}{sets the error level}
\numentry{2}{if 1, the SPA conventions are used}
\numentry{3}{No effect}
\numentry{4}{No effect}
\numentry{7}{Skip two loop Higgs masses}
\numentry{11}{if 1 then the branching ratios of the SUSY and Higgs particles
are calculated, if 0 then this calculation is omitted.}
\numentry{12}{sets minimum value for a branching ratios, so that it appears in
the output}
\numentry{13}{Include possible three-body decays (Note, these calculations can be time consuming!)}
\numentry{21-26}{No effect}
\numentry{31}{sets the value of $M_{\rm GUT}$, otherwise $M_{\rm GUT}$
is determined by the condition $g_1=g_2$}
\numentry{32}{sets strict unification, i.e.\  $g_1=g_2=g_3$}
\numentry{33}{Set fixed renormalization scale (Note, SPA conventions have to be switched off)}
\numentry{34}{sets the relative precision with which the masses are
calculated, default is $10^{-4}$} 
\numentry{35}{sets the maximal number of iterations in the calculation
 of the masses, default  is 40}
\numentry{36}{whether to write out debug information for the loop calculations}
\numentry{38}{this entry sets the loop order of the RGEs: either 1 or 2,
default  is 2, i.e.\ using  2-loop RGEs}
\numentry{41}{sets the width of the Z-boson $\Gamma_Z$,
default is 2.49 GeV}
\numentry{42}{sets the width of the W-boson $\Gamma_W$,
default is 2.06 GeV}
\numentry{{\bf 50}}{if 1, negative fermion masses are rotated to real ones by
using complex mixing matrices; if 0, all mixing matrices for Majorana fermions
are real, but masses can be negative}
\numentry{{\bf 51}}{If set to {\tt 0}, the parameters
\(Y_u,Y_d,T_u,T_d,m_q^2,m_d^2,m_u^2\) are not rotated in SCKM basis in the
output file}
\numentry{{\bf 52}}{if 1, a negative mass squared is always ignored and set 0}
\numentry{{\bf 53}}{if 1, a negative mass squared at $M_Z$ is always ignored and set 0}
\numentry{{\bf 54}}{if 1, the output is written even if there has been a problem during the run}
\numentry{{\bf 55}}{if 0, the loop corrections to the masses are skipped}
\numentry{{\bf 57}}{if 0, the calculation of the low energy observables is skipped}
\numentry{{\bf 58}}{if 0, the calculation of $\delta_{VB}$ in the boundary conditions at the SUSY scale is skipped}
\numentry{{\bf 60}}{if 0, possible effects from kinetic mixing are neglected}
\numentry{{\bf 62}}{if 1, a sign flip in $\mu$ (or other quadratic terms obtained by the tadpole equations) is ignored}
\numentry{{\bf 65}}{X, defined the solution of the tadpole equations if several, independent solutions exist}
\numentry{{\bf 72}}{if 1, the running values of all parameters at the GUT scale are written}
\numentry{{\bf 75}}{if 1, a file containing all parameters in \WHIZARD format is
created}
\numentry{{\bf 76}}{if 1, input files for \HB are written}
\numentry{80}{if not set 0  the program exists with a non-zero value  if
           a problem has occurred}
\numentry{90-92}{No effect}
\section{Decays calculated by \SPheno}
\label{app:SPhenoDecays}
In general, \SPheno version produced with \\SARAH calculate the two-body decays of all SUSY scalars as well as the Higgs fields. In addition, for SUSY fermions also the three-body decays are included at tree-level. 
In addition, the calculation of all decays has been improved by performing an
 RGE evaluation of all couplings from the SUSY scale to the mass scale of the
 decaying particle.  Furthermore, the
 calculation of the loop-induced decays of a Higgs particle into
 two photons and two gluons include also the dominant QCD
 corrections based on the results given in Ref.~\cite{Spira:1995rr}. This leads to the following
 precision precision in the calculations: \\
\begin{itemize}
\item for all  fermionic SUSY particle the two- and 
 three body decays are calculated at tree-level
 \item for squarks, sleptons and additional 
 heavy vector bosons the two-body decays are calculated at tree level 
 \item In the Higgs sector, possible decays 
 into two SUSY or SM particles are calculated at tree-level. In the case of two quarks in the final state the
 dominant QCD corrections due to gluons are included. The loop induced decays into two photons
 and gluons are fully calculated at LO with the dominant NLO corrections as just mentioned. In 
 addition, in the Higgs decays also final states with off-shell gauge bosons ($Z Z^*$, $W W^*$) are 
 also take into account.
\end{itemize}
\section{Input files to generate a \SPheno version for the MSSM and NMSSM}
\label{app:MSSM_NMSSM}
\subsection{MSSM}
\label{app:MSSM_input}
\subsubsection*{{\tt MSSM.m}}
We have already discussed the main parts of the MSSM input file for \\SARAH in
sec.~\ref{sec:MSSM_modelfile}. Therefore, we just show the file here for
completeness again.

\begin{lstlisting}
ModelName = "MSSM";
ModelNameLaTeX ="MSSM";

(* ------------------------- Vector Superfields ------------------------- *)

Gauge[[1]]={B,   U[1], hypercharge, g1,False};
Gauge[[2]]={WB, SU[2], left,        g2,True};
Gauge[[3]]={G,  SU[3], color,       g3,False};

(* ------------------------- Chiral Superfields ------------------------- *)

Fields[[1]] = {{uL,  dL},  3, q,   1/6, 2, 3};  
Fields[[2]] = {{vL,  eL},  3, l,  -1/2, 2, 1};
Fields[[3]] = {{Hd0, Hdm}, 1, Hd, -1/2, 2, 1};
Fields[[4]] = {{Hup, Hu0}, 1, Hu,  1/2, 2, 1};

Fields[[5]] = {conj[dR], 3, d,  1/3, 1, -3};
Fields[[6]] = {conj[uR], 3, u, -2/3, 1, -3};
Fields[[7]] = {conj[eR], 3, e,    1, 1,  1};

(* --------------------------- Superpotential --------------------------- *)

SuperPotential = { {{1, Yu},{u,q,Hu}}, {{-1,Yd},{d,q,Hd}},
                   {{-1,Ye},{e,l,Hd}}, {{1,\[Mu]},{Hu,Hd}}};
  

(* ----------------------- Different eigenstates ------------------------ *)

NameOfStates={GaugeES, EWSB};

(* --------------------- Rotations in gauge sector ---------------------- *)

DEFINITION[EWSB][GaugeSector]= 
{ {{VB,VWB[3]},{VP,VZ},ZZ},
  {{VWB[1],VWB[2]},{VWm,conj[VWm]},ZW},
  {{fWB[1],fWB[2],fWB[3]},{fWm,fWp,fW0},ZfW}
};    
        
(* ---------------------- Decomposition of Scalars ---------------------- *)  

DEFINITION[EWSB][VEVs]= 
  {{SHd0, {vd, 1/Sqrt[2]}, {sigmad, \[ImaginaryI]/Sqrt[2]},{phid,1/Sqrt[2]}},
   {SHu0, {vu, 1/Sqrt[2]}, {sigmau, \[ImaginaryI]/Sqrt[2]},{phiu,1/Sqrt[2]}}};


(* --------------------- Rotations in matter sector --------------------- *)

DEFINITION[EWSB][MatterSector]= 
{    {{SdL, SdR},                 {Sd, ZD}},
     {{SvL},                      {Sv, ZV}},
     {{SuL, SuR},                 {Su, ZU}},
     {{SeL, SeR},                 {Se, ZE}},
     {{phid, phiu},               {hh, ZH}},
     {{sigmad, sigmau},           {Ah, ZA}},
     {{SHdm,conj[SHup]},          {Hpm,ZP}},
     {{fB, fW0, FHd0, FHu0},      {L0, ZN}}, 
     {{{fWm, FHdm}, {fWp, FHup}}, {{Lm,UM}, {Lp,UP}}},
     {{{FeL},{conj[FeR]}},        {{FEL,ZEL},{FER,ZER}}},
     {{{FdL},{conj[FdR]}},        {{FDL,ZDL},{FDR,ZDR}}},
     {{{FuL},{conj[FuR]}},        {{FUL,ZUL},{FUR,ZUR}}}     
       }; 
       
(* -------------------------------- Phases ------------------------------- *)

DEFINITION[EWSB][Phases]=  {    {fG, PhaseGlu}     }; 


(* ---------------------------- Dirac Spinors ---------------------------- *)

DEFINITION[EWSB][DiracSpinors]={
 Fd  -> {FDL, conj[FDR]},
 Fe  -> {FEL, conj[FER]},
 Fu  -> {FUL, conj[FUR]},
 Fv  -> {FvL, 0},
 Chi -> {L0, conj[L0]},
 Cha -> {Lm, conj[Lp]},
 Glu -> {fG, conj[fG]}
};

DEFINITION[GaugeES][DiracSpinors]={
  Bino -> {fB, conj[fB]},
  Wino -> {fWB, conj[fWB]},
  Glu  -> {fG, conj[fG]},
  H0   -> {FHd0, conj[FHu0]},
  HC   -> {FHdm, conj[FHup]},
  Fd1  -> {FdL, 0},
  Fd2  -> {0, FdR},
  Fu1  -> {FuL, 0},
  Fu2  -> {0, FuR},
  Fe1  -> {FeL, 0},
  Fe2  -> {0, FeR},
  Fv   -> {FvL,0}
};
\end{lstlisting}

\subsubsection*{{\tt SPheno.m}}
To generate the \SPheno output for the MSSM, {\tt SPheno.m} should provide the
following information: \\
We want to have mSugra like boundary conditions. Therefore, we define as
minimal set of parameters for the model \(m_0,M_{1/2},A_0,\text{sign}\mu\)
and \(\tan\beta\). These will later on be read from the {\tt MINPAR} block of a
LesHouches input file
\begin{lstlisting}
MINPAR={{1,m0},
        {2,m12},
        {3,TanBeta},
        {4,SignumMu},
        {5,Azero}};
\end{lstlisting}
In general, these parameters are assumed to be complex, i.e. it is possible to
use also the block {\tt IMMINPAR} to define the imaginary part. However, some
Fortran functions like {\tt sin} can't be used with complex numbers, therefore
we have to define \(\tan\beta\) explicitly as real. Also $m_0$ is a real parameter.
\begin{lstlisting}
RealParameters = {TanBeta, m0};
\end{lstlisting}
As usual in the MSSM, the tadpole equations should be solved with respect to
\(\mu\) and \(B_\mu\). That's defined by
\begin{lstlisting}
ParametersToSolveTadpoles = {\[Mu],B[\[Mu]]};
\end{lstlisting}
To study models with a dynamically adjusted SUSY scale, the expressions for the
definition of the SUSY scale can be given. The first expression is used only
before the mass spectrum has been calculated the first time. Note, that these
definitions can easily disabled in the LesHouches input file by flag {\tt
MODSEL 12} and a fixed scale can be used. Also, when SPA conventions are
switched on in the LesHouches input file by the flag {\tt 2} in the block {\tt SPhenoInput}, a fixed scale
of 1 TeV is used. 
\begin{lstlisting}
RenormalizationScaleFirstGuess = m0^2 + 4 m12^2;
RenormalizationScale = MSu[1]*MSu[6];
\end{lstlisting}
As said, we want to use mSugra like boundary conditions. These are
defined by
\begin{lstlisting}
BoundaryHighScale={
  {T[Ye],   Azero*Ye},
  {T[Yd],   Azero*Yd},
  {T[Yu],   Azero*Yu},
  {mq2,     DIAGONAL m0^2},
  {ml2,     DIAGONAL m0^2},
  {md2,     DIAGONAL m0^2},
  {mu2,     DIAGONAL m0^2},
  {me2,     DIAGONAL m0^2},
  {mHd2,    m0^2},
  {mHu2,    m0^2},
  {MassB,   m12},
  {MassWB,  m12},
  {MassG,   m12}
};
\end{lstlisting}
It is also possible to use the generated \SPheno version with a low scale
input. This is enabled by setting {\tt MODSEL 1} to {\tt 0} in the LesHouches 
input file. In that case,
input values for all free parameters of the model are expected. However, it is
possible to define also some boundary conditions to calculate for example the
SUSY VEVs dynamically.
\begin{lstlisting}
BoundaryLowScaleInput={
 {vd,Sqrt[2 mz2/(g1^2+g2^2)]*Sin[ArcTan[TanBeta]]},
 {vu,Sqrt[2 mz2/(g1^2+g2^2)]*Cos[ArcTan[TanBeta]]}
};
\end{lstlisting}
Finally, we define that the code for the calculation of the two and three body
decays is generated for all SUSY and Higgs particles. That's done by using the flag {\tt
Automatic}.
\begin{lstlisting}
ListDecayParticles = Automatic;
ListDecayParticles3B = Automatic;
\end{lstlisting}
\subsection{NMSSM}
\subsubsection*{{\tt NMSSM.m}}
% We have presented in sec.~\ref{sec:ToNMSSM} the steps from the NMSSM to the
% MSSM in \\SARAH. Therefore, we show here again only the entire file.

\begin{lstlisting}
ModelName = "NMSSM";
ModelNameLaTeX ="NMSSM";

(* ------------------------- Vector Superfields ------------------------- *)

Gauge[[1]]={B,   U[1], hypercharge, g1,False};
Gauge[[2]]={WB, SU[2], left,        g2,True};
Gauge[[3]]={G,  SU[3], color,       g3,False};

(* ------------------------- Chiral Superfields ------------------------- *)

Fields[[1]] = {{uL,  dL},  3, q,   1/6, 2, 3};  
Fields[[2]] = {{vL,  eL},  3, l,  -1/2, 2, 1};
Fields[[3]] = {{Hd0, Hdm}, 1, Hd, -1/2, 2, 1};
Fields[[4]] = {{Hup, Hu0}, 1, Hu,  1/2, 2, 1};

Fields[[5]] = {conj[dR], 3, d,  1/3, 1, -3};
Fields[[6]] = {conj[uR], 3, u, -2/3, 1, -3};
Fields[[7]] = {conj[eR], 3, e,    1, 1,  1};
Fields[[8]] = {sR,       1, s,    0, 1,  1};


(* --------------------------- Superpotential --------------------------- *)

SuperPotential = { {{1, Yu},{q,Hu,u}}, {{-1,Yd},{q,Hd,d}},
  {{-1,Ye},{l,Hd,e}}, {{1,\[Lambda]},{Hu,Hd,s}}, {{1/3,\[Kappa]},{s,s,s}}};


(* ----------------------- Different eigenstates ------------------------ *)

NameOfStates={GaugeES, EWSB};


(* --------------------- Rotations in gauge sector ---------------------- *)

DEFINITION[EWSB][GaugeSector]= 
{ {{VB,VWB[3]},{VP,VZ},ZZ},
  {{VWB[1],VWB[2]},{VWm,conj[VWm]},ZW},
  {{fWB[1],fWB[2],fWB[3]},{fWm,fWp,fW0},ZfW}
};    

(* ---------------------- Decomposition of Scalars ---------------------- *)    
   

DEFINITION[EWSB][VEVs]= 
{    {SHd0, {vd, 1/Sqrt[2]}, {sigmad, I/Sqrt[2]}, {phid,1/Sqrt[2]}},
     {SHu0, {vu, 1/Sqrt[2]}, {sigmau, I/Sqrt[2]}, {phiu,1/Sqrt[2]}},
     {SsR, {vS, 1/Sqrt[2]},  {sigmaS, I/Sqrt[2]}, {phiS, 1/Sqrt[2]}}
 
  };


(* --------------------- Rotations in matter sector --------------------- *)

DEFINITION[EWSB][MatterSector]= 
{    {{SdL, SdR},                  {Sd, ZD}},
     {{SvL},                       {Sv, ZV}},
     {{SuL, SuR},                  {Su, ZU}},
     {{SeL, SeR},                  {Se, ZE}},
     {{phid, phiu, phiS},          {hh, ZH}},
     {{sigmad, sigmau,sigmaS},     {Ah, ZA}},
     {{SHdm,conj[SHup]},           {Hpm,ZP}},
     {{fB, fW0, FHd0, FHu0,FsR},   {L0, ZN}}, 
     {{{fWm, FHdm}, {fWp, FHup}},  {{Lm,UM}, {Lp,UP}}},
     {{{FeL},{conj[FeR]}},         {{FEL,ZEL},{FER,ZER}}},
     {{{FdL},{conj[FdR]}},         {{FDL,ZDL},{FDR,ZDR}}},
     {{{FuL},{conj[FuR]}},         {{FUL,ZUL},{FUR,ZUR}}}            
   }; 

(* -------------------------------- Phases ------------------------------- *)

DEFINITION[EWSB][Phases]=  {    {fG, PhaseGlu}     }; 

(* ---------------------------- Dirac Spinors ---------------------------- *)

DEFINITION[EWSB][DiracSpinors]={
 Fd  -> {FDL, conj[FDR]},
 Fe  -> {FEL, conj[FER]},
 Fu  -> {FUL, conj[FUR]},
 Fv  -> {FvL, 0},
 Chi -> {L0, conj[L0]},
 Cha -> {Lm, conj[Lp]},
 Glu -> {fG, conj[fG]}
};

DEFINITION[GaugeES][DiracSpinors]={
  Bino -> {fB, conj[fB]},
  Wino -> {fWB, conj[fWB]},
  Glu  -> {fG, conj[fG]},
  H0   -> {FHd0, conj[FHu0]},
  HC   -> {FHdm, conj[FHup]},
  Fd1  -> {FdL, 0},
  Fd2  -> {0, FdR},
  Fu1  -> {FuL, 0},
  Fu2  -> {0, FuR},
  Fe1  -> {FeL, 0},
  Fe2  -> {0, FeR},
  Fv   -> {FvL,0},
  S    -> {FsR, conj[FsR]}
};
\end{lstlisting}

\subsubsection*{{\tt SPheno.m}}
We define as free parameters \(\kappa,A_\kappa,\lambda,A_\lambda\) as well as
the singlet VEV \(v_S\) in addition to the mSugra parameters of the MSSM. To be
SLHA conform, these parameters are added the the {\tt EXTPAR} block.
\begin{lstlisting}
MINPAR={{1,m0},
        {2,m12},
        {3,TanBeta},
        {5,Azero}};

EXTPAR = {
	   {61,LambdaInput},
	   {62,KappaInput},
	   {63,ALambdaInput},
	   {64,AKappaInput},
           {65,vSInput}
	 };
\end{lstlisting}
Again, \(\tan\beta\) must be defined as real an we choose the same expressions
for the SUSY scale as in the MSSM. 
\begin{lstlisting}
RealParameters = {TanBeta, m0};

RenormalizationScaleFirstGuess = m0^2 + 4 m12^2;
RenormalizationScale = MSu[1]*MSu[6];
\end{lstlisting}
We need new parameters to solve the tadpole equations. The easiest choice is to
use the soft breaking masses for the Higgs and the gauge singlet. 
\begin{lstlisting}
ParametersToSolveTadpoles = {mHd2,mHu2,ms2};
\end{lstlisting}
This time, we want to specify two different boundary conditions:
\begin{enumerate}
 \item Input for \(\kappa\) and \(\lambda\) are used at the SUSY scale, while
\(A_\kappa\) and \(A_\lambda\) are used at the GUT scale.
 \item \(\kappa\), \(\lambda\), \(A_\kappa\) and \(A_\lambda\) are used at the
GUT scale. 
\end{enumerate}
For all other parameters, the boundaries are always set at the GUT scale. Only
the singlet VEV is taken at the SUSY scale. For that reason, we initialize the
arrays
\begin{lstlisting}
BoundarySUSYScale = Table[{},{2}];
BoundaryHighScale = Table[{},{2}];
\end{lstlisting}
First, the two sets of conditions at the SUSY scale.
\begin{lstlisting}
BoundarySUSYScale[[1]] = {
  {vS,        vSInput},  
  {\[Kappa],  KappaInput},
  {\[Lambda], LambdaInput}
};

BoundarySUSYScale[[2]] = {
  {vS,        vSInput}
};
\end{lstlisting}
Second, the GUT scale conditions
\begin{lstlisting}
BoundaryHighScale[[1]]={
  {T[Ye],         Azero*Ye},
  {T[Yd],         Azero*Yd},
  {T[Yu],         Azero*Yu},
  {mq2,           DIAGONAL m0^2},
  {ml2,           DIAGONAL m0^2},
  {md2,           DIAGONAL m0^2},
  {mu2,           DIAGONAL m0^2},
  {me2,           DIAGONAL m0^2},
  {T[\[Kappa]],   AKappaInput*\[Kappa]},
  {T[\[Lambda]],  ALambdaInput*\[Lambda]},
  {MassB,         m12},
  {MassWB,        m12},
  {MassG,         m12}
};

BoundaryHighScale[[2]]={
  {T[Ye],         Azero*Ye},
  {T[Yd],         Azero*Yd},
  {T[Yu],         Azero*Yu},
  {mq2,           DIAGONAL m0^2},
  {ml2,           DIAGONAL m0^2},
  {md2,           DIAGONAL m0^2},
  {mu2,           DIAGONAL m0^2},
  {me2,           DIAGONAL m0^2},
  {\[Kappa],      KappaInput},
  {\[Lambda],     LambdaInput},
  {T[\[Kappa]],   AKappaInput*KappaInput},
  {T[\[Lambda]],  ALambdaInput*LambdaInput},
  {MassB,         m12},
  {MassWB,        m12},
  {MassG,         m12}
};
\end{lstlisting}
As for the MSSM, we calculate the VEVs dynamically when a low energy input is
chosen. 
\begin{lstlisting}
BoundaryLowScaleInput={
 {vd, Sqrt[2 mz2/(g1^2+g2^2)]*Sin[ArcTan[TanBeta]]},
 {vu, Sqrt[2 mz2/(g1^2+g2^2)]*Cos[ArcTan[TanBeta]]}
};
\end{lstlisting}
Also for the calculation of the decay widths and branching ratios we make the
same choice as in the MSSM. 
\begin{lstlisting}
ListDecayParticles = Automatic;
ListDecayParticles3B = Automatic;
\end{lstlisting}
% We have only a bit careful in the pseudo scalar sector of the NMSSM: to
% identify the Goldstone boson in the case \(\kappa = 0\), we use the following
% check (see sec.~\ref{sec:sphenoinputfile} for more information)
% \begin{lstlisting}
% ConditionForMassOrdering={
% {Ah,
% "If
% ((Abs(ZA(1,3)).gt.Abs(ZA(2,3))).And.(MAh2(1).lt.1._dp).And.(MAh2(2).lt.1._dp))
% Then \n
%    MAh2temp = MAh2 \n
%    ZAtemp = ZA \n
%    ZA(1,:) = ZAtemp(2,:) \n
%    ZA(2,:) = ZAtemp(1,:) \n
%    MAh2(1) = MAh2temp(2) \n
%    MAh2(2) = MAh2temp(1) \n
% End If \n \n"}
% };
% \end{lstlisting}

\section{Model files for Seesaw type I -- III}
For the model files, we show only the difference in comparison to the MSSM. For
a discussion of seesaw I -- III see Ref.~\cite{Esteves:2010ff} and references
therein. 
\label{app:Seesaw}
\subsection{Seesaw I}
\subsubsection*{Seesaw1.m}
In the case of  seesaw~I, the particle content is extended by three generations
of a gauge singlet \(\hat{\nu}_R\). 
\begin{lstlisting}
Fields[[8]] = {vR,       3, v,   0,1,  1};
\end{lstlisting}
The new field causes a Yukawa like interaction and a mass term. In addition,
there is an effective operator which gets initialized after integrating out the
right handed neutrino. 
\begin{lstlisting}
SuperPotential =
  {...,{{1,Yv},{v,l,Hu}},{{1/2,Mv},{v,v}},{{1,MNuL},{l,Hu,l,Hu}}};
\end{lstlisting}
Since the field is heavy, it should not be included in the calculation of the
vertices and masses at the SUSY scale. Therefore, we 'delete' it. Note, the low
energy results like masses and vertices are independent of deleted particles.
However, the RGEs are not since they have to be also valid at the GUT scale.
\begin{lstlisting}
DeleteParticles={v};
\end{lstlisting}

\subsubsection*{SPheno.m}
We choose a unification of the soft-breaking mass of the scalar singlet with the
other soft-breaking masses at the GUT scale. In addition, we want to define the
value of the superpotential parameters at the GUT scale as input values in the
LesHouches file. Furthermore, we have also a mSugra like condition for the
trilinear soft-breaking coupling. Since the bilinear soft-breaking term does
not influence the RGE running of the other parameters, we can safely set it to
zero. Now, the additional boundary conditions at the GUT scale are
\begin{lstlisting}
BoundaryHighScale={
  ...,
  {mv2,    DIAGONAL m0^2},
  {Mv,     LHInput[Mv]},
  {Yv,     LHInput[Yv]}
  {B[Mv],  0},
  {T[Yv],  Azero*LHInput[Yv]}
};
\end{lstlisting}
We want to include three threshold scales: each generation of the gauge singlet
should be integrated out at energies similar to their mass. Therefore, a good
choice is
\begin{lstlisting}
Thresholds={
  {Abs[MvIN[1,1]],{v[1]}},
  {Abs[MvIN[2,2]],{v[2]}},
  {Abs[MvIN[3,3]],{v[3]}}
};
\end{lstlisting}
When thresholds are included, it is possible to define for each threshold scale
the boundary conditions separately for running up and down the RGEs. First, it
is necessary to initialize the corresponding arrays
\begin{lstlisting}
BoundaryConditionsUp=Table[{},{Length[Thresholds]}];
BoundaryConditionsDown=Table[{},{Length[Thresholds]}];
\end{lstlisting}
When a gauge singlet is integrated out, the effective operator receives a
contribution of the form
\begin{equation}
\label{eq:mnuSeesawI}
 \kappa = - Y_\nu^T \, M_N^{-1} \, Y_\nu \thickspace . 
\end{equation}
When heavy superfields are integrated, the mass splitting between the fermionic
and scalar component is neglected. The masses are calculated individually at
each threshold scale and saved in arrays with the name 
\begin{lstlisting}
 MassOf <> Name of Superfield
\end{lstlisting}
Therefore, the contributions to the effective operator at the different scales
are given by
\begin{lstlisting}
BoundaryConditionsDown[[1]]={
  {MNuL[index1,index2], - Yv[3,index1] Yv[index2,3]/MassOfv[3]}};
BoundaryConditionsDown[[2]]={
  {MNuL[index1,index2], - Yv[2,index1] Yv[index2,2]/MassOfv[2]}};
BoundaryConditionsDown[[3]]={
  {MNuL[index1,index2], - Yv[1,index1] Yv[index2,1]/MassOfv[1]}};
\end{lstlisting}

\subsection{Seesaw II}
\subsubsection*{Seesaw2.m}
For the seesaw~II, it is necessary to add a scalar \(SU(2)_L\) triplet which
also carries hypercharge \(Y_i\). Such a particle is part of the 15-plet of
\(SU(5)\). Therefore, we add to the \(SU(5)\) invariant superpotential the
interactions of a pair of \({\bf 15}\) and \({\bf \overline{15}}\). After
\(SU(5)\) breaking, the {\bf 15} splits into irreducible representations of
\(SU(3)_C\times SU(2)_L\times U(1)_Y\)  
\begin{equation}
\label{eq:split15}
 {\bf 15} = \hat{S} + \hat{T} + \hat{Z} \thickspace .
\end{equation}
The corresponding quantum numbers are 
\begin{equation}
 \hat{S}: \left(\bf{6}, \bf{1}\right)_{-2/3} \thickspace, \hspace{1cm} \hat{T}:
\left(\bf{1},\bf{3}\right)_1 \thickspace, \hspace{1cm} \hat{Z}:
\left(\bf{3},\bf{2}\right)_{1/6} \thickspace.
\end{equation}
These fields are implemented in \\SARAH by
\begin{lstlisting}
Fields[[8]] = {{{Tpp,1/Sqrt[2] Tp},{1/Sqrt[2] Tp, T0}}, 1, t, 1, 3, 1};
Fields[[9]] = {{{T0b,1/Sqrt[2] Tm},{1/Sqrt[2] Tm, Tmm}}, 1, tb, -1, 3, 1};
Fields[[10]] = {S, 1, s, -2/3, 1, 6};
Fields[[11]] = {conj[Sc], 1, sb, 2/3, 1, -6};
Fields[[12]] = {{z1,z2}, 1, z, 1/6, 2, 3};
Fields[[13]] = {{z1b,z2b}, 1, zb, -1/6, 2, -3};
\end{lstlisting}
The new terms in the superpotential after \(SU(5)\) breaking are
Again, after integrating out the colored Higgs fields, we end up with the
superpotential terms 
\begin{eqnarray}
\nonumber W^{II} &=& \frac{1}{\sqrt{2}} \left(Y_T\, \hat{l}\, \hat{T}\, \hat{l}
+ Y_S\, \hat{d}\, \hat{S}\, \hat{d}\right) + Y_Z\, \hat{d}\, \hat{Z}\, \hat{l}
+ \frac{1}{\sqrt{2}} \lambda_1 \hat{H}_d \hat{T} \hat{H}_d  + \frac{1}{\sqrt{2}}
\lambda_2 \hat{H}_u \hat{\bar{T}} \hat{H}_u \\
 && + M_T \hat{T} \hat{\bar{T}} + M_Z \hat{Z} \hat{\bar{Z}} + M_S \hat{S}
\hat{\bar{S}} 
\end{eqnarray}
This reads in \\SARAH
\begin{lstlisting}
SuperPotential = { ...,
 {{1/Sqrt[2],Yt},{l,t,l}},{{1/Sqrt[2],Ys},{d,s,d}},{{1,Yz},{d,z,l}},
 {{1/Sqrt[2],L1},{Hd,t,Hd}},{{1/Sqrt[2],L2},{Hu,tb,Hu}},
 {{1,MT},{t,tb}}, {{1,MZ},{z,zb}},{{1,MS},{s,sb}}, {{1,MNuL},{l,Hu,l,Hu}}};
\end{lstlisting}
All components of the 15-plet should be removed at the low scale, therefore we
add 
\begin{lstlisting}
DeleteParticles={t,tb,s,sb,z,zb};
\end{lstlisting}

\subsubsection*{SPheno.m}
We define as new parameters which can be adjusted by the LesHouches input file
the values for \(\lambda_1\), \(\lambda_2\) and \(M_T\)
\begin{lstlisting}
EXTPAR={{200,Lambda1IN},
        {201,Lambda2IN},
        {210,MTScaleIN}};
\end{lstlisting}
In addition, we use again mSugra boundary conditions which are also \(SU(5)\)
invariant with respect to the new parameters. It is again possible to neglect
the bilinear soft-breaking terms corresponding to the mass terms in the
superpotential.  
\begin{lstlisting}
BoundaryHighScale={
  ...,
  {mt2,   m0^2},
  {mtb2,  m0^2},
  {ms2,   m0^2},
  {msb2,  m0^2},
  {mz2,   m0^2},
  {mzb2,  m0^2},
  {MT,    MTScaleIN},
  {MZ,    MT},
  {MS,    MT},
  {B[MZ], 0},
  {B[MS], 0},
  {B[MT], 0},
  {Ys,    Yt},
  {Yz,    Yt},
  {T[Yt], Azero*Yt},
  {T[Ys], Azero*Yt},
  {T[Yz], Azero*Yt},
  {T[L1], Azero*L1},
  {T[L2], Azero*L2}
};
\end{lstlisting}
Since, we have only one 15-plet, we need only one threshold scale.
\begin{lstlisting}
Thresholds={
  {Abs[MTScaleIN],{s,sb,t,tb,z,zb}}
};
\end{lstlisting}
As boundary conditions, we choose this time that the input values of \(Y_t\),
\(\lambda_1\) and \(\lambda_2\) are used at the threshold scale. Of course, we
initialize again the Weinberg operator when crossing the threshold. This time,
the analytical expression is
\begin{equation}
  \kappa = \frac{1}{2} \lambda_2 M_T^{-1} Y_T 
\end{equation}
The corresponding lines in the input file are
\begin{lstlisting}
BoundaryConditionsUp=Table[{},{Length[Thresholds]}];
BoundaryConditionsDown=Table[{},{Length[Thresholds]}];

BoundaryConditionsUp[[1]] = {
  {Yt, LHInput[Yt]},
  {L1, Lambda1IN},
  {L2, Lambda2IN}
}; 

BoundaryConditionsDown[[1]] = {
  {MNuL, -L2 Yt/MT }
}; 
\end{lstlisting}

\subsection{Seesaw III}
\subsubsection*{Seesaw3.m}
The type~III seesaw is based on  additional fields belonging to the adjoint
representation of \(SU(2)_L\). Hence, we add particles sitting in the 24-plet,
the adjoint representation of \(SU(5)\), to the spectrum. It is not sufficient
to add just one generation of 24-plets to explain all neutrino data if we assume
\(SU(5)\) invariant boundary conditions at the GUT scale: the induced mass
splitting between the different generations of neutrinos won't be large enough.
Therefore, we will add three generations of 24-plets. \\
The {\bf 24} has the same gauge quantum numbers as the gauge bosons of \(SU(5)\)
and can be decomposed in SM representations by 
\begin{equation}
 24_M = \hat{G}_M + \hat{W}_M + \hat{B}_M + \hat{X}_M + \hat{\bar{X}}_M 
\end{equation}
with
\begin{equation}
  \hat{G}_M: \left(\bf{8},\bf{1}\right)_0, \thickspace \hat{W}_M:
\left(\bf{1},\bf{3}\right)_0, \thickspace \hat{B}_M:
\left(\bf{1},\bf{1}\right)_0, \hat{X}_M: \left(\bf{3},\bf{2} \right)_{-5/6},
\thickspace \hat{\bar{X}}_M: \left(\bf{\bar{3}},\bf{2} \right)_{5/6} \thickspace
.
\end{equation}
Hence, we have to add to the \\SARAH input file the following field definitions
\begin{lstlisting}
Fields[[8]] = {{{HW0/Sqrt[2],HWp},{HWm, -HW0/Sqrt[2]}}, 3, Hw3, 0, 3, 1};
Fields[[9]] =  {HG, 3, Hg3, 0, 1, 8};
Fields[[10]] = {HB, 3, Hb3, 0, 1, 1};
Fields[[11]] = {{HXu,HXd}, 3, Hx3, 5/6, 2, -3};
Fields[[12]] = {{HXub,HXdb}, 3, Hxb3, -5/6, 2, 3};
\end{lstlisting}
The new terms in the superpotential are
 \begin{eqnarray}
\nonumber W^{III} &=& Y_W \hat{H}_u \hat{W}_M \hat{l} -\sqrt{\frac{3}{10}} Y_B
\hat{H}_u \hat{B}_M \hat{l} + Y_X \hat{H}_u \hat{\bar{X}}_M \hat{d} + \\
&& \frac{1}{2} M_B \hat{B}_M \hat{B}_M
+ \frac{1}{2} M_G \hat{G}_M \hat{G}_M + \frac{1}{2} M_W \hat{W}_M \hat{W}_M +
M_X \hat{X}_M \hat{\bar{X}}_M \thickspace .
\end{eqnarray}
These terms and the Weinberg operator read in \\SARAH
\begin{lstlisting}
SuperPotential = { ..,
 {{Sqrt[6/20],Yb3},{Hu,Hb3,l}},{{1,Yw3},{Hu,Hw3,l}},{{1,Yx3},{Hu,Hxb3,d}},      
 {{1,MXM3},{Hx3,Hxb3}},{{1/2,MWM3},{Hw3,Hw3}},{{1/2,MGM3},{Hg3,Hg3}},
 {{1/2,MBM3},{Hb3,Hb3}}, {{1,MNuL},{l,Hu,l,Hu}}};
\end{lstlisting}
Again, all additional fields are removed at the low scale
\begin{lstlisting}
DeleteParticles={Hw3,Hg3,Hb3,Hx3,Hxb3};
\end{lstlisting}

\subsubsection*{SPheno.m}
We use here the same kind of boundary conditions at the GUT scale as for type I
and II: mSugra-like and \(SU(5)\) invariant.
\begin{lstlisting}
BoundaryHighScale={
...,
  {mHw32,   DIAGONAL m0^2},
  {mHx32,   DIAGONAL m0^2},
  {mHxb32,  DIAGONAL m0^2},
  {mHg32,   DIAGONAL m0^2},
  {MWM3,    LHInput[MWM3]},
  {MXM3,    MWM3},
  {MBM3,    MWM3},
  {MGM3,    MWM3},
  {B[MWM3], 0},
  {B[MXM3], 0},
  {B[MBM3], 0},
  {B[MGM3], 0},
  {Yb3,     LHInput[Yb3]},
  {Yw3,     Yb3},
  {Yx3,     Yb3},
  {T[Yw3],  Azero*Yb3},
  {T[Yx3],  Azero*Yb3},
  {T[Yb3],  Azero*Yb3}
};
\end{lstlisting}
These boundary conditions have the effect that the mass splitting between the
different members of the 24-plet is not very large. Therefore, we can integrate
one generation of 24-plets at each scale:
\begin{lstlisting}
Thresholds={
  {Abs[MWM3IN[1,1]],{Hx3[1],Hxb3[1],Hg3[1],Hb3[1],Hw3[1]}},
  {Abs[MWM3IN[2,2]],{Hx3[2],Hxb3[2],Hg3[2],Hb3[2],Hw3[2]}},
  {Abs[MWM3IN[3,3]],{Hx3[3],Hxb3[3],Hg3[3],Hb3[3],Hw3[3]}}
};
\end{lstlisting}
Finally, we need the boundary conditions. Since the value of the Weinberg
operator at the threshold scale is the sum of seesaw I and III contributions,
we have
\begin{equation}
 \kappa = -  \left(\frac{1}{2} Y^T_W M_W^{-1} Y_W +  \frac{3}{10}
Y^T_B M_B^{-1} Y_B \right) \thickspace . 
\end{equation}
That's equivalent to the the following definitions in {\tt SPheno.m}
\begin{lstlisting}
BoundaryConditionsUp=Table[{},{Length[Thresholds]}];
BoundaryConditionsDown=Table[{},{Length[Thresholds]}];

BoundaryConditionsDown[[1]]={
  {MNuL[index1,index2],MNuL[index1,index2] +1/2 Yw3[3,index1]
     Yw3[index2,3]/MassOfHw3[3] +3/10 Yb3[3,index1] Yb3[index2,3]/MassOfHb3[3]}
};

BoundaryConditionsDown[[2]]={
  {MNuL[index1,index2],MNuL[index1,index2] + 1/2 Yw3[2,index1]
     Yw3[index2,2]/MassOfHw3[2] +3/10 Yb3[2,index1] Yb3[index2,2]/MassOfHb3[2]}
};

BoundaryConditionsDown[[3]]={
  {MNuL[index1,index2], 1/2 Yw3[1,index1] Yw3[index2,1]/MassOfHw3[1] +
     3/10 Yb3[1,index1] Yb3[index2,1]/MassOfHb3[1]}
};
\end{lstlisting}
\section{Implementation of a model with a gauge symmetry breaking scale in
\\SARAH and \SPheno}
\label{app:Omega} 
We show here the implementation of a left-right supersymmetric model
in \SPheno and \\SARAH in detail. This discussion is based on the model
presented in \cite{Esteves:2010si}. Since we are here mainly interested in the
implementation in \\SARAH, it is sufficient to simplify the model a bit by just
choosing one threshold scale and not two as discussed in \cite{Esteves:2010si}.
Adding the second threshold scale is straightforward, but would lead to some
redundancy in the following. 

\subsection{Summary of the model} 
We give here only a short summary about the model and refer for more details
to \cite{Esteves:2010si} and references therein. As mentioned, we use in the
following only one threshold scale at which \(\times SU(2)_R \times
U(1)_{B-L}\) gets broken
\begin{equation}
 SU(2)_L \times SU(2)_R \times U(1)_{B-L} \rightarrow  SU(2)_L \times U(1)_Y
\end{equation}

\paragraph*{From GUT scale to $SU(2)_R \times U(1)_{B-L}$ breaking scale}
The MSSM particle content above the threshold is extended by the presence of
four fields which are triplets under \(SU(2)_L\) or \(SU(2)_R\) and which carry
a \(B-L\) charge. In addition, there are two triplets which are uncharged under
\(B-L\). Furthermore, the right handed neutrino are part of the spectrum and
the Higgs fields are arranged in so called bi-doublets \(\Phi\). To get a
non-trivial CKM matrix, we need at least two generations of \(\Phi\) fields.
The particle content is summarized as follows \\
\begin{center}
\begin{tabular}{c c c c c c}
\hline
Superfield & generations & $SU(3)_c$ & $SU(2)_L$ & $SU(2)_R$ & $U(1)_{B-L}$ \\
\hline
$Q$ & 3 & 3 & 2 & 1 & $\frac{1}{3}$ \\
$Q^c$ & 3 & $\bar{3}$ & 1 & 2 & $-\frac{1}{3}$ \\
$L$ & 3 & 1 & 2 & 1 & -1 \\
$L^c$ & 3 & 1 & 1 & 2 & 1 \\
$\Phi$ & 2 & 1 & 2 & 2 & 0 \\
$\Delta$ & 1 & 1 & 3 & 1 & 2 \\
$\bar{\Delta}$ & 1 & 1 & 3 & 1 & -2 \\
$\Delta^c$ & 1 & 1 & 1 & 3 & -2 \\
$\bar{\Delta}^c$ & 1 & 1 & 1 & 3 & 2 \\
$\Omega$ & 1 & 1 & 3 & 1 & 0 \\
$\Omega^c$ & 1 & 1 & 1 & 3 & 0 \\
\hline
\end{tabular} 
\end{center}
The superpotential for the model reads
\begin{eqnarray} \label{eq:Wsuppot1}
{\cal W} &=& Y_Q Q \Phi Q^c 
          +  Y_L L \Phi L^c 
          - \frac{\mu}{2} \Phi \Phi
          +  f L \Delta L
          +  f^* L^c \Delta^c L^c \nonumber \\
         &+& a \Delta \Omega \bar{\Delta}
          +  a^* \Delta^c \Omega^c \bar{\Delta}^c
          + \alpha \Omega \Phi \Phi
          +  \alpha^* \Omega^c \Phi \Phi \nonumber \\
         &+& M_\Delta \Delta \bar{\Delta}
          +  M_\Delta^* \Delta^c \bar{\Delta}^c
          +  M_\Omega \Omega \Omega
          +  M_\Omega^* \Omega^c \Omega^c \thickspace.
\end{eqnarray}

\paragraph*{Below $SU(2)_R \times U(1)_{B-L}$ breaking scale} Here, we are
left with the MSSM plus the effective Weinberg operator which causes neutrino
masses after EWSB.  

\paragraph*{Boundary conditions} To link both scales, we need the following set
of boundary conditions:
\begin{eqnarray}
\label{eq:boundary_LR_1}
Y_d = Y_Q^1 \cos \theta_1 - Y_Q^2 \sin \theta_1 \thickspace,
&\qquad& Y_u = - Y_Q^1 \cos \theta_2 + Y_Q^2 \sin \theta_2  \thickspace, \\
Y_e = Y_L^1 \cos \theta_1 - Y_L^2 \sin \theta_1 \thickspace,
&\qquad& Y_\nu = - Y_L^1 \cos \theta_2 + Y_L^2 \sin \theta_2 \thickspace,
\end{eqnarray}
where $R = \sin (\theta_1 - \theta_2)$. For the soft-trilinear couplings, we
must just replace \(Y\) by \(T\) in the expressions. 
For the sfermionic soft masses, we have 
\begin{eqnarray}
m_{q}^2 =m_{u^c}^2 = m_{d^c}^2 &=& m_{Q^c}^2 \thickspace,\\ 
m_{l}^2=m_{e^c}^2  &=& m_{L^c}^2 \thickspace, \\ 
M_L = M_R &=& M_2 \thickspace.
\end{eqnarray}
while we need in the Higgs sector the relations
\begin{eqnarray}
m_{H_d}^2 &=& \cos^2 \theta_1 (m_\Phi^2)_{11} + \sin^2 \theta_1
(m_\Phi^2)_{22} - \sin \theta_1 \cos \theta_1 \left[ (m_\Phi^2)_{12} +
(m_\Phi^2)_{21} \right] \thickspace,  \\ 
m_{H_u}^2 &=& \cos^2 \theta_2
(m_\Phi^2)_{11} + \sin^2 \theta_2 (m_\Phi^2)_{22} - \sin \theta_2 \cos
\theta_2 \left[ (m_\Phi^2)_{12} + (m_\Phi^2)_{21} \right] \thickspace, 
\end{eqnarray}
In the gauge sector, we have to express the hypercharge coupling and the
corresponding gaugino by
\begin{eqnarray}
g_1 & = & \frac{\sqrt{5} g_2 g_{BL}}{\sqrt{2 g_2^2 + 3 g_{BL}^2}} \thickspace ,
\\
\label{eq:boundary_LR_2}
M_1 & = & \frac{2 g_2^2 M_1 + 3 g_{BL}^2 M_R}{2 g_2^2 + 3 g_{BL}^2} \thickspace.
\end{eqnarray}

\subsection{Model files for \\SARAH}
We present in the following the input files for \\SARAH to define the model at
the different scales. For shortness, we concentrate here on the parts necessary
for the \SPheno output and skip the gauge fixing terms and definition of Dirac
spinors above the threshold scale. 
\subsubsection*{From GUT scale to $SU(2)_R \times U(1)_{B-L}$ breaking scale} 
The vector and chiral superfields of the highest scale define the gauge sector
and particle content are
\lstset{basicstyle=\scriptsize,
frame=shadowbox}
\begin{lstlisting}
Gauge[[1]]={B,   U[1], bminl,       gBL,False};
Gauge[[2]]={WL, SU[2], left,        g2,True};
Gauge[[3]]={WR, SU[2], right,       g2,True};
Gauge[[4]]={G,  SU[3], color,       g3,False};

Fields[[1]] = {{uL,  dL},                           3, qL,       1/6, 2, 1, 3};
Fields[[2]] = {{conj[dR], - conj[uR]},              3, qR,      -1/6, 1, 2,-3}; 
Fields[[3]] = {{vL,  eL},                           3, lL,      -1/2, 2, 1, 1};
Fields[[4]] = {{conj[eR],  - conj[vR]},             3, lR,       1/2, 1, 2, 1};
Fields[[5]] = {{{Hd0, Hup},{Hdm, Hu0}},             2, Phi,        0, 2,-2, 1};
Fields[[6]] = {{{deltaLp/Sqrt[2], deltaLpp},
                {deltaL0, - deltaLp/Sqrt[2]}},      1, deltaL,     1, 3, 1, 1};
Fields[[7]] = {{{deltaLbarm/Sqrt[2], deltaLbar0},
              {deltaLbarmm, - deltaLbarm/Sqrt[2]}}, 1, deltaLbar, -1, 3, 1, 1};
Fields[[8]] = {{{deltaRm/Sqrt[2], deltaR0},
              {deltaRmm, - deltaRm/Sqrt[2]}},       1, deltaR,    -1, 1, 3, 1};
Fields[[9]] = {{{deltaRbarp/Sqrt[2], deltaRbarpp},
              {deltaRbar0, - deltaRbarp/Sqrt[2]}},  1, deltaRbar,  1, 1, 3, 1};
Fields[[10]] = {{{omegaL0/Sqrt[2], omegaLp},
               {omegaLm, - omegaL0/Sqrt[2]}},       1, omegaL,     0, 3, 1, 1};
Fields[[11]] = {{{omegaR0/Sqrt[2], omegaRp},
               {omegaRm, - omegaR0/Sqrt[2]}},       1, omegaR,     0, 1, 3, 1};
\end{lstlisting}
The superpotential reads
\begin{lstlisting}
SuperPotential = { {{1, YQ},           {qL,qR,Phi}},
		   {{1, YL},           {lL,lR,Phi}}, 
		   {{1,  f},           {lL,deltaL,lL}},
		   {{1,conj[f]},       {lR,deltaR,lR}}, 
		   {{1,Mdelta},        {deltaL,deltaLbar}},
		   {{1,conj[Mdelta]},  {deltaR,deltaRbar}},
		   {{-1/2,Mu3},        {Phi,Phi}},
                   {{1,Momega},        {omegaL,omegaL}},
                   {{1,conj[Momega]},  {omegaR,omegaR}},
                   {{1,a},             {deltaL,omegaL,deltaLbar}},
                   {{1,conj[a]},       {deltaR,omegaR,deltaRbar}},
                   {{1,AlphaOm},       {omegaL,Phi,Phi}},
                   {{1,conj[AlphaOm]}, {omegaR,Phi,Phi}}   };
\end{lstlisting}
The gauge bosons and gauginos of the right sector decompose into
\begin{lstlisting}
DEFINITION[RSB][GaugeSector]= 
{ {{VWR[1],VWR[2]},{VWRm,conj[VWRm]},ZW},
  {{VB,VWR[3]},{VBY,VZ2},ZZ},
  {{fWR[1],fWR[2],fWR[3]},{fWRm,fWRp,fWR0},ZfW}
};
\end{lstlisting}
after the omega and delta fields have received their VEV
\begin{lstlisting}
DEFINITION[RBLSB][VEVs]= 
{ {SomegaR0,    {vR,1/Sqrt[2]},  {sigmaOmR,I/Sqrt[2]}, {phiOmR,1/Sqrt[2]}},
  {SdeltaR0,    {vBL,1/Sqrt[2]}, {sigmaR,I/Sqrt[2]},   {phiR,1/Sqrt[2]}},
  {SdeltaRbar0, {vBL,1/Sqrt[2]}, {sigmaRbar,I/Sqrt[2]},{phiRbar,1/Sqrt[2]}}};
\end{lstlisting}
Finally, we need the rotations in the matter sector to the new mass eigenstates
\begin{lstlisting}
DEFINITION[RSB][MatterSector]= 
{    {{SdeltaRm, conj[SdeltaRbarp]},             {Hpm1R1,ZC1}},
     {{SomegaRm, conj[SomegaRp]},                {Hpm2R1,ZC2}},
     {{fB,fWR0,FdeltaR0, FdeltaRbar0, FomegaR0}, {L0, ZN}},
     {{{fWRm, FomegaRm}, {fWRp, FomegaRp}},      {{Lm,UM}, {Lp,UP}}},
     {{phiR, phiRbar, phiOmR},                   {hhR2, ZH}},
     {{sigmaR, sigmaRbar, sigmaOmR},             {AhR2, ZP}},
     {{FvL, conj[FvR]},                          {N0, Znu}},
     {{SHd0,conj[SHu0]},                         {SH0r1,UH0}},
     {{SHdm,conj[SHup]},                         {SHCr1,UHC}}, 
     {{SomegaLm,conj[SomegaLp]},                 {SO1r1,UO1}},
     {{SdeltaLp,conj[SdeltaLbarm]},              {SDLpR1,UDLp}},
     {{SdeltaLpp,conj[SdeltaLbarmm]},            {SDLppR1,UDLpp}},
     {{SdeltaL0,conj[SdeltaLbar0]},              {SDL0r1,UDL0}},
     {{SdeltaRmm,conj[SdeltaRbarpp]},            {SDRmmR1,UDRmm}},
     {{SdeltaR0,conj[SdeltaRbar0]},              {SDR0r1,UDR0}}
      };
\end{lstlisting}

\subsubsection*{Below $SU(2)_R \times U(1)_{B-L}$ breaking scale} 
Here, we have just the particles and gauge groups of the MSSM. Therefore, the
model file nearly the same in the one presented in sec.~\ref{app:MSSM_NMSSM}.
We point only out the difference:
\begin{itemize}
 \item The superpotential contains the Weinberg operator
 \begin{lstlisting}
SuperPotential = {...,{{1,WOp},{l,Hu,l,Hu}}  }; 
\end{lstlisting}
 \item The neutrinos are massive and mix among each other
\begin{lstlisting}                                     
DEFINITION[EWSB][MatterSector]= { ...,{{FvL}, {FV, UV}}}; 
\end{lstlisting}
 \item These states form Majorana spinors
\begin{lstlisting}
dirac[[4]] = {Fv,  FV, conj[FV]};
\end{lstlisting}
\end{itemize}
\subsection{Model files for \SPheno output}
We need for both scales the corresponding {\tt SPheno.m} files for \\SARAH
to create the Fortran source code. While the first one is rather short, the
second one includes all necessary boundary conditions.
\subsubsection*{Above $SU(2)_R \times U(1)_{B-L}$ breaking scale} 
First, it is necessary to tell \\SARAH the number of the regime (counted from
GUT to low scale) and that it is an intermediate scale.
\begin{lstlisting}
RegimeNr = 1;
IntermediateScale = True; 
\end{lstlisting}
Afterwards, we give a list with all particles which are integrated out at the
threshold scale after gauge symmetry breaking.
\begin{lstlisting}
HeavyFields = {Hpm1R1, ChiR1, Cha1r1, hhR1, AhR1,
               FvR1, SVRr1, SH0r1[3], SHCr1[3],
               SO1r1, SDLpR1, SDLppR1,
               SDL0r1, SDRmmR1, DR3r1,
               DL1r1, DL2r1, DL3r1, H0r1, HCr1};
\end{lstlisting}
The number in bracket coming with {\tt SH0r1} and {\tt SHCr1} means  that only
the third generation and above is integrated out. \\
To calculate the finite shifts of the gauge couplings and gaugino masses, it is
necessary to define the gauge sector of the next scale {\tt NextGauge} as well
as the quantum number of the fields which are integrated out with respect to
that gauge groups {\tt NextQN}.
\begin{lstlisting}
NextGauge=           {U[1], SU[2], SU[3]};
NextQN = { {Hpm1R1,     -1,     1,    1},
           {ChiR1,       0,     1,    1},
	   {Cha1r1,     -1,     1,    1},
           {hhR1,        0,     1,    1},
           {AhR1,        0,     1,    1},
           {FvR1,        0,     1,    1},
           {SVRr1,       0,     1,    1},
	   {SH0r1,    -1/2,     1,    1},
	   {SHCr1,     1/2,     2,    1},
	   {SO1r1,       0,     1,    1},
	   {SDLpR1,      1,     1,    1},
	   {SDLppR1,     2,     1,    1},
	   {SDL0r1,      1,     3,    1},
	   {SDRmmR1,    -2,     1,    1},
	   {DR3r1,      -2,     1,    1},
	   {DL1r1,       1,     1,    1},
	   {DL2r1,       1,     1,    1},
	   {DL3r1,       1,     3,    1},
	   {H0r1,     -1/2,     1,    1},
	   {HCr1,      1/2,     2,    1}
};
\end{lstlisting}
Finally, it is necessary to define the information about the vacuum conditions.
We have here two different VEVs and therefore need parameter which should be
fixed by the tadpole equations. Since, there is no closed analytical solution
for that, we give an approximated expression by neglecting the soft-breaking
terms.
\begin{lstlisting}
ParametersToSolveTadpoles = {Mdelta, Momega};
UseGivenTadpoleSolution = True;

SubSolutionsTadpolesTree={
  Mdelta -> - SignumMdelta ac1 vR/Sqrt[2],
  Momega -> - SignumMomega ac1 vBL^2/(2 Sqrt[2] vR)  
};
SubSolutionsTadpolesLoop={}; 
\end{lstlisting}

\subsubsection*{Below $SU(2)_R \times U(1)_{B-L}$ breaking scale} 
The second scale is not an intermediate scale. Hence,
\begin{lstlisting}
RegimeNr = 2;
IntermediateScale = False; 
\end{lstlisting}
We make the following choice of free parameters of that model: in addition to
the standard mSugra parameters (\(m_0,M_{1/2},A_0,\tan\beta,\text{sign}\mu\)),
we need input values for \(B_0\), the superpotential parameter \(a\), the signs
of \(M_\Omega\) and \(M_\Delta\), the two VEVs \(v_R\) and \(v_{BL}\) as well as
the threshold scale. These are defined in the blocks {\tt MINPAR} and {\tt
EXTPAR}.
\begin{lstlisting}
MINPAR= {
         {1,  m0},
         {2,  m12},
         {3,  TanBeta},
         {4,  SignumMu},
         {5,  Azero},
         {6,  Bzero},
         {7,  SignumMomega},
         {8,  SignumMdelta},
         {9,  aInput}};

EXTPAR = {
          {100, vRinput},
	  {101, vBLinput},
	  {200, TScale}};
\end{lstlisting}
As in the usual MSSM we solve the tadpole equations for \(v_d\) and \(v_u\)
with respect to \(\mu\) and \(B_\mu\). Furthermore, we use also the some
definition for the SUSY scale. 
\begin{lstlisting}
ParametersToSolveTadpoles = {\[Mu],B[\[Mu]]}; 

RenormalizationScaleFirstGuess = m0^2 + 4 m12^2;
RenormalizationScale = MSu[1]*MSu[6];
\end{lstlisting}
At the GUT scale we use boundary conditions motivated by minimal supergravity:
all scalar soft-breaking masses are proportional to \(m_0\), the gaugino masses
are proportional to \(M_{1/2}\), the trilinear soft-breaking couplings are
given by the corresponding superpotential parameter times \(A_0\) and the
bilinear soft-breaking couplings are \(B_0\) times the superpotential
parameter. In addition, the values for the coupling matrices \(f\),
\(\alpha\)  as well as for \(\mu\) are also read from the LesHouches input file.
\begin{lstlisting}
BoundaryHighScale={
   {T[YQ],       Azero*YQ},
   {T[YL],       Azero*YL},
   {f,           LHInput[f]},
   {T[f],        Azero*LHInput[fm]},
   {AlphaOm,     LHInput[AlphaOm]},
   {T[AlphaOm],  Azero*LHInput[AlphaOm]},
   {T[a],        Azero*aInput},
   {B[Mdelta],   Bzero*Mdelta},
   {B[Momega],   Bzero*Momega},
   {B[Mu3],      Bzero*LHInput[Mu3]},
   {mqL2,        DIAGONAL m0^2},
   {mqR2,        DIAGONAL m0^2},
   {mlL2,        DIAGONAL m0^2},
   {mlR2,        DIAGONAL m0^2},
   {mPhi2,       DIAGONAL m0^2},
   {mdeltaL2,    m0^2},
   {mdeltaLbar2, m0^2},
   {mdeltaR2,    m0^2},
   {mdeltaRbar2, m0^2},
   {momegaL2,    m0^2},
   {momegaR2,    m0^2},
   {MassB,       m12},
   {MassWL,      m12},
   {MassG,       m12}
}; 
\end{lstlisting}
To glue the both regimes, we need to define the appropriate boundary
conditions. First, we initialize the arrays
\begin{lstlisting}
ThresholdScales = {TSCALE};

BoundaryConditionsUp = Table[{},{Length[ThresholdScales]}];
BoundaryConditionsDown = Table[{},{Length[ThresholdScales]}]; 
\end{lstlisting}
and use afterwards the equations
eqs.~(\ref{eq:boundary_LR_1})-(\ref{eq:boundary_LR_2}). In order to keep
the code short, we define
\begin{lstlisting}
  ST1  = Sin[Theta1];
  CT1  = Cos[Theta1];
  ST2  = Sin[Theta2];
  CT2  = Cos[Theta2];
  ST21 = Sin[Theta2-Theta1];
  CT21 = Cos[Theta2-Theta1];
\end{lstlisting}
Using these abbreviation, the boundary conditions can be written as
\begin{lstlisting}
BoundaryConditionsUp[[1]] = { 
 {YQ[index1,index2,1],   (Yu[index1,index2] ST1 + Yd[index1,index2]ST2)/ST21 },
 {YQ[index1,index2,2],   (Yu[index1,index2] CT1 + Yd[index1,index2]CT2)/ST21 },
 {YL[index1,index2,1],   (Yv[index1,index2] ST1 + Ye[index1,index2]ST2)/ST21 },
 {YL[index1,index2,2],   (Yv[index1,index2] CT1 + Ye[index1,index2]CT2)/ST21 },
 {gBL,                    Sqrt[2] g1 g2 /Sqrt[5 g2^2 -3 g1^2]},
 {Yv,                      LHInput[Yv]}
};

BoundaryConditionsDown[[1]] = {
 {vR,           vRinput}, 
 {vBL,          vBLinput},
 {a,            aInput},
 {Theta1,       ArcTan[RealPart[((vR*AlphaOm[1,2])/2 + Mu3[1,2])/Mu3[2,2]]]},
 {Theta2,       ArcTan[RealPart[(-(vR*AlphaOm[1,2])/2 + Mu3[1,2])/Mu3[2,2]]]},
 {g1,           Sqrt[5] g2 gBL/Sqrt[2 g2^2 + 3 gBL^2]},
 {MassB,        (2 g2^2 MassB + 3 gBL^2 MassWL)/(2 g2^2 + 3 gBL^2)},
 {MassWB,       MassWL},
 {Yd[index1,index2], YQ[index1,index2,1] CT1 - YQ[index1,index2,2] ST1},
 {Yu[index1,index2], - YQ[index1,index2,1] CT2 + YQ[index1,index2,2] ST2},
 {Ye[index1,index2], YL[index1,index2,1] CT1 - YL[index1,index2,2] ST1},
 {Yv[index1,index2], - YL[index1,index2,1] CT2 + YL[index1,index2,2] ST2},
 {T[Yd][index1,index2], T[YQ][index1,index2,1] CT1 -T[YQ][index1,index2,2]ST1},
 {T[Yu][index1,index2], -T[YQ][index1,index2,1] CT2+T[YQ][index1,index2,2] ST2},
 {T[Ye][index1,index2], T[YL][index1,index2,1] CT1 -T[YL][index1,index2,2] ST1},
 {mu2,          mqR2},
 {md2,          mqR2},
 {mq2,          mqR2},
 {me2,          mlR2},
 {ml2,          mlR2},
 {mHd2, CT1^2 mPhi2[1,1] + ST1^2 mPhi2[2,2] - ST1 CT1(mPhi2[1,2] + mPhi2[2,1])},
 {mHu2, CT2^2 mPhi2[1,1] + ST2^2 mPhi2[2,2] - ST2 CT2(mPhi2[1,2] + mPhi2[2,1])},
 {WOp,          MatMul2[MatMul2[Yv,InverseMatrix[f],FortranFalse],Transpose[Yv],
                   FortranFalse]/vR}
 };
\end{lstlisting}
Note, \(Y_\nu\) does not appear in one of the model files. Therefore, it is
necessary to define the dimension of that matrix by hand
\begin{lstlisting}
AdditionalVariablesSPheno={Yv[3,3]};
\end{lstlisting}
There are some parameters involved which must be real values. It might be also
helpful to choose some initialization values for some parameters to stabilize
the numerics in the first iteration
\begin{lstlisting}
RealParameters = {TanBeta, vRinput,vBLinput,Theta1,Theta2,TScale};

InitializationValues = {
 {Mu3IN[1,1], (Mu3IN[1,2]^2 - AlphaOmIN[1,2]^2 vRInput^2/4)/Mu3IN[2,2]},
 {Theta1,   ArcTan[RealPart[-(Mu3IN[1,2]+AlphaOmIN[1,2]vRInput/2)/Mu3IN[2,2]]]},
 {Theta2,   ArcTan[RealPart[(Mu3IN[1,2]-AlphaOmIN[1,2]vRInput/2)/Mu3IN[2,2]]]},
 {Mdelta,   aInput*SignumMdelta*vRinput/2 },
 {Momega,   SignumMomega*(aInput^2*vBLinput^2)/(8 Mdelta)}
}; 
\end{lstlisting}

\chapter{The minimal supersymmetric standard model}
\label{sec:MSSM}
\section{Conventions}
\subsection{Vector Superfields} 
\begin{center} 
\begin{tabular}{|c|c|c|c|c|c|} 
\hline \hline 
SF & Spin \(\frac{1}{2}\) & Spin 1 & \(SU(N)\) & Coupling & Name \\ 
 \hline 
\(\hat{B}\) & \(\lambda_{\tilde{B}}\) & \(B\) & \(U(1)\) & \(g_1\)
&\text{hypercharge}\\ 
\(\hat{W}\) & \(\lambda_{\tilde{W}}\) & \(W\) & \(\text{SU}(2)\) & \(g_2\)
&\text{left}\\ 
\(\hat{g}\) & \(\lambda_{\tilde{g}}\) & \(g\) & \(\text{SU}(3)\) & \(g_3\)
&\text{color}\\ 
\hline \hline
\end{tabular} 
\end{center} 
\subsection{Chiral Superfields} 
\begin{center} 
\begin{tabular}{|c|c|c|c|c|c|} 
\hline \hline 
SF & Spin 0 & Spin \(\frac{1}{2}\) & Generations & \((U(1)\otimes\,
\text{SU}(2)\otimes\, \text{SU}(3))\) \\ 
\hline 
\(\hat{q}\) & \(\tilde{q}\) & \(q\) & 3 & \((\frac{1}{6},{\bf 2},{\bf 3}) \) \\ 
\(\hat{l}\) & \(\tilde{l}\) & \(l\) & 3 & \((-\frac{1}{2},{\bf 2},{\bf 1}) \)
\\ 
\(\hat{H}_d\) & \(H_d\) & \(\tilde{H}_d\) & 1 & \((-\frac{1}{2},{\bf 2},{\bf 1})
\) \\ 
\(\hat{H}_u\) & \(H_u\) & \(\tilde{H}_u\) & 1 & \((\frac{1}{2},{\bf 2},{\bf 1})
\) \\ 
\(\hat{d}\) & \(\tilde{d}_R^*\) & \(d_R^*\) & 3 & \((\frac{1}{3},{\bf 1},{\bf
\overline{3}}) \) \\ 
\(\hat{u}\) & \(\tilde{u}_R^*\) & \(u_R^*\) & 3 & \((-\frac{2}{3},{\bf 1},{\bf
\overline{3}}) \) \\ 
\(\hat{e}\) & \(\tilde{e}_R^*\) & \(e_R^*\) & 3 & \((1,{\bf 1},{\bf 1}) \) \\ 
\hline \hline
\end{tabular} 
\end{center} 
\subsection{Superpotential and Lagrangian} 
\paragraph{Superpotential} 
\begin{align} 
W = & \,  Y_u\,\hat{u}\,\hat{q}\,\hat{H}_u\,- Y_d
\,\hat{d}\,\hat{q}\,\hat{H}_d\,- Y_e
\,\hat{e}\,\hat{l}\,\hat{H}_d\,+\mu\,\hat{H}_u\,\hat{H}_d\,\end{align} 
\paragraph{Softbreaking terms} 
\begin{align} 
- L_{SB,W} = \, & - H_d^0 H_u^0 B_{\mu} +H_d^- H_u^+ B_{\mu} +H_d^0
\tilde{d}^*_{R,{i \alpha}} \delta_{\alpha\beta} \tilde{d}_{L,{j \beta}} T_{d,{i
j}} - H_d^- \tilde{d}^*_{R,{i \alpha}} \delta_{\alpha\beta} \tilde{u}_{L,{j
\beta}} T_{d,{i j}} \nonumber \\ 
 &+H_d^0 \tilde{e}^*_{R,{i}} \tilde{e}_{L,{j}} T_{e,{i j}} - H_d^-
\tilde{e}^*_{R,{i}} \tilde{\nu}_{L,{j}} T_{e,{i j}} - H_u^+ \tilde{u}^*_{R,{i
\alpha}} \delta_{\alpha\beta} \tilde{d}_{L,{j \beta}} T_{u,{i j}} +H_u^0
\tilde{u}^*_{R,{i \alpha}} \delta_{\alpha\beta} \tilde{u}_{L,{j \beta}} T_{u,{i
j}} + \mbox{h.c.} \\ 
- L_{SB,\phi} = \, & +m_{H_d}^2 |H_d^0|^2 +m_{H_d}^2 |H_d^-|^2 +m_{H_u}^2
|H_u^0|^2 +m_{H_u}^2 |H_u^+|^2 +\tilde{d}^*_{L,{j \beta}} \delta_{\alpha\beta}
m_{q,{i j}}^{2} \tilde{d}_{L,{i \alpha}} \nonumber \\ 
 &+\tilde{d}^*_{R,{i \alpha}} \delta_{\alpha\beta} m_{d,{i j}}^{2}
\tilde{d}_{R,{j \beta}} +\tilde{e}^*_{L,{j}} m_{l,{i j}}^{2} \tilde{e}_{L,{i}}
+\tilde{e}^*_{R,{i}} m_{e,{i j}}^{2} \tilde{e}_{R,{j}} +\tilde{u}^*_{L,{j
\beta}} \delta_{\alpha\beta} m_{q,{i j}}^{2} \tilde{u}_{L,{i \alpha}} \nonumber
\\ 
 &+\tilde{u}^*_{R,{i \alpha}} \delta_{\alpha\beta} m_{u,{i j}}^{2}
\tilde{u}_{R,{j \beta}} +\tilde{\nu}^*_{L,{j}} m_{l,{i j}}^{2}
\tilde{\nu}_{L,{i}} \\ 
- L_{SB,\lambda} = \, & \frac{1}{2} \left( \lambda_{\tilde{B}}^{2} M_1  + M_2
\lambda_{{\tilde{W}},{i}}^{2}  + M_3 \lambda_{{\tilde{g}},{i}}^{2} + \mbox{h.c.}
\right) 
\end{align} 
\subsection{Gauge fixing terms} 
\paragraph{Gauge fixing terms for gauge eigenstates } 
\begin{align} 
L_{GF} = \, &-\frac{1}{2 \xi_{G}} \partial_{\mu}g_\alpha   -\frac{1}{2 \xi_{W}}
\partial_{\mu}W^i 
\end{align} 
\paragraph{Gauge fixing terms for mass eigenstates after EWSB } 
\begin{align} 
L_{GF} = \, &-\frac{1}{2 \xi_{P}} \partial_{\mu}\gamma  -\frac{1}{2 \xi_{G}}
\partial_{\mu}g_\alpha   -\frac{1}{2 \xi_{Z}}  \Big(- A^0_{{1}} m_{Z} \xi_{Z}  +
\partial_{\mu}Z\Big) - \frac{1}{\xi_{W}} \Big(i H^-_{{1}} m_{W^-} \xi_{W}  +
\partial_{\mu}W^-\Big)
\end{align} 
\subsection{Vacuum expectation values}
\label{sec:VEVsMSSM}
\begin{align} 
H_d^0 =  \, \frac{1}{\sqrt{2}} \left( \phi_{d}  + i \sigma_{d}  +  v_d  \right)
\, , \hspace{1cm} 
H_u^0 =  \, \frac{1}{\sqrt{2}} \left( \phi_{u}    + i  \sigma_{u}  +  v_u
\right) 
\end{align}

\subsection{Rotations of vector bosons and gauginos after EWSB} 
\label{sec:RotGaugeMSSM}
\begin{align} 
W^-_{{1 \rho}} = & \,\frac{1}{\sqrt{2}} W^-_{{\rho}}  + \frac{1}{\sqrt{2}}
W^+_{{\rho}} \, , \hspace{1cm} 
W^-_{{2 \rho}} =  \,-i \frac{1}{\sqrt{2}} W^-_{{\rho}}  + i \frac{1}{\sqrt{2}}
W^+_{{\rho}} \\ 
W^-_{{3 \rho}} = & \,\cos\Theta_W  Z_{{\rho}}  + \sin\Theta_W  \gamma_{{\rho}}
\, , \hspace{1cm}  
B_{{\rho}} =  \,\cos\Theta_W  \gamma_{{\rho}}  - \sin\Theta_W  Z_{{\rho}} \\ 
\lambda_{{\tilde{W}},{1}} = & \,\frac{1}{\sqrt{2}} \tilde{W}^-  +
\frac{1}{\sqrt{2}} \tilde{W}^+ \, , \hspace{1cm}  
\lambda_{{\tilde{W}},{2}} =  \,-i \frac{1}{\sqrt{2}} \tilde{W}^-  + i
\frac{1}{\sqrt{2}} \tilde{W}^+ \, , \hspace{1cm}  
\lambda_{{\tilde{W}},{3}} =  \,\tilde{W}^0
\end{align} 

\subsection{Rotations in matter sector to mass eigenstates after EWSB}
\label{sec:RotMatterMSSM}
In the following, Greek letters \(\alpha_i,\beta_i\) refer to color indices and
\(o_i, p_i\) to generations indices.  
\begin{enumerate} 
\item {\bf Mass matrix for neutralinos}, basis: \( \left(\lambda_{\tilde{B}},
\tilde{W}^0, \tilde{H}_d^0, \tilde{H}_u^0\right)\)
\begin{equation} 
m_{\tilde{\chi}^0} = \left( 
\begin{array}{cccc}
M_1 &0 &-\frac{1}{2} g_1 v_d  &\frac{1}{2} g_1 v_u \\ 
0 &M_2 &\frac{1}{2} g_2 v_d  &-\frac{1}{2} g_2 v_u \\ 
-\frac{1}{2} g_1 v_d  &\frac{1}{2} g_2 v_d  &0 &- \mu \\ 
\frac{1}{2} g_1 v_u  &-\frac{1}{2} g_2 v_u  &- \mu  &0\end{array} 
\right) 
\end{equation} 
This matrix is diagonalized by \(N\): 
\begin{equation} 
N m_{\tilde{\chi}^0} N^{\dagger} = m^{dia}_{\tilde{\chi}^0} 
\end{equation} 
with 
\begin{align} 
\lambda_{\tilde{B}} = \sum_j N^*_{j 1}\lambda^0_{{j}}\,, \hspace{1cm} 
\tilde{W}^0 = \sum_j N^*_{j 2}\lambda^0_{{j}} \,, \hspace{1cm}  
\tilde{H}_d^0 = \sum_j N^*_{j 3}\lambda^0_{{j}}\,, \hspace{1cm}  
\tilde{H}_u^0 = \sum_j N^*_{j 4}\lambda^0_{{j}}
\end{align} 
\item {\bf Mass matrix for charginos}, basis: \( \left(\tilde{W}^-,
\tilde{H}_d^-\right)/\left(\tilde{W}^+, \tilde{H}_u^+\right) \) 
 
\begin{equation} 
m_{\tilde{\chi}^-} = \left( 
\begin{array}{cc}
M_2 &\frac{1}{\sqrt{2}} g_2 v_u \\ 
\frac{1}{\sqrt{2}} g_2 v_d  &\mu\end{array} 
\right) 
\end{equation} 
This matrix is diagonalized by \(U\) and \(V\) 
\begin{equation} 
U^* m_{\tilde{\chi}^-} V^{\dagger} = m^{dia}_{\tilde{\chi}^-} 
\end{equation} 
with 
\begin{align} 
\tilde{W}^- = \sum_j U^*_{j 1}\lambda^-_{{j}}\,, \hspace{1cm} 
\tilde{H}_d^- = \sum_j U^*_{j 2}\lambda^-_{{j}} \,, \hspace{1cm}  
\tilde{W}^+ = \sum_j V^*_{1 j}\lambda^+_{{j}}\,, \hspace{1cm} 
\tilde{H}_u^+ = \sum_j V^*_{2 j}\lambda^+_{{j}}
\end{align} 
\item {\bf Mass matrix for leptons}, basis: \(
\left(e_{L,{{o_1}}}\right)/\left(e^*_{R,{{p_1}}}\right) \) 
 
\begin{equation} 
m_{e} = \left( 
\begin{array}{c}
\frac{1}{\sqrt{2}} v_d Y_{e,{{p_1} {o_1}}} \end{array} 
\right) 
\end{equation} 
This matrix is diagonalized by \(U^e_L\) and \(U^e_R\) 
\begin{equation} 
U^{e,*}_L m_{e} U_{R}^{e,\dagger} = m^{dia}_{e} 
\end{equation} 
with 
\begin{align} 
e_{L,{i}} = \sum_j U^{e,*}_{L,{j i}}E_{L,{j}}\,, \hspace{1cm} 
e_{R,{i}} = \sum_j U_{R,{i j}}^{e}E^*_{R,{j}}
\end{align} 
\item {\bf Mass matrix for down-quarks}, basis: \( \left(d_{L,{{o_1}
{\alpha_1}}}\right)/\left(d^*_{R,{{p_1} {\beta_1}}}\right) \) 
 
\begin{equation} 
m_{d} = \left( 
\begin{array}{c}
\frac{1}{\sqrt{2}} v_d \delta_{{\alpha_1}{\beta_1}} Y_{d,{{p_1} {o_1}}}
\end{array} 
\right) 
\end{equation} 
This matrix is diagonalized by \(U^d_L\) and \(U^d_R\) 
\begin{equation} 
U^{d,*}_L m_{d} U_{R}^{d,\dagger} = m^{dia}_{d} 
\end{equation} 
with 
\begin{align} 
d_{L,{i \alpha}} = \sum_{t_2}U^{d,*}_{L,{j i}}D_{L,{j \alpha}}\,, \hspace{1cm}  
d_{R,{i \alpha}} = \sum_{t_2}U_{R,{i j}}^{d}D^*_{R,{j \alpha}}
\end{align} 
\item {\bf Mass matrix for up-quarks}, basis: \( \left(u_{L,{{o_1}
{\alpha_1}}}\right)/\left(u^*_{R,{{p_1} {\beta_1}}}\right) \) 
 
\begin{equation} 
m_{u} = \left( 
\begin{array}{c}
\frac{1}{\sqrt{2}} v_u \delta_{{\alpha_1}{\beta_1}} Y_{u,{{p_1} {o_1}}}
\end{array} 
\right) 
\end{equation} 
This matrix is diagonalized by \(U^u_L\) and \(U^u_R\) 
\begin{equation} 
U^{u,*}_L m_{u} U_{R}^{u,\dagger} = m^{dia}_{u} 
\end{equation} 
with 
\begin{align} 
u_{L,{i \alpha}} = \sum_{t_2}U^{u,*}_{L,{j i}}U_{L,{j \alpha}}\,, \hspace{1cm}  
u_{R,{i \alpha}} = \sum_{t_2}U_{R,{i j}}^{u}U^*_{R,{j \alpha}}
\end{align} 
\item {\bf Mass matrix for down-squarks}, basis: \( \left(\tilde{d}_{L,{{o_1}
{\alpha_1}}}/\tilde{d}_{R,{{o_2} {\alpha_2}}}\right),
\left(\tilde{d}^*_{L,{{p_1} {\beta_1}}}, \tilde{d}^*_{R,{{p_2}
{\beta_2}}}\right) \) 
\begin{align} 
m_{11} &= \frac{1}{24} \delta_{{\alpha_1}{\beta_1}} \Big(12 \Big(2 m_{q,{{o_1}
{p_1}}}^{2}  + v_{d}^{2} \sum_{a=1}^{3}Y^*_{d,{a {p_1}}} Y_{d,{a {o_1}}}  \Big)
- \Big(3 g_{2}^{2}  + g_{1}^{2}\Big)\Big(- v_{u}^{2}  +
v_{d}^{2}\Big)\delta_{{o_1}{p_1}} \Big)\\
m_{12} &=  \frac{1}{\sqrt{2}} \delta_{{\alpha_1}{\beta_2}} \Big(v_d T_{d,{{p_2}
{o_1}}}  - v_u \mu^* Y_{d,{{p_2} {o_1}}} \Big) \\
m_{22} &= \frac{1}{12} \delta_{{\alpha_2}{\beta_2}} \Big(6 \Big(2 m_{d,{{p_2}
{o_2}}}^{2}  + v_{d}^{2} \sum_{a=1}^{3}Y^*_{d,{{o_2} a}} Y_{d,{{p_2} a}}  \Big)
+ g_{1}^{2} \Big(- v_{d}^{2}  + v_{u}^{2}\Big)\delta_{{o_2}{p_2}} \Big)
\end{align} 
This matrix is diagonalized by \(Z^D\): 
\begin{equation} 
Z^D m^2_{\tilde{d}} Z^{D,\dagger} = m^{dia}_{2,\tilde{d}} 
\end{equation} 
with 
\begin{align} 
\tilde{d}_{L,{i \alpha}} = \sum_{t_2}Z^{D,*}_{j i}\tilde{d}_{{j \alpha}}\,,
\hspace{1cm} 
\tilde{d}_{R,{i \alpha}} = \sum_{t_2}Z^{D,*}_{j i}\tilde{d}_{{j \alpha}}
\end{align} 
\item {\bf Mass matrix for sneutrinos}, basis: \(
\left(\tilde{\nu}_{L,{{o_1}}}\right)/ \left(\tilde{\nu}^*_{L,{{p_1}}}\right) \) 
 
\begin{equation} 
m^2_{\tilde{\nu}} = \left( 
\begin{array}{c}
\frac{1}{8} \Big(8 m_{l,{{o_1} {p_1}}}^{2}  + \Big(g_{1}^{2} +
g_{2}^{2}\Big)\Big(- v_{u}^{2}  + v_{d}^{2}\Big)\delta_{{o_1}{p_1}}
\Big)\end{array} 
\right) 
\end{equation} 
This matrix is diagonalized by \(Z^V\): 
\begin{equation} 
Z^V m^2_{\tilde{\nu}} Z^{V,\dagger} = m^{dia}_{2,\tilde{\nu}} 
\end{equation} 
with 
\begin{align} 
\tilde{\nu}_{L,{i}} = \sum_{t_2}Z^{V,*}_{j i}\tilde{\nu}_{{j}}
\end{align} 
\item {\bf Mass matrix for up-squarks}, basis: \( \left(\tilde{u}_{L,{{o_1}
{\alpha_1}}}, \tilde{u}_{R,{{o_2} {\alpha_2}}}\right)/
\left(\tilde{u}^*_{L,{{p_1} {\beta_1}}}, \tilde{u}^*_{R,{{p_2}
{\beta_2}}}\right) \) 
\begin{align} 
m_{11} &= \frac{1}{24} \delta_{{\alpha_1}{\beta_1}} \Big(12 \Big(2 m_{q,{{o_1}
{p_1}}}^{2}  + v_{u}^{2} \sum_{a=1}^{3}Y^*_{u,{a {p_1}}} Y_{u,{a {o_1}}}  \Big)
- \Big(-3 g_{2}^{2}  + g_{1}^{2}\Big)\Big(- v_{u}^{2}  +
v_{d}^{2}\Big)\delta_{{o_1}{p_1}} \Big)\\ 
m_{12} &= \frac{1}{\sqrt{2}} \delta_{{\alpha_1}{\beta_2}} \Big(- v_d \mu^*
Y_{u,{{p_2} {o_1}}}  + v_u T_{u,{{p_2} {o_1}}} \Big) \\
m_{22} &= \frac{1}{6} \delta_{{\alpha_2}{\beta_2}} \Big(3 v_{u}^{2}
\sum_{a=1}^{3}Y^*_{u,{{o_2} a}} Y_{u,{{p_2} a}}   + 6 m_{u,{{p_2} {o_2}}}^{2}  +
g_{1}^{2} \Big(- v_{u}^{2}  + v_{d}^{2}\Big)\delta_{{o_2}{p_2}} \Big)
\end{align} 
This matrix is diagonalized by \(Z^U\): 
\begin{equation} 
Z^U m^2_{\tilde{u}} Z^{U,\dagger} = m^{dia}_{2,\tilde{u}} 
\end{equation} 
with 
\begin{align} 
\tilde{u}_{L,{i \alpha}} = \sum_{t_2}Z^{U,*}_{j i}\tilde{u}_{{j \alpha}}\,,
\hspace{1cm} 
\tilde{u}_{R,{i \alpha}} = \sum_{t_2}Z^{U,*}_{j i}\tilde{u}_{{j \alpha}}
\end{align} 
\item {\bf Mass matrix for sleptons}, basis: \( \left(\tilde{e}_{L,{{o_1}}},
\tilde{e}_{R,{{o_2}}}\right)/\left(\tilde{e}^*_{L,{{p_1}}},
\tilde{e}^*_{R,{{p_2}}}\right) \) 
\begin{align} 
m_{11} &= \frac{1}{8} \Big(4 v_{d}^{2} \sum_{a=1}^{3}Y^*_{e,{a {p_1}}} Y_{e,{a
{o_1}}}   + 8 m_{l,{{o_1} {p_1}}}^{2}  + \Big(- g_{2}^{2}  +
g_{1}^{2}\Big)\Big(- v_{u}^{2}  + v_{d}^{2}\Big)\delta_{{o_1}{p_1}} \Big)\\ 
m_{12} &= \frac{1}{\sqrt{2}} \Big(v_d T_{e,{{p_2} {o_1}}}  - v_u \mu^*
Y_{e,{{p_2} {o_1}}} \Big) \\
m_{22} &= \frac{1}{4} \Big(2 v_{d}^{2} \sum_{a=1}^{3}Y^*_{e,{{o_2} a}}
Y_{e,{{p_2} a}}   + 4 m_{e,{{p_2} {o_2}}}^{2}  + g_{1}^{2} \Big(- v_{d}^{2}  +
v_{u}^{2}\Big)\delta_{{o_2}{p_2}} \Big)
\end{align} 
This matrix is diagonalized by \(Z^E\): 
\begin{equation} 
Z^E m^2_{\tilde{e}} Z^{E,\dagger} = m^{dia}_{2,\tilde{e}} 
\end{equation} 
with 
\begin{align} 
\tilde{e}_{L,{i}} = \sum_{t_2}Z^{E,*}_{j i}\tilde{e}_{{j}}\,, \hspace{1cm} 
\tilde{e}_{R,{i}} = \sum_{t_2}Z^{E,*}_{j i}\tilde{e}_{{j}}
\end{align} 
\item {\bf Mass matrix for scalar Higgs}, basis: \( \left(\phi_{d},
\phi_{u}\right)\)
\begin{equation} 
m^2_{h} = \left( 
\begin{array}{cc}
m_{H_d}^2  +  |\mu|^2  +\frac{1}{8} \Big(g_{1}^{2} + g_{2}^{2}\Big)\Big(3
v_{d}^{2}  - v_{u}^{2} \Big) & -{\Re\Big(B_{\mu}\Big)}  -
\frac{1}{4}\Big(g_{1}^{2} + g_{2}^{2}\Big)v_d v_u \\ 
-{\Re\Big(B_{\mu}\Big)}  - \frac{1}{4}\Big(g_{1}^{2} + g_{2}^{2}\Big)v_d v_u
&m_{H_u}^2  +  |\mu|^2  - \frac{1}{8}\Big(g_{1}^{2} + g_{2}^{2}\Big)\Big(-3
v_{u}^{2}  + v_{d}^{2}\Big)\end{array} 
\right) 
\end{equation} 
This matrix is diagonalized by \(Z^H\): 
\begin{equation} 
Z^H m^2_{h} Z^{H,\dagger} = m^{dia}_{2,h} 
\end{equation} 
with 
\begin{align} 
\phi_{d} = \sum_{t_2}Z_{{j 1}}^{H}h_{{j}}\,, \hspace{1cm} 
\phi_{u} = \sum_{t_2}Z_{{j 2}}^{H}h_{{j}}
\end{align} 
The mixing matrix can be parametrized by 
\begin{equation} 
Z^H= \, \left( 
\begin{array}{cc} 
- \sin\alpha   & \cos\alpha  \\ 
 \cos\alpha  & \sin\alpha \end{array} 
\right) 
\end{equation} 
\item {\bf Mass matrix for pseudo scalar Higgs}, basis: \( \left(\sigma_{d},
\sigma_{u}\right)\) 
\begin{equation} 
m^2_{A^0} = \left( 
\begin{array}{cc}
m_{H_d}^2  +  |\mu|^2  + \frac{1}{8}  \Big(g_{1}^{2} + g_{2}^{2}\Big)\Big(-
v_{u}^{2}  + v_{d}^{2}\Big) &{\Re\Big(B_{\mu}\Big)}\\ 
{\Re\Big(B_{\mu}\Big)} & m_{H_u}^2  +  |\mu|^2  -\frac{1}{8}  \Big(g_{1}^{2} +
g_{2}^{2}\Big)\Big(- v_{u}^{2}  + v_{d}^{2}\Big)\end{array} 
\right) 
\end{equation} 
This matrix is diagonalized by \(Z^A\): 
\begin{equation} 
Z^A m^2_{A^0} Z^{A,\dagger} = m^{dia}_{2,A^0} 
\end{equation} 
with 
\begin{align} 
\sigma_{d} = \sum_{t_2}Z_{{j 1}}^{A}A^0_{{j}}\,, \hspace{1cm} 
\sigma_{u} = \sum_{t_2}Z_{{j 2}}^{A}A^0_{{j}}
\end{align} 
The mixing matrix can be parametrized by 
\begin{equation} 
Z^A= \, \left( 
\begin{array}{cc} 
- \cos\beta   & \sin\beta  \\ 
 \sin\beta  & \cos\beta \end{array} 
\right) 
\end{equation} 
\item {\bf Mass matrix for charged Higgs}, basis: \( \left(H_d^-,
H_u^{+,*}\right)\) 
\begin{equation} 
m^2_{H^-} = \left( 
\begin{array}{cc}
 m_{H_d}^2  +  |\mu|^2  +  \frac{1}{8} \Big(g_{1}^{2} + g_2^2\Big)\Big(v_{d}^{2}
 -  v_{u}^{2}\Big) &\frac{1}{4} g_{2}^{2} v_d v_u  + B_{\mu}\\ 
\frac{1}{4} g_{2}^{2} v_d v_u  + B_{\mu}^* & m_{H_u}^2  + |\mu|^2  + \frac{1}{8}
\Big(g_2^2 - g_{1}^{2}\Big)\Big(v_{d}^{2}  -  v_{u}^{2}\Big)\end{array} 
\right) 
\end{equation} 
This matrix is diagonalized by \(Z^+\): 
\begin{equation} 
Z^+ m^2_{H^-} Z^{+,\dagger} = m^{dia}_{2,H^-} 
\end{equation} 
with 
\begin{align} 
H_d^- = \sum_{t_2}Z^{+,*}_{j 1}H^-_{{j}}\,, \hspace{1cm} 
H_u^+ = \sum_{t_2}Z_{{j 2}}^{+}H^+_{{j}}
\end{align} 
The mixing matrix can be parametrized by 
\begin{equation} 
Z^+= \, \left( 
\begin{array}{cc} 
- \cos\beta   & \sin\beta  \\ 
 \sin\beta  & \cos\beta \end{array} 
\right) 
\end{equation} 
\end{enumerate} 
\subsection{Tadpole equations}
\begin{align} 
\frac{\partial V}{\partial v_d} &= \frac{1}{8} \Big(8 v_d |\mu|^2  -8 v_u
{\Re\Big(B_{\mu}\Big)}  + v_d \Big(8 m_{H_d}^2  + g_{1}^{2} v_{d}^{2}  -
g_{1}^{2} v_{u}^{2}  + g_{2}^{2} v_{d}^{2}  - g_{2}^{2} v_{u}^{2} \Big)\Big)\\ 
\frac{\partial V}{\partial v_u} &= \frac{1}{8} \Big(-8 v_d
{\Re\Big(B_{\mu}\Big)}  + 8 v_u |\mu|^2  + v_u \Big(8 m_{H_u}^2  - g_{1}^{2}
v_{d}^{2}  + g_{1}^{2} v_{u}^{2}  - g_{2}^{2} v_{d}^{2}  + g_{2}^{2} v_{u}^{2}
\Big)\Big)
\end{align}

\section{Implementation in \SARAH}
\label{sec:MSSM_modelfile}
\lstset{frame=shadowbox}
In the following, we showcase the different components of the MSSM
implementation in
\SARAH (see \cite{Staub:2010jh} for a summary of our conventions).
\begin{enumerate}
\item The gauge sector is \(U(1)\times SU(2)\times SU(3)\) and is defined
by declaring the corresponding vector superfields.
\begin{lstlisting}
Gauge[[1]]={B,   U[1], hypercharge, g1, False};
Gauge[[2]]={WB, SU[2], left,        g2, True};
Gauge[[3]]={G,  SU[3], color,       g3, False};
\end{lstlisting}
First, the name of the vector superfield is given. The second entry defines the
dimension of the group, the third one is the name of the gauge group and the
forth one the name of the corresponding gauge coupling. If the last entry is
set to {\tt True}, the sum over the charge induces is processed, otherwise the
charges are used as variable. In that case, the color charges are written as
indices, while the sum over isospins is expanded. \\
Note, \SARAH adds for every vector superfield automatically a soft-breaking
gaugino mass.
\item The next step is to define the matter sector. That's done by the array
{\tt Fields}. The conventions are the following. First, the root of the names
for the component fields is given (e.g. {\tt X}): the derived names of the
fermionic components start with {\tt F} in front (i.e. {\tt FX}), while for
scalars a {\tt S} is used (i.e. {\tt SX}). At second position the number of
generations is defined and the third entry is the name of the entire
superfield. The remaining entries are the transformation properties with
respect  to the different gauge groups. \\
Using these conventions, the doublet superfields \(\hat{q},\hat{l}, \hat{H}_d,
\hat{H}_u\) are added by
\begin{lstlisting}
Fields[[1]] = {{uL,  dL},  3, q,   1/6, 2, 3};  
Fields[[2]] = {{vL,  eL},  3, l,  -1/2, 2, 1};
Fields[[3]] = {{Hd0, Hdm}, 1, Hd, -1/2, 2, 1};
Fields[[4]] = {{Hup, Hu0}, 1, Hu,  1/2, 2, 1};
\end{lstlisting}
While for the singlet superfields \(\hat{d}, \hat{u}, \hat{e}\) 
\begin{lstlisting}
Fields[[5]] = {conj[dR], 3, d,  1/3, 1, -3};
Fields[[6]] = {conj[uR], 3, u, -2/3, 1, -3};
Fields[[7]] = {conj[eR], 3, e,    1, 1,  1};
\end{lstlisting}
is used. \\
Note, \SARAH adds also for scalars automatically the soft masses. 
\item The  superpotential of the MSSM is
\begin{equation}
\label{superpotential_MSSM}
W =  \hat{q} Y_u \hat{u} \hat{H}_u -  \hat{q} Y_d \hat{d} \hat{H}_d  - \hat{l}
   Y_e \hat{e} \hat{H}_d  +\mu \hat{H}_u \hat{H}_d
\end{equation}
and represented in \SARAH by
\begin{lstlisting}
SuperPotential = { {{1, Yu},{u,q,Hu}}, {{-1,Yd},{d,q,Hd}},
                   {{-1,Ye},{e,l,Hd}}, {{1,\[Mu]},{Hu,Hd}}  };
\end{lstlisting}
\item There are two different sets of eigenstates: the gauge eigenstates before
EWSB and the mass eigenstates after EWSB. The internal names are
\begin{lstlisting}
NameOfStates={GaugeES, EWSB};
\end{lstlisting}
\item The vector bosons and gauginos rotate after EWSB as follows
\begin{lstlisting}
DEFINITION[EWSB][GaugeSector]= 
{ {{VB,VWB[3]},{VP,VZ},ZZ},
  {{VWB[1],VWB[2]},{VWm,conj[VWm]},ZW},
  {{fWB[1],fWB[2],fWB[3]},{fWm,fWp,fW0},ZfW}
};    
\end{lstlisting}
The rotation matrices $Z^{\gamma Z}$ (\verb"ZZ"), $Z^W$ (\verb"ZW") and $Z^{\tilde{W}}$ (\verb"ZfW") are defined in the parameter file of the corresponding model as
\begin{equation}
Z^{\gamma Z} = \left(\begin{array}{cc} \cos\Theta_W & - \sin\Theta_W \\ \sin\Theta_W & \cos\Theta_W \end{array}\right)\,, \hspace{0.5cm} Z^W = Z^{\tilde{W}} = \frac{1}{\sqrt{2}}\left(\begin{array}{cc} 1 & 1 \\ -i & i \end{array}\right) 
\end{equation}
This encodes the common mixing of vector bosons and gauginos after EWSB
\begin{align} 
W_{{1 \rho}} = & \,\frac{1}{\sqrt{2}} W^-_{{\rho}}  + \frac{1}{\sqrt{2}}
W^+_{{\rho}} \, , \hspace{1cm} 
W_{{2 \rho}} =  \,-i \frac{1}{\sqrt{2}} W^-_{{\rho}}  + i \frac{1}{\sqrt{2}}
W^+_{{\rho}} \\ 
W_{{3 \rho}} = & \,\cos\Theta_W  Z_{{\rho}}  + \sin\Theta_W  \gamma_{{\rho}}
\, , \hspace{1cm}  
B_{{\rho}} =  \,\cos\Theta_W  \gamma_{{\rho}}  - \sin\Theta_W  Z_{{\rho}} \\ 
\lambda_{{\tilde{W}},{1}} = & \,\frac{1}{\sqrt{2}} \tilde{W}^-  +
\frac{1}{\sqrt{2}} \tilde{W}^+ \, , \hspace{1cm}  
\lambda_{{\tilde{W}},{2}} =  \,-i \frac{1}{\sqrt{2}} \tilde{W}^-  + i
\frac{1}{\sqrt{2}} \tilde{W}^+ \, , \hspace{1cm}  
\lambda_{{\tilde{W}},{3}} =  \,\tilde{W}^0
\end{align} 

\item The neutral components of the scalar Higgs receive vacuum expectation
values (VEVs) \(v_d\)/\(v_u\) and split into scalar and pseudo scalar components
\begin{align} 
H_d^0 =  \, \frac{1}{\sqrt{2}} \left( \phi_{d}  + i \sigma_{d}  +  v_d  \right)
\, , \hspace{1cm} 
H_u^0 =  \, \frac{1}{\sqrt{2}} \left( \phi_{u}    + i  \sigma_{u}  +  v_u
\right) 
\end{align} 
This is encoded in \SARAH by
\begin{lstlisting}
DEFINITION[EWSB][VEVs]= 
{{SHd0,{vd,1/Sqrt[2]},{sigmad,I/Sqrt[2]},{phid,1/Sqrt[2]}},
 {SHu0,{vu,1/Sqrt[2]},{sigmau,I/Sqrt[2]},{phiu,1/Sqrt[2]}}};
\end{lstlisting}
\item The particles mix after EWSB to new mass eigenstates
\begin{lstlisting}
DEFINITION[EWSB][MatterSector]= 
{{{SdL, SdR           }, {Sd, ZD}},
 {{SuL, SuR           }, {Su, ZU}},
 {{SeL, SeR           }, {Se, ZE}},
 {{SvL                }, {Sv, ZV}},
 {{phid, phiu         }, {hh, ZH}},
 {{sigmad, sigmau     }, {Ah, ZA}},
 {{SHdm, conj[SHup]   }, {Hpm,ZP}},
 {{fB, fW0, FHd0, FHu0}, {L0, ZN}}, 
 {{{fWm, FHdm}, {fWp, FHup}}, {{Lm,U},  {Lp,V}}},
 {{{FeL},       {conj[FeR]}}, {{FEL,ZEL},{FER,ZER}}},
 {{{FdL},       {conj[FdR]}}, {{FDL,ZDL},{FDR,ZDR}}},
 {{{FuL},       {conj[FuR]}}, {{FUL,ZUL},{FUR,ZUR}}} }; 
\end{lstlisting}
This defines the mixings to the mass eigenstates: first, a list with gauge
eigenstates is given, then the name of the new mass eigenstates and the
mixing matrix follows. Hence, the first line is interpreted as
\begin{align} 
d_{L,{i \alpha}} = \sum_j U^{d,*}_{L,{j i}}D_{L,{j \alpha}}\,, \hspace{1cm}  
d_{R,{i \alpha}} = \sum_j U_{R,{i j}}^{d}D^*_{R,{j \alpha}} \, ,
\end{align} 
while the 8th line defines the mixing in the chargino sector
\begin{align} 
\tilde{W}^- = \sum_j U^*_{j 1}\lambda^-_{{j}}\,, \hspace{1cm} 
\tilde{H}_d^- = \sum_j U^*_{j 2}\lambda^-_{{j}} \,, \hspace{1cm}  
\tilde{W}^+ = \sum_j V^*_{1 j}\lambda^+_{{j}}\,, \hspace{1cm} 
\tilde{H}_u^+ = \sum_j V^*_{2 j}\lambda^+_{{j}}
\end{align} 
\item No particles should be integrated out or deleted
\begin{lstlisting}
IntegrateOut={};
DeleteParticles={};
\end{lstlisting}
\item The Dirac spinors for the mass eigenstates are 
\begin{lstlisting}
DEFINITION[EWSB][DiracSpinors]={
 Fd - > {FDL, conj[FDR]},
 Fe  -> {FEL, conj[FER]},
 Fu  -> {FUL, conj[FUR]},
 Fv  -> {FvL, 0},
 Chi -> {L0, conj[L0]},
 Cha -> {Lm, conj[Lp]},
 Glu -> {fG, conj[fG]}
};
\end{lstlisting}
That leads to the replacements
\begin{equation}
d \rightarrow \left(\begin{array}{c} d_L \\ d_R \end{array} \right)\,, \dots \,,
\tilde{\chi}^- \rightarrow \left(\begin{array}{c} \lambda^- \\
(\lambda^+)^* \end{array} \right)\,, 
\tilde{g} \rightarrow \left(\begin{array}{c} \lambda_g \\ \lambda_g^*
\end{array} \right)\
\end{equation}
when going from four- to two-component formalism.
\end{enumerate}

\chapter{Verification of Output}
For details about our checks to verify and validate the output of \SARAH we refer to the following references:
\begin{itemize}
\item Renormalization group equations and self-energies see Ref.~\cite{Staub:2010jh,Staub:2010ty}
 \item \FeynArts/\FormCalc as well as \CalcHep/\CompHep see Ref.~\cite{Staub:2009bi}
 \item \WHIZARD and \SPheno, output see Ref.~\cite{Staub:2011dp,Staub:2010ty}
 \item UFO output, see Ref.~\cite{Staub:2012pb}
 \end{itemize}

\chapter{Evaluation Time}

To give an impression for the needed evaluation time for different routines and
models, we collected some values in Table~\ref{time}. This times were measured
under Mathematica 5.2 running on an a Lenovo Thinkpad X220 with a Intel i7-2620M (2.70GHz) core. 

% \begin{table}[h]
% \begin{tabular}{|c|ccccc|}
% \hline
% Command & MSSM (No FV) & MSSM-CKM & NMSSM \footnotemark[1] & \(\mu\nu\)SSM &
% MSSM + \(U(1)\) \\
% \hline
% \verb"Start" & 12.75 & 18.03 & 19.02 & 27.06 & 16.14\\
% \verb"ModelOutput[EWSB]" & 74.83 & 78.70 & 94.64 & 115.08 & 110.47 \\
% \verb"MakeFeynArts" & 0.74 & 3.58 & 1.12 & 0.98 & 0.48\\
% \verb"MakeCalcHep[]" &  6.03 & 22.74 & 15.57 & 47.08 & 6.7 \\
% \verb"MakeTeX[]" & 0.81 & 5.79  & 1.25 & 1.38 & 1.48 \\
% \hline
% \verb"CalcRGEs[]" & 50.72 & 50.8  & 91.07 & 265.29 & 68.18 \\
% \verb"CalcLoopCorrections[EWSB]" & 7.07 & 28.44 & 8.14 & 7.98 & 8.84  \\
% \hline 
% \verb"Vertex[{bar[Fd],Fd,VP}]" & 0.05 & 0.06  & 0.06 & 0.06 & 0.05\\
% \verb"Vertex[{Sd,Sd*,Su,Su*}]" & 0.54 & 0.61 & 0.56 & 0.56 & 0.52 \\
% \verb"Vertex[{VG,VG,VG,VG}]" & 0.03 & 0.03 & 0.04 & 0.03 & 0.03\\
% \verb"Vertex[{Se,Se,VZ}]" & 0.37 & 0.37 & 0.45 & 1.13  & 0.37\\
% \verb"Vertex[{bar[gWmC],gP,VWm*}]" & 0.02 & 0.02  & 0.02 & 0.02  & 0.02\\
% \verb"Vertex[{Hpm,VWm*,VZ}]" & 0.16 & 0.17 & 0.26 & 0.23 & 0.21\\
% \hline
% \verb"TreeMass[VZ,EWSB]" & 0.15 & 0.19 & 0.20 & 0.25 & 0.48\\
% \verb"TreeMass[Su,EWSB]" & 0.25 & 0.27  & 0.29 & 0.34 & 0.23\\
% \hline
% \end{tabular}
% \caption{Time needed in seconds to evaluate several commands of \SARAH in
% Mathematica 5.2}
% \label{time}
% \end{table}

\begin{table}[h]
\begin{tabular}{|c|ccccc|}
\hline
Command & MSSM  &  NMSSM  & B-L-SSM & Seesaw 3 \\
\hline
{\tt Start}                     & 8.9   & 10.3 & 34.0  & 18.5 \\
{\tt ModelOutput[EWSB]}         & 49.7  & 62.6 & 445.3 & 83.3  \\
{\tt CalcRGEs[]}                & 11.2  & 16.9 & 82.5  & 60.3 \\ 
{\tt CalcLoopCorrections[EWSB]} & 5.9   & 6.9  & 55.0  & 9.8  \\
{\tt MakeFeynArts[]}            & 0.5   & 0.6  & 5.5   & 1.7  \\
{\tt MakeCHep[]}                & 10.8  & 12.7 & 98.7  & 19.3  \\
{\tt MakeUFO[]}                 & 38.5  & 62.8 & 887.5 &  80.6 \\
{\tt MakeWHIZARD[]}             & 274.8 & 538.9& 4326.5& 486.8  \\
{\tt MakeSPheno[]}              & 134.1 & 196.5& 1024.7& 292.3  \\
{\tt MakeTeX[]}                 & 3.9   & 6.3  & 61.6  & 7.1  \\
\hline
\end{tabular}
\caption{Time needed in seconds to evaluate several commands of \SARAH in
Mathematica 5.2}
 \label{time}
\end{table}
\chapter{Models included in the public version}

\begin{itemize}
 \item Minimal supersymmetric standard model (see Ref.~\cite{Martin:1997ns} and references therein):  
    \begin{itemize} 
     \item With general flavor and CP structure ({\tt MSSM})
     \item Without flavor violation ({\tt MSSM/NoFV})
     \item With explicit CP violation in the Higgs sector ({\tt MSSM/CPV})
     \item In SCKM basis ({\tt MSSM/CKM})
    \end{itemize}
   \item Singlet extensions: 
   \begin{itemize}
    \item Next-to-minimal supersymmetric standard model ({\tt NMSSM},
 {\tt NMSSM/NoFV}, {\tt NMSSM/CPV}, {\tt NMSSM/CKM}) (see Refs.~\cite{Maniatis:2009re,Ellwanger:2009dp} and references therein)
    \item near-to-minimal supersymmetric standard model
 ({\tt near-MSSM}) \cite{Barger:2006dh}
    \item General singlet extended, supersymmetric standard model ({\tt SMSSM})  \cite{Barger:2006dh,Ross:2012nr}
  \end{itemize}
  \item Triplet extensions 
  \begin{itemize} 
    \item Triplet extended MSSM ({\tt TMSSM}) \cite{DiChiara:2008rg}
    \item Triplet extended NMSSM ({\tt TNMSSM}) \cite{Agashe:2011ia}
  \end{itemize}
   \item Models with $R$-parity violation  \cite{Hall:1983id,Dreiner:1997uz,Allanach:2003eb,Bhattacharyya:1997vv,Barger:1989rk,Allanach:1999ic,Hirsch:2000ef,Barbier:2004ez}
  \begin{itemize}
    \item bilinear RpV ({\tt MSSM-RpV/Bi}) 
    \item Lepton number violation ({\tt MSSM-RpV/LnV})
    \item Only trilinear lepton number violation ({\tt MSSM-RpV/TriLnV})
    \item Baryon number violation ({\tt MSSM-RpV/BnV})  
    \item $\mu\nu$SSM ({\tt munuSSM}) \cite{LopezFogliani:2005yw,Bartl:2009an}
  \end{itemize}
   \item Additional $U(1)'s$ 
  \begin{itemize}
    \item $U(1)$-extended MSSM ({\tt UMSSM})  \cite{Barger:2006dh}
    \item secluded MSSM ({\tt secluded-MSSM}) \cite{Chiang:2009fs}
    \item minimal $B-L$ model ({\tt B-L-SSM})  \cite{Khalil:2007dr,FileviezPerez:2010ek,O'Leary:2011yq,Basso:2012gz}
    \item minimal singlet-extended $B-L$ model ({\tt N-B-L-SSM})
  \end{itemize}
   \item SUSY-scale seesaw extensions
    \begin{itemize}
      \item inverse seesaw ({\tt inverse-Seesaw}) \cite{Malinsky:2005bi,Abada:2012cq,Dev:2012ru}
      \item linear seesaw ({\tt LinSeesaw}) \cite{Malinsky:2005bi,DeRomeri:2012qd}
      \item singlet-extended inverse seesaw ({\tt inverse-Seesaw-NMSSM})
      \item inverse seesaw with $B-L$ gauge group ({\tt B-L-SSM-IS}) 
      \item minimal $U(1)_R \times U(1)_{B-L}$ model with inverse seesaw
 ({\tt BLRinvSeesaw}) \cite{Hirsch:2011hg,Hirsch:2012kv}
\end{itemize}
 \item Models with Dirac Gauginos
   \begin{itemize}
    \item MSSM/NMSSM with Dirac Gauginos ({\tt DiracGauginos}) \cite{Belanger:2009wf,Benakli:2010gi,Benalki:2012}
    \item minimal R-Symmstric SSM ({\tt MRSSM}) \cite{Kribs:2007ac}
   \end{itemize}
 \item High-scale extensions
\begin{itemize}
 \item Seesaw 1 - 3 ($SU(5)$ version) ,
 ({\tt Seesaw1},{\tt Seesaw2},{\tt Seesaw3}) \cite{Borzumati:2009hu,Rossi:2002zb,Hirsch:2008dy,Esteves:2009vg,Esteves:2010ff}
 \item Left/right model ($\Omega$LR) ({\tt Omega}) \cite{Esteves:2010si,Esteves:2011gk}
\end{itemize}
\item Non-SUSY models:
\begin{itemize}
\item SM ({\tt SM}, {\tt SM/CKM}) (see for instance  Ref.~\cite{Hollik:2010id} and references therein)
\item inert doublet model ({\tt Inert}) \cite{LopezHonorez:2006gr}
\end{itemize}
\end{itemize}
\newpage

\addcontentsline{toc}{chapter}{References}
\bibliographystyle{hunsrt}
\bibliography{sarah}

\end{appendix}

\end{document}